\documentclass{aa}  
\usepackage{psfig,longtable,lscape,amssym,natbib}  
\newcommand{\ltsima} {$\; \buildrel < \over \sim \;$}  
\newcommand{\gtsima} {$\; \buildrel > \over \sim \;$}  
\newcommand{\lta} {\lower.5ex\hbox{\ltsima}}  
\newcommand{\gta} {\lower.5ex\hbox{\gtsima}}  
\newcommand{\Ha} {H$\alpha$}  
\newcommand{\Hb} {H$\beta$}

\begin{document}

\title{An optical spectroscopic survey of the 3CR sample of radio galaxies
  with $z<0.3$. I.  Presentation of the data\thanks{Based on observations made
    with the Italian Telescopio Nazionale Galileo operated on the island of La
    Palma by the Centro Galileo Galilei of INAF (Istituto Nazionale di
    Astrofisica) at the Spanish Observatorio del Roque del los Muchachos of
    the Instituto de Astrofısica de Canarias.}}

\titlerunning{Optical spectra of 3CR sources} \authorrunning{S. Buttiglione et
  al.}
  
\author{Sara Buttiglione \inst{1} \and Alessandro Capetti \inst{2} \and
  Annalisa Celotti \inst{1} \and David J. Axon \inst{3} \and Marco Chiaberge
  \inst{4} \and F. Duccio Macchetto \inst{4} \and William B. Sparks \inst{4} }
   
\offprints{S. Buttiglione}
     
\institute{SISSA-ISAS, Via Beirut 2-4, I-34014 Trieste, Italy\\
  \email{buttigli@sissa.it} \and INAF - Osservatorio Astronomico di Torino,
  Strada Osservatorio 20, I-10025 Pino Torinese, Italy \and Department of
  Physics, Rochester Institute of Technology, 85 Lomb Memorial Drive,
  Rochester, NY 14623, USA \and Space Telescope Science Institute, 3700 San
  Martin Drive, Baltimore, MD 21218, U.S.A.}

\date{}

\abstract{ We present a homogeneous and 92 \% complete dataset of optical
  nuclear spectra for the 113 3CR radio sources with redshifts $<$ 0.3,
  obtained with the Telescopio Nazionale Galileo.  For these sources we could
  obtain uniform and uninterrupted coverage of the key spectroscopic optical
  diagnostics. The observed sample, including powerful classical FR~II
    radio-galaxies and FR~I, together spanning four orders of magnitude
    in radio-luminosity, provides a broad representation of the spectroscopic
    properties of radio galaxies. In this first paper we present an atlas of
  the spectra obtained, provide measurements of the diagnostic emission line
  ratios, and identify active nuclei with broad line emission. These data will
  be used in follow-up papers to address the connection between the optical
  spectral characteristics and the multiwavelength properties of the sample.

  }

\maketitle
  
\section{Introduction}
\label{introduction}

Radio galaxies (RG) are an important class of extragalactic objects for many
reasons. They are extremely interesting in their own right, representing one
of the most energetic astrophysical phenomena. The intense nuclear emission
and collimated jets can influence the star formation history, excitation, and
ionization of the ISM, representing a fundamental ingredient for the formation
and evolution of their hosts but also of their larger scale environment. The
host galaxies are giant early-type galaxies, often situated centrally in
galaxy clusters; indeed, with their symbiotic relationship, central dominant
ellipticals and the AGN phenomenon encompass many of the major cosmological
and astrophysical issues we confront today. These include the ubiquity and
growth of supermassive black holes, their relationship to the host, the onset
of nuclear activity, the nature of their jets and the interaction with the
interstellar medium, and the physics of the central engine itself.  In
particular, studies of radio-loud AGN are the key to understanding the
processes leading to the ejection of material in relativistic jets, its
connection with gas accretion onto the central black holes, the way that
different levels of accretion are related to the process of jets launching,
the origin of the AGN onset, and its lifetime.

The attractiveness of the 3CR catalog of radio sources as a basis for such a
study is obvious, being one of the best studied sample of RG. Its
selection criteria are unbiased with respect to optical properties and
orientation, and it spans a relatively wide range in redshift and radio power.
A vast suite of observations is available for this sample, from multi-band HST
imaging, to observations with Chandra, Spitzer and the VLA, that can be used
to address the issues listed above in a robust multiwavelength approach.

Quite surprisingly, despite the great interest of the community, the
available optical spectroscopic data for the 3CR sample are sparse and
incomplete. 

To fill this gap, we carried out a homogeneous and complete survey of optical
spectroscopy. We targeted the complete subsample of 113 3CR radio sources
with z$<$0.3, for which we can obtain uniform uninterrupted coverage of the
key spectroscopic optical diagnostics. The observed sources include a
significant number of powerful classical FR~II RG, as well as the more common
(at low redshift) FR~Is \citep{fanaroff74}, spanning four orders of
magnitude in radio luminosity, thus providing a broad representation of the
spectroscopic properties of radio-loud AGN.

The primary goals of this survey are (we defer the detailed list of
bibliographical references to the follow-up papers dealing with the specific
issues):

\begin{itemize}
\item[--] to measure luminosities and ratios of the key diagnostic emission
  line; these will be used to prove the possible existence (and to
  characterize the properties) of different spectroscopic
  subpopulations of radio-loud AGN;
\item[--] to explore the presence of broad Balmer lines in the different
  subclasses of RGs, separated on the basis of e.g. radio morphology and
  diagnostic line ratios;
\item[--] to perform a detailed test on the unified models for RL AGN, taking
  advantage of the uniform dataset, of the sample completeness, and of the
  accurate spectral characterization;
\item[--] to compare the spectral properties of radio-loud and radio-quiet
  AGN;
\item[--] to analyze the host galaxy properties from the point of view of its
  stellar component, looking for connections between the nuclear
  activity (and its onset) and the star formation history of the host;
\item[-- ] to address the connection between the optical spectral
  characteristics and the multiwavelength properties of the sample, derived 
  combining HST, Chandra, Spitzer and VLA data;
\item[--] to study the environment of radio sources (using the 
target acquisition images); 
\item[--] to explore the properties of the off-nuclear gas (e.g. spatially
  resolved line ratios and velocity fields) and stellar component by using the
  full two dimensional dataset.
\end{itemize}

The paper is organized as follows: in Sect. \ref{sect1} we present the
sample, the observational procedure and the data reduction,
leading to an atlas of calibrated spectra. In
Sect. \ref{sect3} we provide measurements of the emission line fluxes,
obtained after separating the starlight from the emission of the
active nucleus. In Sect. \ref{sect4} we describe the quality of the resulting
measurements and describe in more details the RGs showing broad emission
lines. A summary is given in Sect. \ref{summary}.

Throughout, we have used $H_o = 71$ km s$^{-1}$ Mpc$^{-1}$, $\Omega_{\Lambda}
= 0.73$ and $\Omega_m = 0.27$.

\section{Observations and data reduction}
\label{sect1}

\subsection{Sample selection}
\label{sample} 

The Third Cambridge (3C) Catalogue of radio sources was compiled by
\citet{edge59}: they used the Cambridge double Interferometer to detect all
the radio emitters at a frequency of 159 MHz, resulting in 471 sources
catalogued.\\
\citet{bennett62b, bennett62a} revised the catalog creating the 3CR version.
He obtained new observational data to provide a complete and reliable list for
all sources with a flux density greater than F$_r$ $>$ 9 Jy at a frequency of
178 MHz and for which the declination is $\delta > -05^{\circ}$ degrees (apart
from some limitations for sources of low surface brightness).  The revision
excluded some sources from the catalog (as being below the flux limit or as
resolved blends of adjacent sources) and included other sources (as
extended sources with diameters up to 1$^o$). 328 sources are catalogued in
the 3CR.  \citet{spinrad85} identified the optical counterparts and measured
the redshift of the 298 extragalactic sources in the 3CR.  Our sample
comprises all of the 113 3CR sources from \citet{spinrad85} with $z < 0.3$.  We
only excluded 3C~231 (M82, a starburst galaxy) and 3C~71 (NGC~1068, a
Seyfert galaxy). \\
Here we present the optical spectra of 104 3CR objects.

\subsection{Observations}
\label{obs}

All but 18 optical spectra were taken with the Telescopio Nazionale Galileo
(TNG), a 3.58 m optical/infrared telescope located on the Roque de los
Muchachos in La Palma Canary Island (Spain).  The observations were made using
the DOLORES (Device Optimized for the LOw RESolution) spectrograph
installed at the Nasmyth B focus of the telescope.  The observations were
organized into four runs of 4-5 nights per run starting from November 2006
and ending on January 2008. The chosen long-slit width is 2$\arcsec$: it
is centered on the nucleus of the sources and it is aligned along the
parallactic angle in order to minimize light losses due to atmospheric
dispersion.

\begin{table}
  \begin{center}
    \caption{Detectors used for the survey}
    \label{ccd}
    \begin{tabular}{c | c c c c}
      \hline \hline
      CCD  & type & pixel size & scale & FOV \\
      2100x2100 pix  &  & $\mu$m & $\arcsec$ pix$^{-1}$ & $\arcmin$ \\
      \hline
      1 &Loral   & 15.0 &  0.275 & 9.40\\
      2 &E2V4240 & 13.5 &  0.252 & 8.60\\
      \hline
    \end{tabular}
  \end{center}
\end{table}

For the first run (22-25 November 2006) and the second run (12 April 2007),
the detector was a Loral thinned and back-illuminated CCD.  For the third run
(6-18 August 2007) and the fourth run (4-8 January 2008) a new thinned,
back-illuminated, E2V 4240 CCD had been installed. It has a smaller pixel size
and the field of view is decreased by $\sim$1 arcmin.
More details on the CCD detectors are listed in Table \ref{ccd}.\\

\begin{table}
  \begin{center}
    \caption{Grisms used for the observations}
    \label{grism}
    \begin{tabular}{l c c c c}
      \hline \hline
Grism & dispersion  & Resol. & $\lambda$ range & $\lambda$ range \\
       & (\AA\ pixels$^{-1}$)  &(2$\arcsec$ slit) & CCD 1 & CCD 2 \\
      \hline
      LR-B  & 2.52 & 1200 & 3500-8050 & 3500-7700 \\
      VHR-R & 0.70 & 5000 & 6050-7800 & 6100-7700 \\
      VHR-I & 0.68 & 6000 & 7200-8900 & 7250-8800 \\
      \hline
    \end{tabular}
  \end{center}
\end{table}

\begin{table}
  \begin{center}
    \caption{Observational strategy}
    \label{obstrat}
    \begin{tabular}{r c l c}
      \hline \hline
      redshift range & T$_{\rm exp(LRB)}$ (s) &HR grism & T$_{\rm exp(HR)}$ (s) \\
      \hline
      $z<0.100$       & 500& VHR-R &1000 \\
      $0.100<z<0.175$ & 750& VHR-R &1500 \\
      $0.175<z<0.200$ & 750& VHR-I &1500 \\
      $0.200<z<0.300$ &1000& VHR-I &2000 \\
      \hline
    \end{tabular}
  \end{center}
\end{table}

For each target we took:
\begin{itemize}
\item[--] one acquisition image (240 sec) in R band;
\item[--] one (or two) low resolution spectrum with the LR-B grism
  ($\sim$3500-8000 \AA) with a resolution of $\sim$20 \AA;
\item[--] two high resolution spectra with the VHR-R (6100-7800 \AA) or VHR-I
  (7200-8900 \AA) grisms with a resolution of $\sim$5 \AA.
\end{itemize}
The 
spectral coverages of the spectra are reported in Table \ref{grism}. 
The grism wavelength ranges change slightly between the two CCDs because of
the small difference in the pixel size.  The observational
strategy is summarized in Table \ref{obstrat}.  
In order to compensate for the galaxies' dimming with $(1+z)^4$, the exposure
times depend on the source redshifts. We divided the targets into three groups:
\begin{itemize}
\item[--] for $z<0.1$ t$_{exp}$(LR-B)$=$500 sec;
\item[--] for $0.1<z<0.2$ t$_{exp}$(LR-B)$=$750 sec;
\item[--] for $0.2<z<0.3$ t$_{exp}$(LR-B)$=$1000 sec.
\end{itemize}
The 1000 sec exposures were divided into two subexposures of 500 sec.
The high resolution spectra have an exposure time twice the low resolution
ones. For each target, the VHR-R or VHR-I grisms are chosen in such a way
that the [O~I]$\lambda\lambda$6300,64 \AA\ lines are observable, i.e.:
\begin{itemize}
\item[--] VHR-R for $z<0.175$;
\item[--] VHR-I for $z>0.175$.
\end{itemize}
The combination of the LR-B and VHR ranges of wavelengths enables us to cover
the most relevant emission lines of the optical spectrum and in particular the
key diagnostic lines H$\beta$, [O~III]$\lambda\lambda$4959,5007,
[O~I]$\lambda\lambda$6300,64, H$\alpha$, [N~II]$\lambda\lambda$6548,84,
[S~II]$\lambda\lambda$6716,31.  The high resolution spectra are a sort of {\it
  zoom} on the H$\alpha$ region with the aim of resolving the  
H$\alpha$ from the [N~II] doublet, as well as the two lines of the [S~II]
doublet.

Table \ref{logoss} provides the journal of observations for the four runs and
the main information on the sources: (1) name of the source; (2) and (3) J2000
coordinates; (4) redshift; (5) date of observation; (6) CCD used in reference
to Table \ref{ccd}; (7) number of low resolution spectra; (8) exposure time
for each low resolution spectrum; (9) high resolution grism used; (10) number
of high resolution spectra; (11) exposure time for each high resolution
spectrum. In some cases the exposure times were increased because of bad sky
conditions (i.e. seeing $>$ 2.5$\arcsec$): these sources are identified with
an ``a'' in the note column of Table \ref{logoss}.

Nine sources of our sample (namely 3C~020, 3C~063, 3C~132, 3C~288, 3C~346,
3C~349, 3C~403.1, 3C~410, 3C~458) could not be observed due to scheduling
problems and time constraints. Since these sources were excluded for random
reasons from the observing list we believe that they do not induce any bias in
the sample. We verified a posteriori that they are apparently randomly
distributed in the sky, as well as within the distributions of radio power and
redshift of the whole sample.

\subsection{Data reduction}
\label{dred}
The LONGSLIT package of NOAO's IRAF\footnote{IRAF is the Image Reduction and
  Analysis Facility of the National Optical Astronomy Observatories, which are
  operated by AURA, Inc., under contract with the U.S. National Science
  Foundation. It is also available at http://iraf.noao.edu/} reduction
software was used in order to:
\begin{itemize}
\item[--] subtract the bias from each science image (we used the over-scan
  region of each image);
\item[--] divide the science images for the flat field (created from the
  halogen lamp emission);
\item[--] subtract the background; most observations were split into two
  subexposures (with only the exception of the LR-B spectra of low redshift
  galaxies, see Table \ref{logoss}) obtained moving the target by 25$\arcsec$
  or 50$\arcsec$ along the slit. Subtraction of the two subexposures removes
  most of the sky background. The residual background (due to changes in sky
  brightness or for the single LR-B spectra) was subtracted measuring the
  average on each pixel along the dispersion direction in spatial regions
  immediately surrounding the source spectrum;
\item[--] wavelength calibrate the images using the calibration lamps (Ar, Ne
  and He);
\item[--] correct the optical distortions by
fitting polynomial functions along both
the horizontal and vertical directions of the CCD to the lamp spectra;
\item[--] flux calibrate the spectra using spectrophotometric standard stars.
  We observed two or three standard stars per night using the same telescope
  configuration used for the targets.
\end{itemize}

We then extracted and summed a region of 2$\arcsec$ along the spatial
direction, resulting in a region covered by our spectra of 2$\arcsec$ x
2$\arcsec$.  For each galaxy, multiple exposures taken with the
same setting were averaged into a single spectrum.\\
The accuracy of the relative flux calibration was estimated from the residuals
of the calibration of the standard stars; comparison of the calibrated spectra
obtained using the different standard stars
observed during the same night implies that our flux calibration agrees to
within a $<$5\% level. \\
More than 75\% of the nights were photometric and we 
therefore expect a rather
stable absolute flux calibration. This result was checked for 20 objects of the sample randomly selected using their HST R-band
images 

\begin{table*}
  \begin{center}
    \caption{Log of the observations}
    \label{logoss}
    \begin{tabular}{l| c c| c| c c| c c| c c c|c}
      \hline \hline
Name &\multicolumn{2}{|c|}{ Coordinates J2000}& z &\multicolumn{2}{|c|}{  Telescope information}  & \multicolumn{2}{|c|}{ Low Res.}
&\multicolumn{3}{|c|}{  High Res. }  & Notes \\
\hline
     &  $\alpha$  & $\delta$ &  & Date Obs. & CCD & n& T$_{exp}$ & HR & n &  T$_{exp}$ &  \\ \hline 
3C~015   &  00 37 04.15  &  --01 09 08.10&  0.073    &  09Aug07     &  2  &  1 & 500    & HRR &  2 &500 & i \\
3C~017   &  00 38 20.48  & --02 07 40.21 &  0.2198   &  06Jan08     &  2  &  2 & 500    & HRI &  2 &1000& f,g \\
3C~018	 &  00 40 50.51  &  +10 03 27.68 &  0.188    &  22Nov06     &  1  &  1 & 900    & HRI &  2 &900 & c,f,g \\
3C~028	 &  00 55 50.65  &  +26 24 36.93 &  0.1952   &  25Nov06     &  1  &  1 & 750    & HRI &  2 &750 &  \\
3C~029	 &  00 57 34.88  & --01 23 27.55 &  0.0448   &  23Nov06     &  1  &  1 & 500    & HRR &  2 &500 & l \\
3C~031   &  01 07 24.99  &  +32 24 45.02 &  0.0167   &  10Aug07     &  2  &  1 & 500    & HRR &  2 &500 & i \\
3C~033	 &  01 08 52.87  &  +13 20 14.52 &  0.0596   &  22Nov06     &  1  &  1 & 600    & HRR &  2 &600 & c \\
3C~033.1 &  01 09 44.27  &  +73 11 57.20 &  0.1809   &  08Aug07     &  2  &  1 & 750    & HRI &  2 &750 & a,f,h \\
3C~035	 &  01 12 02.29  &  +49 28 35.33 &  0.0670   &  25Nov06     &  1  &  1 & 500    & HRR &  2 &500 &  \\
3C~040	 &  01 26 00.62  & --01 20 42.43 &  0.0185   &  25Nov06     &  1  &  2 & 500    & HRR &  4 &500 & a,i \\
3C~052	 &  01 48 28.90  &  +53 32 27.90 &  0.2854   &  23Nov06     &  1  &  2 & 500    & HRI &  2 &1000& i \\
3C~061.1 &  02 22 36.00  &  +86 19 08.00 &  0.184    &  05Jan08     &  2  &  1 & 900    & HRI &  2 &900 & a \\
3C~066B	 &  02 23 11.46  &  +42 59 31.34 &  0.0215   &  25Nov06     &  1  &  1 & 500    & HRR &  2 &500 & i \\
3C~075N	 &  02 57 41.55  &  +06 01 36.58 &  0.0232   &  23Nov06     &  1  &  1 & 500    & HRR &  2 &500 & i \\
3C~076.1 &  03 03 15.00  &  +16 26 19.85 &  0.0324   &  05Jan08     &  2  &  1 & 500    & HRR &  2 &500 & e,i \\
3C~078	 &  03 08 26.27  &  +04 06 39.40 &  0.0286   &  04Jan08     &  2  &  1 & 500    & HRR &  2 &500 & e \\
3C~079	 &  03 10 00.10  &  +17 05 58.91 &  0.2559   &  23Nov06     &  1  &  2 & 500    & HRI &  2 &1000&  \\
3C~083.1 &  03 18 15.80  &  +41 51 28.00 &  0.0255   &  08Jan08     &  2  &  1 & 500    & HRR &  2 &500 & i \\
3C~084	 &  03 19 48.22  &  +41 30 42.10 &  0.0176   &  06Jan08     &  2  &  1 & 500    & HRR &  2 &500 & h \\
3C~088	 &  03 27 54.17  &  +02 33 41.82 &  0.0302   &  25Nov06     &  1  &  1 & 500    & HRR &  2 &500 &  \\
3C~089   &  03 34 15.57  & --01 10 56.40 &  0.1386   &  06Jan08     &  2  &  1 & 750    & HRI &  2 &750 &  \\
3C~093.1 &  03 48 46.90  &  +33 53 15.00 &  0.2430   &  05Jan08     &  2  &  2 & 500    & HRI &  2 &1000&  \\
3C~098   &  03 58 54.43  &  +10 26 03.00 &  0.0304   &  22Nov06     &  1  &  1 & 500    & HRR &  2 &500 &  \\
3C~105	 &  04 07 16.46  &  +03 42 25.68 &  0.089    &  06Jan08     &  2  &  1 & 500    & HRR &  2 &500 &  \\
3C~111	 &  04 18 21.05  &  +38 01 35.77 &  0.0485   &  23Nov06     &  1  &  1 & 500    & HRR &  2 &500 & f,h \\
3C~123	 &  04 37 04.40  &  +29 40 13.20 &  0.2177   &  06Jan08     &  2  &  2 & 500    & HRI &  2 &1000&  \\
3C~129	 &  04 49 09.07  &  +45 00 39.00 &  0.0208   &  25Nov06     &  1  &  1 & 500    & HRR &  2 &500 &  \\
3C~129.1 &  04 50 06.70  &  +45 03 06.00 &  0.0222   &  22Nov06     &  1  &  1 & 500    & HRR &  2 &500 &  \\
3C~130	 &  04 52 52.78  &  +52 04 47.53 &  0.1090   &  04Jan08     &  2  &  1 & 750    & HRR &  2 &750 &  \\
3C~133	 &  05 02 58.40  &  +25 16 28.00 &  0.2775   &  05Jan08     &  2  &  2 & 500    & HRI &  2 &1000&  \\
3C~135	 &  05 14 08.30  &  +00 56 32.00 &  0.1253   &  25Nov06     &  1  &  1 & 500    & HRR &  2 &500 &  \\
3C~136.1 &  05 16 03.16  &  +24 58 25.24 &  0.064    &  23Nov06     &  1  &  1 & 500    & HRR &  2 &500 &  \\
3C~153   &  06 09 32.50  &  +48 04 15.50 &  0.2769   &  06Jan08     &  2  &  3 & 500    & HRI &  3 &1000& a \\
3C~165	 &  06 43 06.60  &  +23 19 03.00 &  0.2957   &  08Jan08     &  2  &  2 & 500    & HRI &  2 &1000&  \\
3C~166	 &  06 45 24.10  &  +21 21 51.00 &  0.2449   &  25Nov06     &  1  &  2 & 500    & HRI &  2 &1000&  \\
3C~171	 &  06 55 14.72  &  +54 08 58.27 &  0.2384   &  22Nov06     &  1  &  2 & 500    & HRI &  2 &1000&  \\ 
3C~173.1 &  07 09 24.34  &  +74 49 15.19 &  0.2921   &  08Jan08     &  2  &  2 & 500    & HRI &  2 &1000&  \\
3C~180   &  07 27 04.77  & --02 04 30.97 &  0.22     &  12Apr07     &  1  &  2 & 500    & HRI &  2 &1000&  \\	 
3C~184.1 &  07 43 01.28  &  +80 26 26.30 &  0.1182   &  04Jan08     &  2  &  1 & 750    & HRR &  2 &750 & f \\
3C~192	 &  08 05 35.00  &  +24 09 50.00 &  0.0598   &  \multicolumn{2}{c|}{SDSS}  & -- & --     & --  & -- & -- &  \\
3C~196.1 &  08 15 27.73  & --03 08 26.99 &  0.198    &  05Jan08     &  2  &  2 & 750+375& HRI &  3 &750 & a \\
3C~197.1 &  08 21 33.70  &  +47 02 37.00 &  0.1301   &  \multicolumn{2}{c|}{SDSS}  &    & --     & --  & -- & -- & f,g \\
3C~198	 &  08 22 31.90  &  +05 57 07.00 &  0.0815   &  \multicolumn{2}{c|}{SDSS}  & -- & --     & --  & -- & -- &  \\
3C~213.1 &  09 01 05.30  &  +29 01 46.00 &  0.194    &  \multicolumn{2}{c|}{SDSS}  & -- & --     & --  & -- & -- &  \\
3C~219	 &  09 21 08.64  &  +45 38 56.49 &  0.1744   &  \multicolumn{2}{c|}{SDSS}  & -- & --     & --  & -- & -- & f,g \\
3C~223	 &  09 39 52.76  &  +35 53 59.12 &  0.1368   &  \multicolumn{2}{c|}{SDSS}  & -- & --     & --  & -- & -- &  \\
3C~223.1 &  09 41 24.04  &  +39 44 42.39 &  0.107    &  \multicolumn{2}{c|}{SDSS}  & -- & --     & --  & -- & -- &  \\
3C~227	 &  09 47 45.14  &  +07 25 20.33 &  0.0861   &  \multicolumn{2}{c|}{SDSS}  & -- & --     & --  & -- & -- & f,h \\
3C~234   &  10 01 49.50  &  +28 47 09.00 &  0.1848   &  \multicolumn{2}{c|}{SDSS}  & -- & --     & --  & -- & -- & f \\
3C~236	 &  10 06 01.70  &  +34 54 10.00 &  0.1005   &  \multicolumn{2}{c|}{SDSS}  & -- & --     & --  & -- & -- &  \\ 
3C~258   &  11 24 43.80  &  +19 19 29.30 &  0.165    &  12Apr07     &  1  &  1 & 750    & HRR &  2 &750 & e \\
3C~264   &  11 45 05.07  &  +19 36 22.60 &  0.0217   &  12Apr07     &  1  &  1 & 500    & HRR &  2 &500 & i \\
3C~270	 &  12 19 23.29  &  +05 49 30.60 &  0.0075   &  \multicolumn{2}{c|}{SDSS} & -- & --     & --  & -- & -- &  \\
3C~272.1 &  12 25 03.80  &  +12 53 12.70 &  0.0035   &  04Jan08     &  2  &  1 & 500    & HRR &  2 &500 & i \\
3C~273   &  12 29 06.69  &  +02 03 08.60 &  0.1583   &  12Apr07     &  1  &  1 & 750    & HRR &  1 &750 & d,f,h \\
3C~274	 &  12 30 49.49  &  +12 23 27.90 &  0.0044   &  05Jan08     &  2  &  1 & 500    & HRR &  2 &500 & i \\ 
3C~277.3 &  12 54 12.06  &  +27 37 32.66 &  0.0857   &  08Jan08     &  2  &  1 & 500    & HRR &  2 &500 &  \\
3C~284   &  13 11 04.70  &  +27 28 08.00 &  0.2394   &  06Jan08     &  2  &  1 & 500    & HRI &  1 &1000& b \\
\hline
  \multicolumn{12}{c}{{Continued on Next Page}} \\
    \end{tabular}
  \end{center}
\end{table*}

\addtocounter{table}{-1}
\begin{table*}
  \begin{center}
    \caption{Continued}
    \begin{tabular}{l| c c| c| c c| c c| c c c|c}
      \hline \hline
Name &\multicolumn{2}{|c|}{ Coordinates J2000}& z &\multicolumn{2}{|c|}{  Telescope information}  & \multicolumn{2}{|c|}{ Low Res.}
&\multicolumn{3}{|c|}{  High Res. }  & Notes \\
\hline
     &  $\alpha$  & $\delta$ &  & Date Obs. & CCD & n& T$_{exp}$ & HR & n &  T$_{exp}$ &  \\ \hline
3C~285	 &  13 21 17.80  &  +42 35 15.00 &  0.0794   &  \multicolumn{2}{c|}{SDSS} & -- & --     & --  & -- & -- &  \\
3C~287.1 &  13 32 53.27  &  +02 00 44.73 &  0.2159   &  \multicolumn{2}{c|}{SDSS} & -- & --     & --  & -- & -- & f,g \\
3C~293   &  13 52 17.91  &  +31 26 46.50 &  0.0450   &  12Apr07     &  1  &  2 & 500    & HRI &  2 &1000& e \\
3C~296   &  14 16 52.94  &  +10 48 27.27 &  0.0240   &  SDSS        &  -- & -- & --     & --  & -- & -- &  \\
3C~300   &  14 22 59.86  &  +19 35 36.99 &  0.27     &  04Jan08     &  2  &  1 & 500    & HRR &  2 &500 &  \\
3C~303	 &  14 43 02.74  &  +52 01 37.50 &  0.141    &  \multicolumn{2}{c|}{SDSS} & -- &  --    & --  & -- & -- & f,g \\
3C~303.1 &  14 43 14.67  &  +77 07 26.59 &  0.267    &  12Apr07     &  1  &  2 & 500    & HRI &  2 &1000&  \\
3C~305	 &  14 49 21.77  &  +63 16 14.10 &  0.0416   &  \multicolumn{2}{c|}{SDSS} & -- &  --    & --  & -- & -- &  \\
3C~310	 &  15 04 57.18  &  +26 00 56.87 &  0.0535   &  08Aug07     &  2  &  1 & 500    & HRR &  2 &500 &  \\
3C~314.1 &  15 10 23.12  &  +70 45 53.40 &  0.1197   &  07Aug07     &  2  &  1 & 750    & HRR &  2 &750 &  \\
3C~315   &  15 13 40.00  &  +26 07 27.00 &  0.1083   &  09Aug07     &  2  &  1 & 750    & HRR &  2 &750 &  \\
3C~317	 &  15 16 44.57  &  +07 01 16.50 &  0.0345   &  11Aug07     &  2  &  1 & 500    & HRR &  2 &500 &  \\
3C~318.1 &  15 21 51.88  &  +07 42 31.73 &  0.0453   &  05Jan08     &  2  &  1 & 500    & HRR &  2 &500 & e \\
3C~319   &  15 24 04.88  &  +54 28 06.45 &  0.192    &  09Aug07     &  2  &  1 & 750    & HRI &  2 &750 &  \\
3C~321   &  15 31 43.40  &  +24 04 19.00 &  0.096    &  12Apr07     &  1  &  2 & 500    & HRR &  2 &500 & e \\
3C~323.1 &  15 47 43.53  &  +20 52 16.48 &  0.264    &  \multicolumn{2}{c|}{SDSS} & -- &  --    & --  & -- & -- & h \\
3C~326	 &  15 52 09.15  &  +20 05 23.70 &  0.0895   &  08Aug07     &  2  &  1 & 500    & HRR &  2 &500 &  \\
3C~327   &  16 02 27.40  &  +01 57 55.48 &  0.1041   &  07Aug07     &  2  &  1 & 500    & HRR &  2 &500 &  \\
3C~332   &  16 17 42.48  &  +32 22 32.74 &  0.1517   &  \multicolumn{2}{c|}{SDSS} & -- &  --    & --  & -- & -- & h \\
3C~338   &  16 28 38.38  &  +39 33 04.80 &  0.0303   &  06Aug07     &  2  &  1 & 500    & HRR &  2 &500-100& a,i \\ 
3C~348   &  16 51 08.16  &  +04 59 33.84 &  0.154    &  10Aug07     &  2  &  1 & 750    & HRR &  2 &750 &  \\
3C~353	 &  17 20 28.16  & --00 58 47.06 &  0.0304   &  11Aug07     &  2  &  1 & 500    & HRR &  2 &500 &  \\
3C~357   &  17 28 20.12  &  +31 46 02.22 &  0.1662   &  09Aug07     &  2  &  1 & 750    & HRR &  2 &750 & l \\
3C~371   &  18 06 50.60  &  +69 49 28.00 &  0.0500   &  10Aug07     &  2  &  1 & 500    & HRR &  2 &500 & l,h \\
3C~379.1 &  18 24 32.53  &  +74 20 58.64 &  0.256    &  11Aug07     &  2  &  2 & 500    & HRI &  2 &1000&  \\
3C~381   &  18 33 46.29  &  +47 27 02.90 &  0.1605   &  07Aug07     &  2  &  1 & 750    & HRR &  2 &750 & l \\
3C~382   &  18 35 03.45  &  +32 41 46.18 &  0.0578   &  10Aug07     &  2  &  1 & 500    & HRR &  2 &500 & h \\
3C~386	 &  18 38 26.27  &  +17 11 49.57 &  0.0170   &  08Aug07     &  2  &  1 & 500    & HRR &  2 &500 &  \\
3C~388   &  18 44 02.40  &  +45 33 30.00 &  0.091    &  10-18Aug07  &  2  &  2 & 500    & HRR &  4 &500 & a,i \\
3C~390.3 &  18 42 09.00  &  +79 46 17.00 &  0.0561   &  08Jan08     &  2  &  1 & 500    & HRR &  2 &500 & h \\
3C~401   &  19 40 25.14  &  +60 41 36.85 &  0.2010   &  09Aug07     &  2  &  2 & 500    & HRI &  2 &1000&  \\
3C~402	 &  19 41 46.00  &  +50 35 44.90 &  0.0239   &  08Aug07     &  2  &  1 & 500    & HRR &  2 &500 & i \\
3C~403	 &  19 52 15.81  &  +02 30 24.40 &  0.0590   &  11Aug07     &  2  &  1 & 500    & HRR &  2 &500 &  \\
3C~424   &  20 48 12.12  &  +07 01 17.50 &  0.127    &  10Aug07     &  2  &  1 & 750    & HRR &  2 &750 &  \\
3C~430   &  21 18 19.15  &  +60 48 06.88 &  0.0541   &  09Aug07     &  2  &  1 & 500    & HRR &  2 &500 &  \\
3C~433	 &  21 23 44.60  &  +25 04 28.50 &  0.1016   &  08Aug07     &  2  &  1 & 750    & HRR &  2 &750 &  \\
3C~436   &  21 44 11.74  &  +28 10 18.67 &  0.2145   &  07Aug07     &  2  &  2 & 500    & HRI &  2 &1000&  \\
3C~438   &  21 55 52.30  &  +38 00 30.00 &  0.290    &  18Aug07     &  2  &  2 & 500    & HRI &  2 &1000&  \\
3C~442	 &  22 14 46.88  &  +13 50 27.22 &  0.0263   &  11Aug07     &  2  &  1 & 500    & HRR &  2 &500 & i \\
3C~445   &  22 23 49.57  &  -02 06 13.08 &  0.0562   &  06Aug07     &  2  &  1 & 500    & HRR &  2 &500 & h \\
3C~449	 &  22 31 20.63  &  +39 21 30.07 &  0.0171   &  11Aug07     &  2  &  1 & 500    & HRR &  2 &500 & i \\
3C~452	 &  22 45 48.90  &  +39 41 14.47 &  0.0811   &  11Aug07     &  2  &  1 & 500    & HRR &  2 &500 &  \\
3C~456   &  23 12 28.11  &  +09 19 23.59 &  0.2330   &  10Aug07     &  2  &  2 & 500    & HRI &  2 &1000&  \\
3C~459	 &  23 16 35.24  &  +04 05 18.29 &  0.2199   &  23Nov06     &  1  &  2 & 500    & HRI &  2 &1000&  \\
3C~460   &  23 21 28.74  &  +23 46 49.66 &  0.268    &  05Jan08     &  2  &  2 & 500    & HRI &  2 &1000&  \\
3C~465   &  23 38 29.41  &  +27 01 53.03 &  0.0303   &  07Aug07     &  2  &  1 & 500    & HRR &  2 &500 & i \\	 
\hline
    \end{tabular}
  \end{center}

  Column description: (1) 3C name of the source; (2) and (3) J2000 coordinates 
  (right ascension and declination); (4) redshift; for the TNG observations:
  (5) UT night of observation; (6) CCD used; (7) number of low resolution
  spectra; (8) exposure time for each low resolution spectrum; (9) high
  resolution grism used; (10) number of high resolution spectra; (11) exposure
  time for each high resolution spectrum.
  Column (12): (a) longer exposure times because of bad seeing conditions; (b)
  exposure shortened because of bad weather conditions and closure
  telescope; (c) longer exposure times because of first night; (d) observed one
  HR spectrum; (e) seeing $>2\arcsec$; (f) broad components; (g) broad ssp
  range; (h) no starlight subtraction; (i) off-nuclear starlight subtraction;
  (l) telluric correction.
\end{table*}

\clearpage

\begin{figure*}[htbp]
\centerline{
\psfig{figure=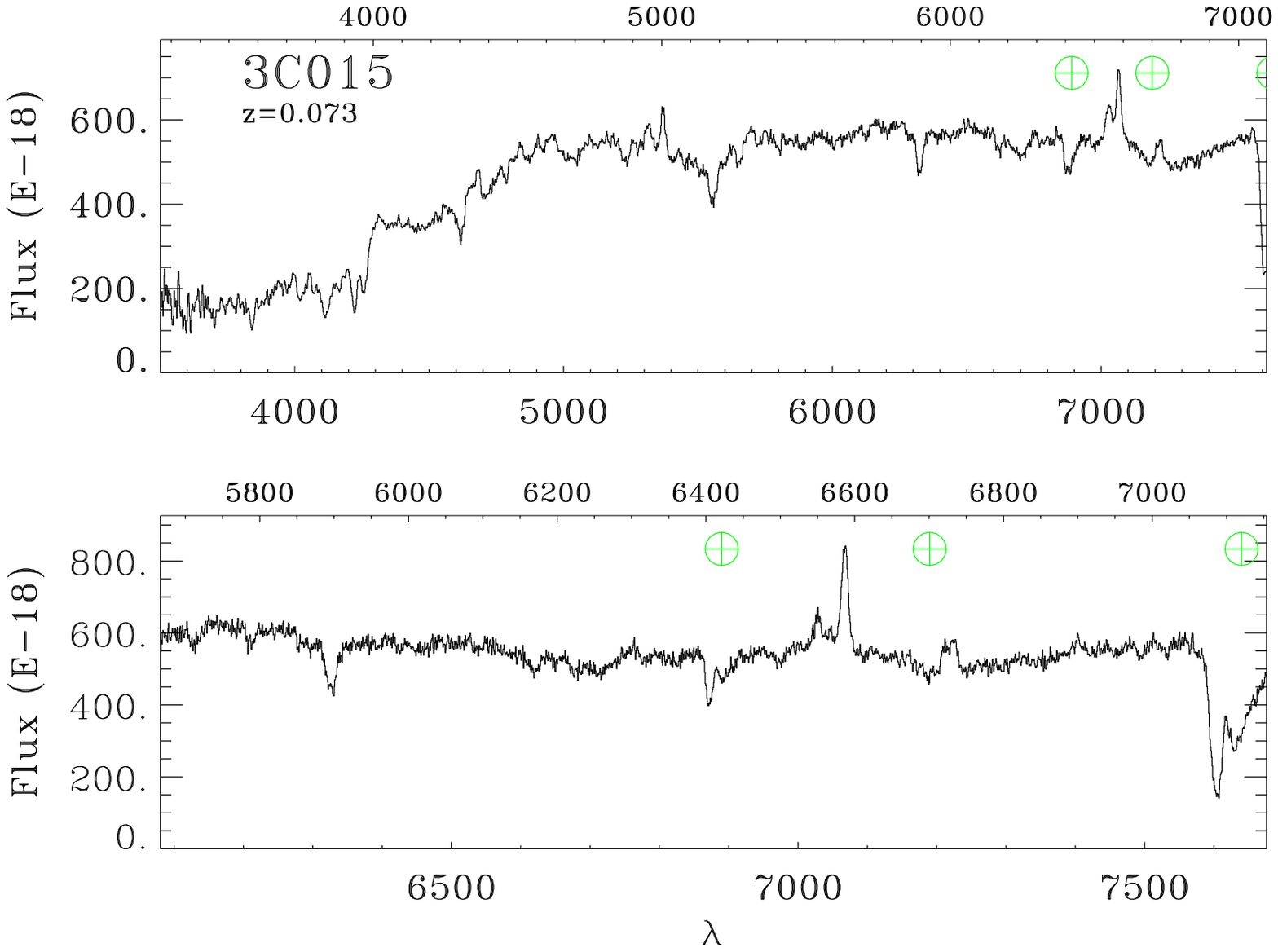,width=0.45\linewidth}
\psfig{figure=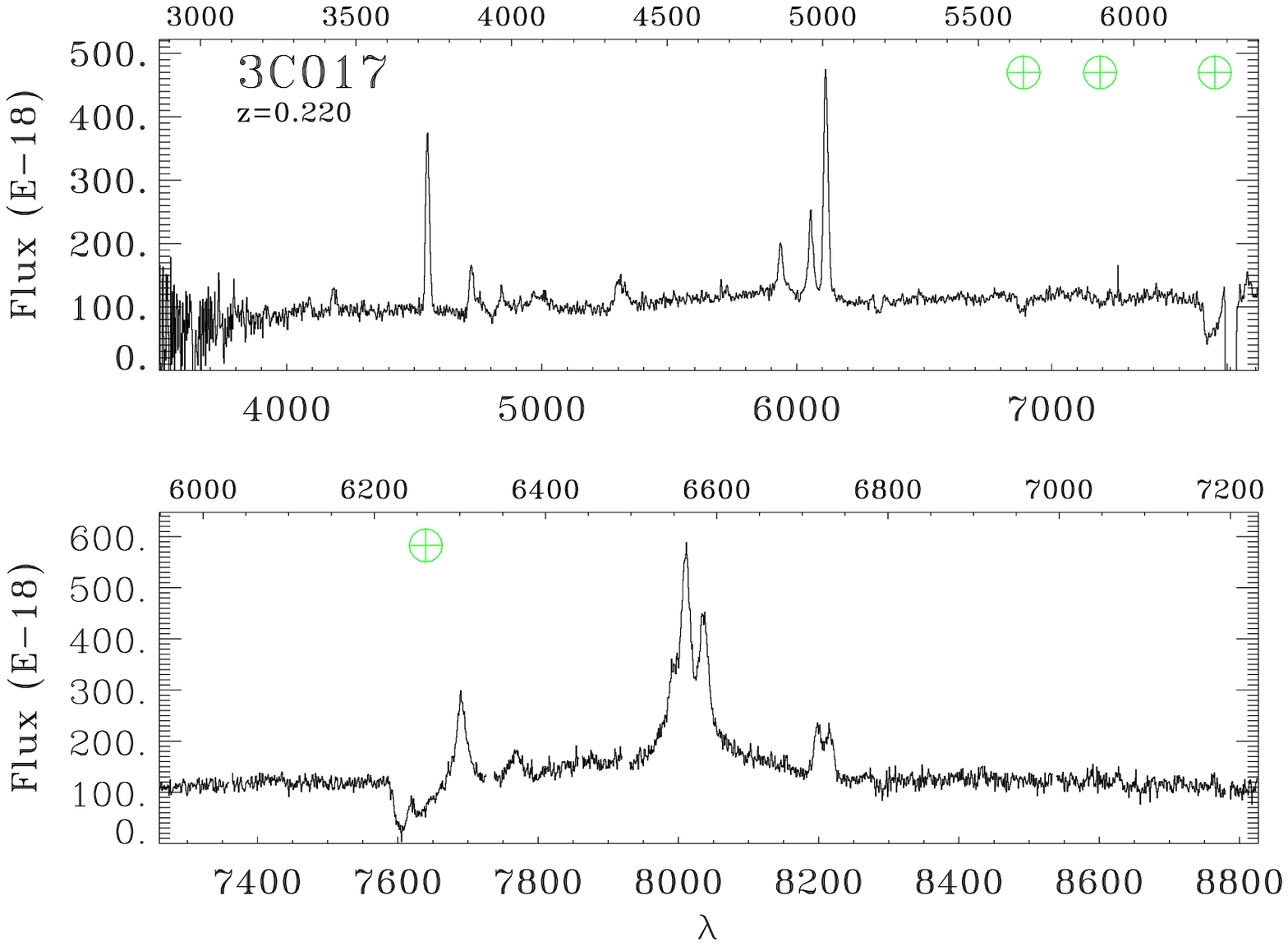,width=0.45\linewidth}}
\centerline{
\psfig{figure=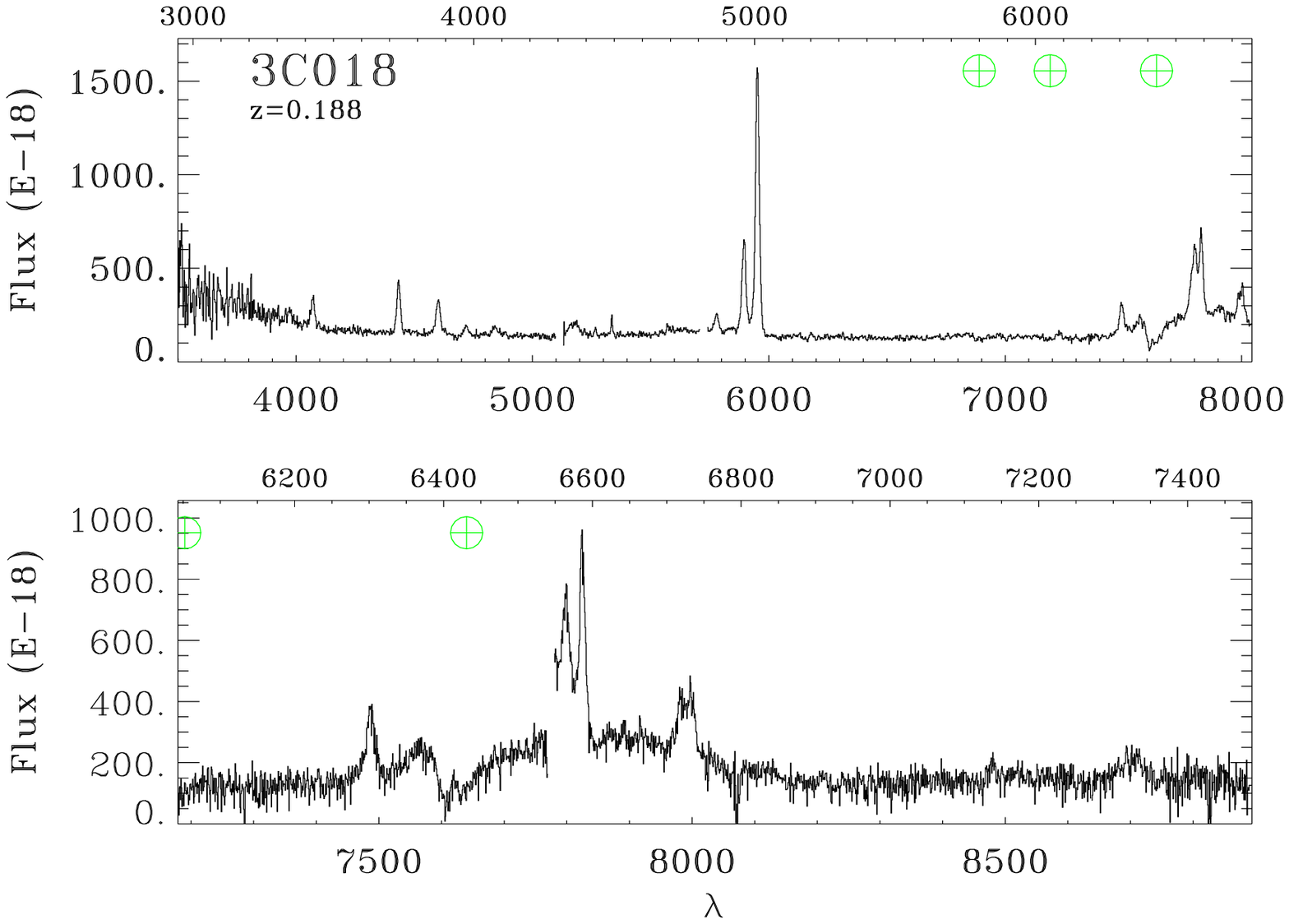,width=0.45\linewidth}
\psfig{figure=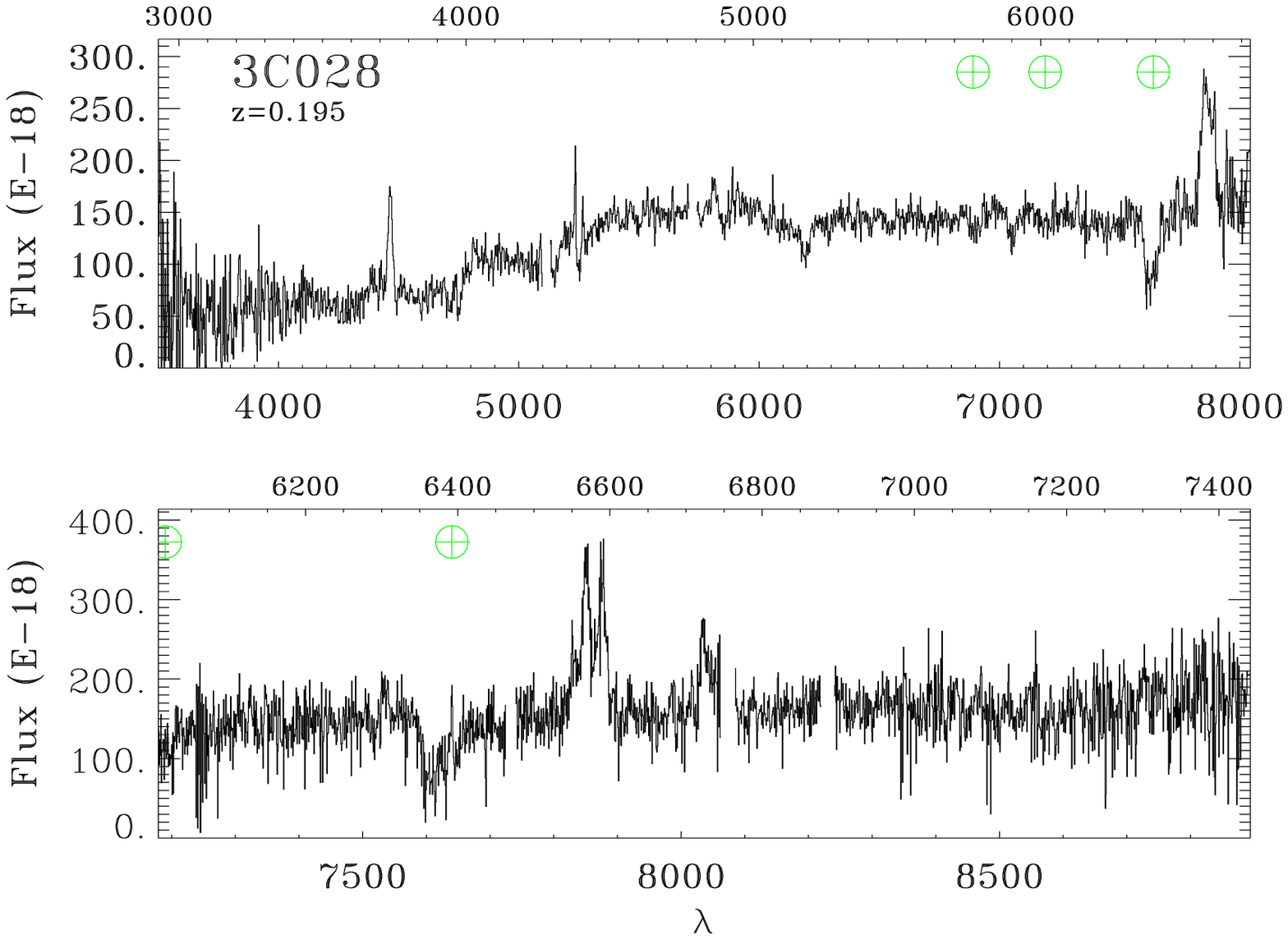,width=0.45\linewidth}}
\centerline{
\psfig{figure=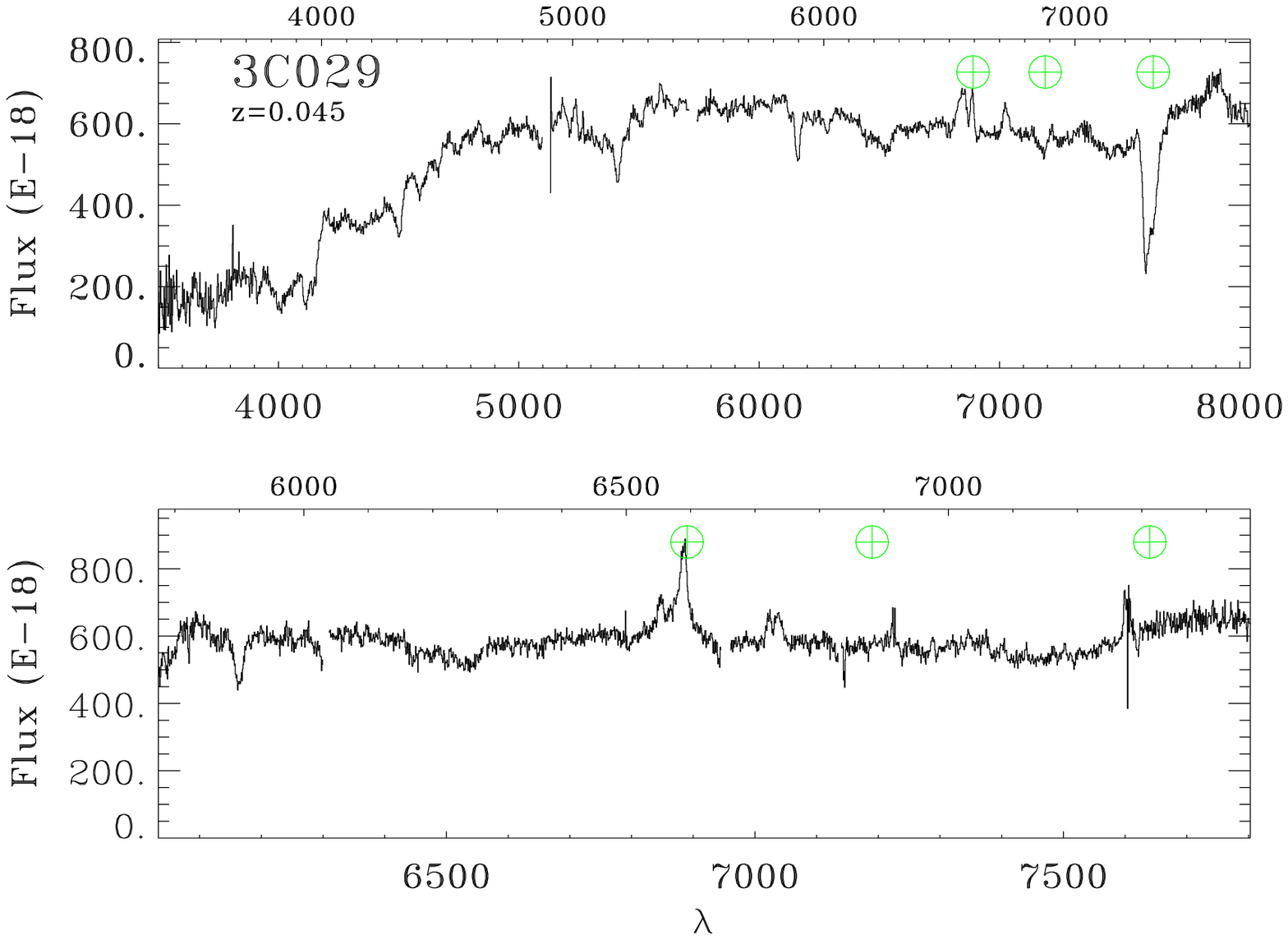,width=0.45\linewidth}
\psfig{figure=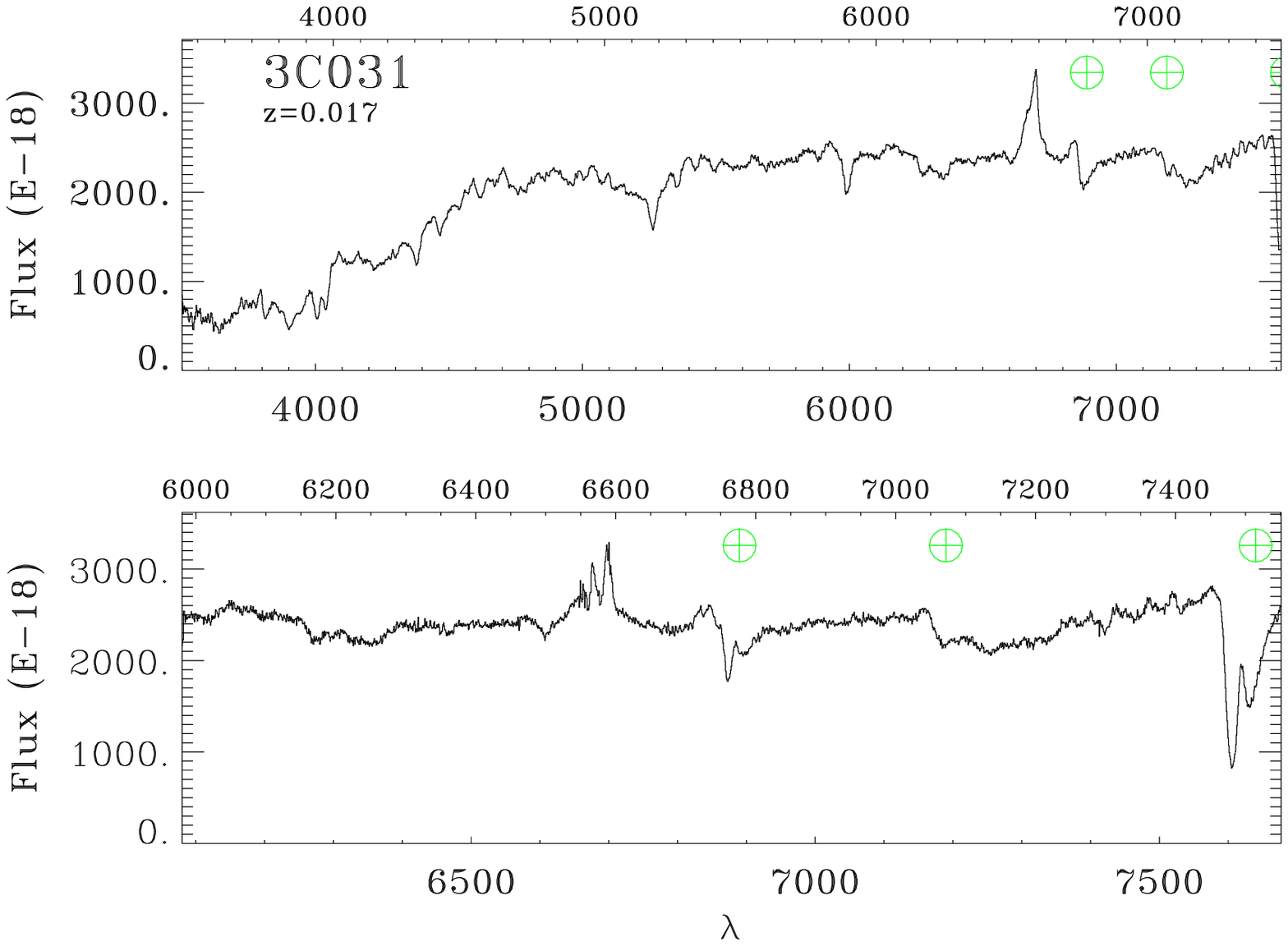,width=0.45\linewidth}}
\centerline{
\psfig{figure=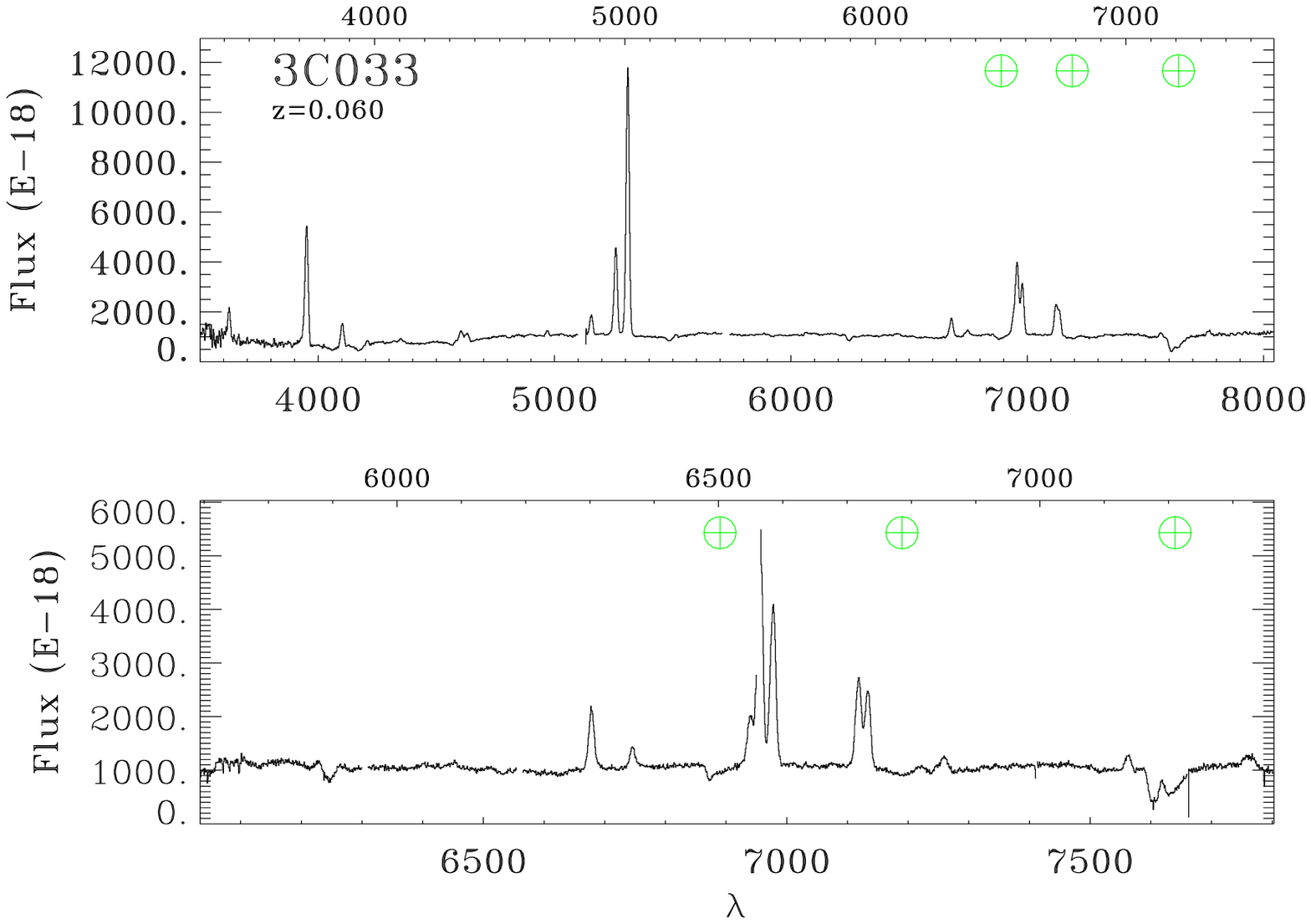,width=0.45\linewidth}
\psfig{figure=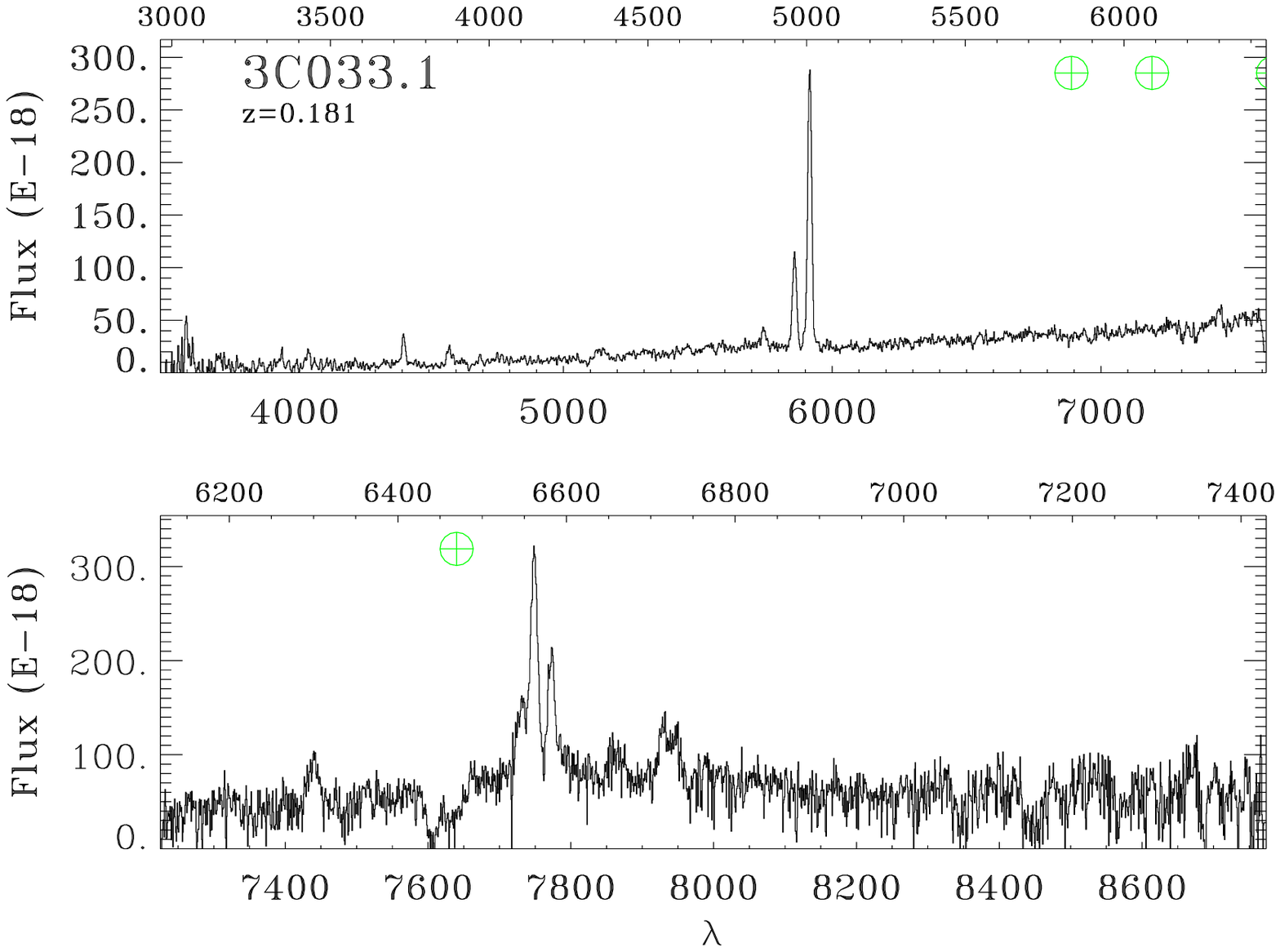,width=0.45\linewidth}}
\caption{\label{spectra} For each source the low (upper panels) 
  and high resolution (bottom panels) spectra are shown. The fluxes are
  in units of erg cm$^{-2}$ s$^{-1}$ \AA$^{-1}$ while the wavelengths are in
  \AA. The lower axes of the spectra show the observed wavelengths
  while the upper axes show the rest frame wavelengths. The three main
  telluric absorption bands are indicated with circled crosses. For the SDSS
  spectra we report the whole spectrum in the upper panel and a zoom of $\sim$
  2000 \AA\ around the H$\alpha$ line in the bottom panel. The remaining
  spectra are available as gif files.}
\end{figure*}

\noindent
\citep{martel99, dekoff96}. The procedure
used for the comparison is
the following: we smoothed the HST image to the seeing measured in the
acquisition TNG image, extracted a region corresponding to 2$\arcsec$x
2$\arcsec$, and measured its flux based on the HST absolute flux calibration.
We convolved the LR-B TNG spectra with the filter transmission used to obtain
the HST images (the broad band F702W filter of the Wide Field Planetary Camera
2) and measured the flux within this spectral band. We
compared these two
measurements and found an
agreement within $\sim$ 20\%. The absolute flux calibration of high resolution
spectra was
obtained scaling them to the low resolution calibrated spectra.

The telluric absorption bands were usually left uncorrected except in the few
cases in which an emission line of interest fell into these bands. In these
cases we corrected the atmospheric absorptions using the associated standard
stars as templates.

Spectra for 18 3CR objects are available from the Sloan Digital Sky Survey
(SDSS) database \citep{york00, st02, yip04}, Data Release 4,5,6.  Since they
have comparable signal to noise, resolution and wavelength coverage (3800-9200
\AA) of the TNG spectra, we decided not to observe them at the TNG and to use
the SDSS spectra for our analysis.
 
Fig. \ref{spectra} shows, for all the observed targets, the low resolution
spectrum (upper image) and the high resolution one (bottom image). The
calibrated spectra are in units of $10^{-18}$ erg cm$^{-2}$ s$^{-1}$
\AA$^{-1}$. The wavelengths (in \AA\ units) are in the observer frame in
the axes below the images while they are in the source frame in the axes above
them. The three main telluric absorption bands are indicated as circles with a
cross inside.

\section{Data analysis}
\label{sect3}

\subsection{Starlight subtraction}
\label{data}

The extraction apertures contain, in addition to the emission produced by the
active nucleus, a substantial contribution from the host galaxy starlight. 
In order to proceed in our analysis it is necessary to separate these two
components by modeling the stellar emission. 

In order to prepare the spectra for the stellar removal, we corrected for
reddening due to the Galaxy \citep{burstein82,burstein84} using the extinction
law of \citet{cardelli89}. The galactic extinction E(B-V) used for each object
was taken from the NASA Extragalactic Database (NED) database and is listed in
Table \ref{bigtable}. We also deredshifted the spectra using the value of
redshift from NED.

\begin{landscape}
\begin{figure}[ht]

  \centerline{ 
    \psfig{figure=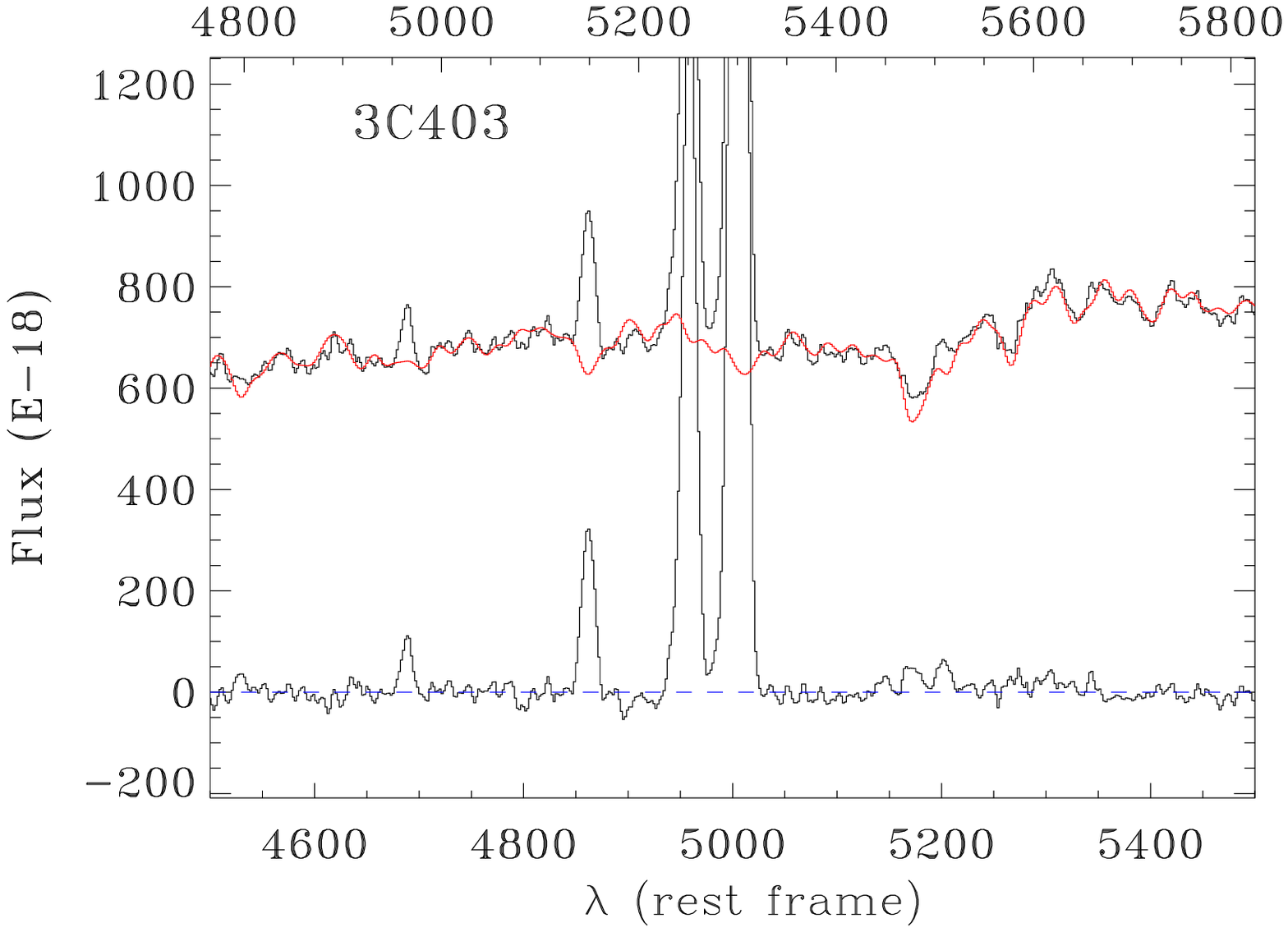,width=0.25\linewidth}
    \psfig{figure=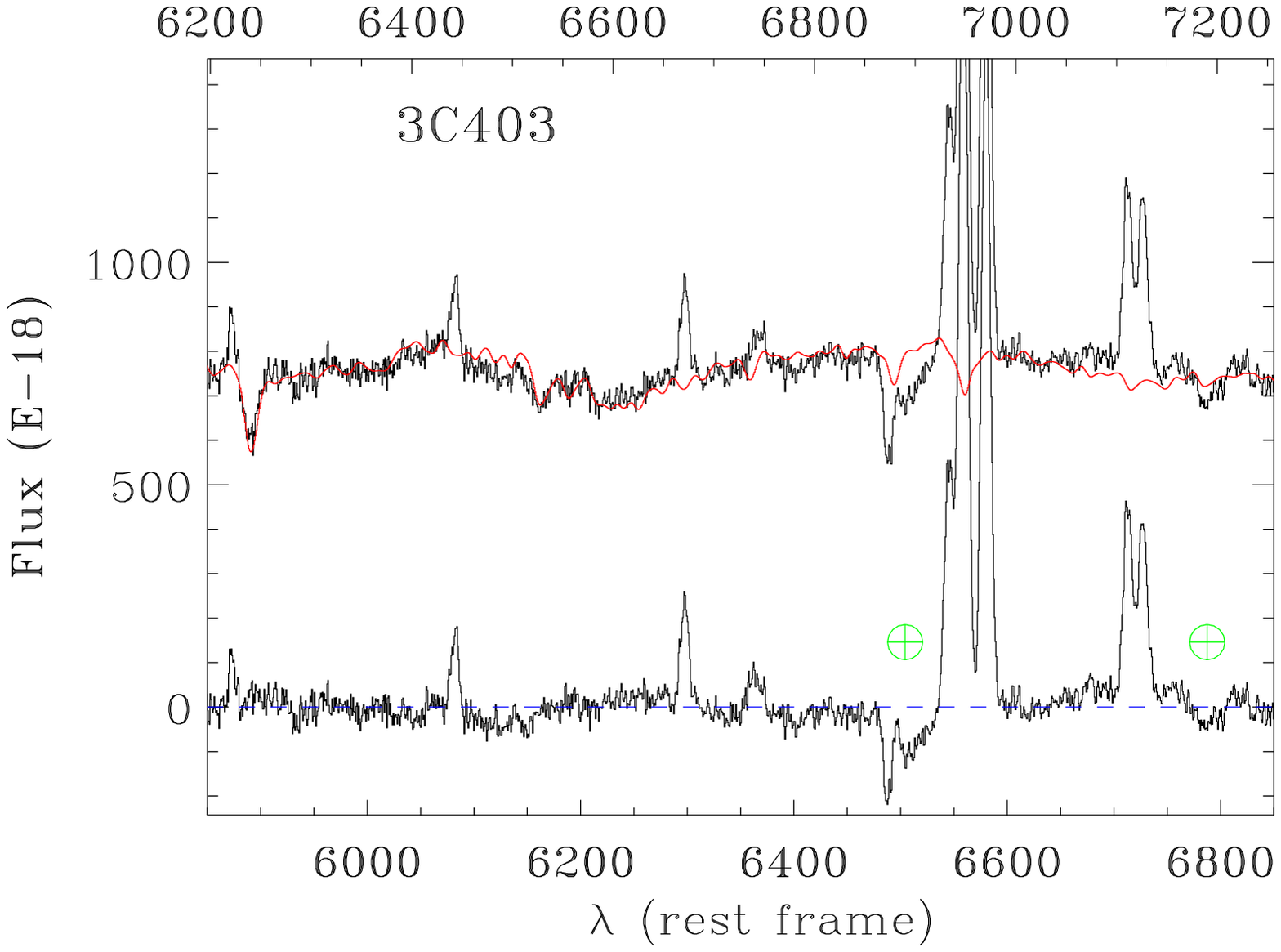,width=0.25\linewidth}
    \psfig{figure=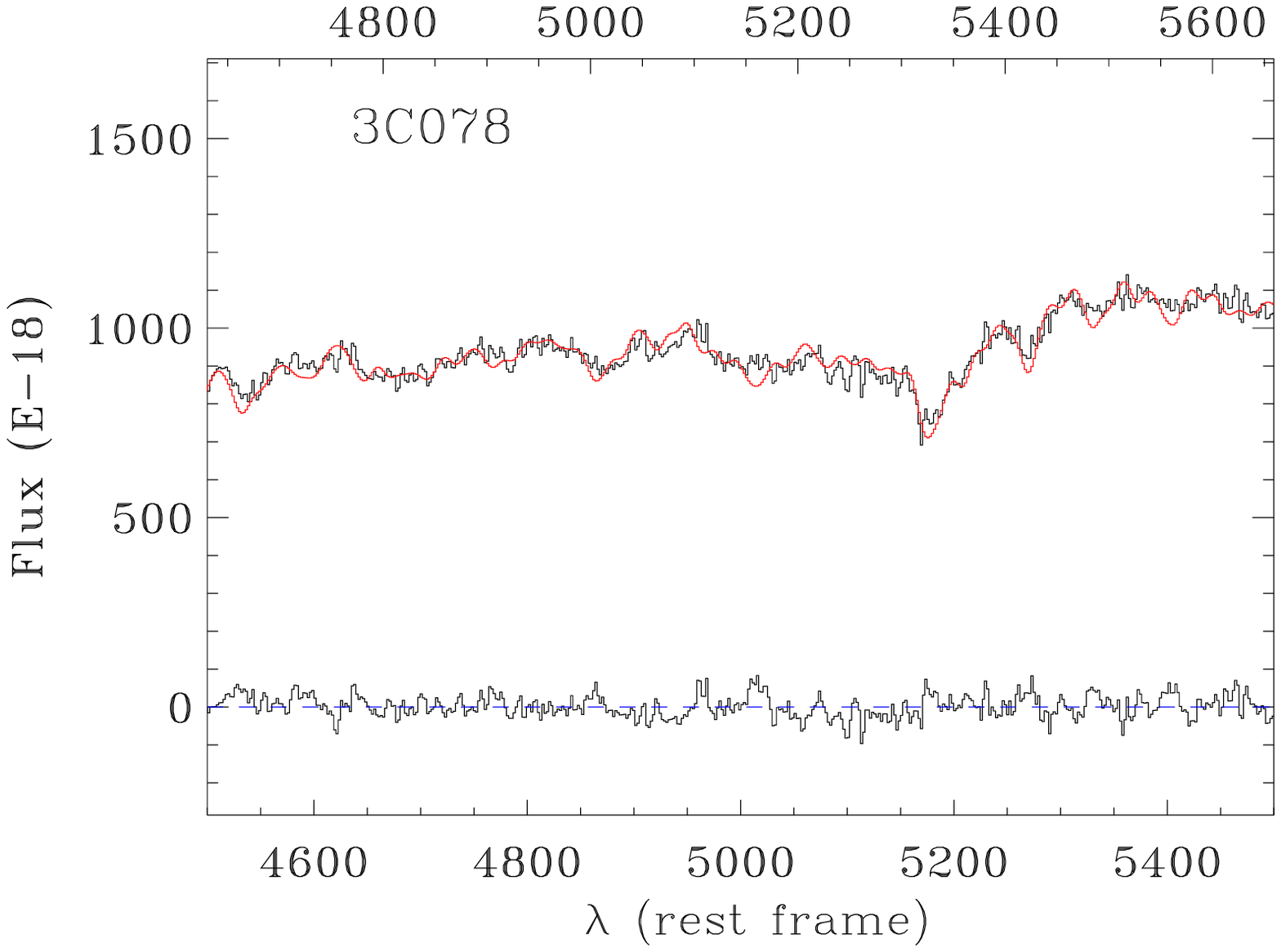,width=0.25\linewidth} 
    \psfig{figure=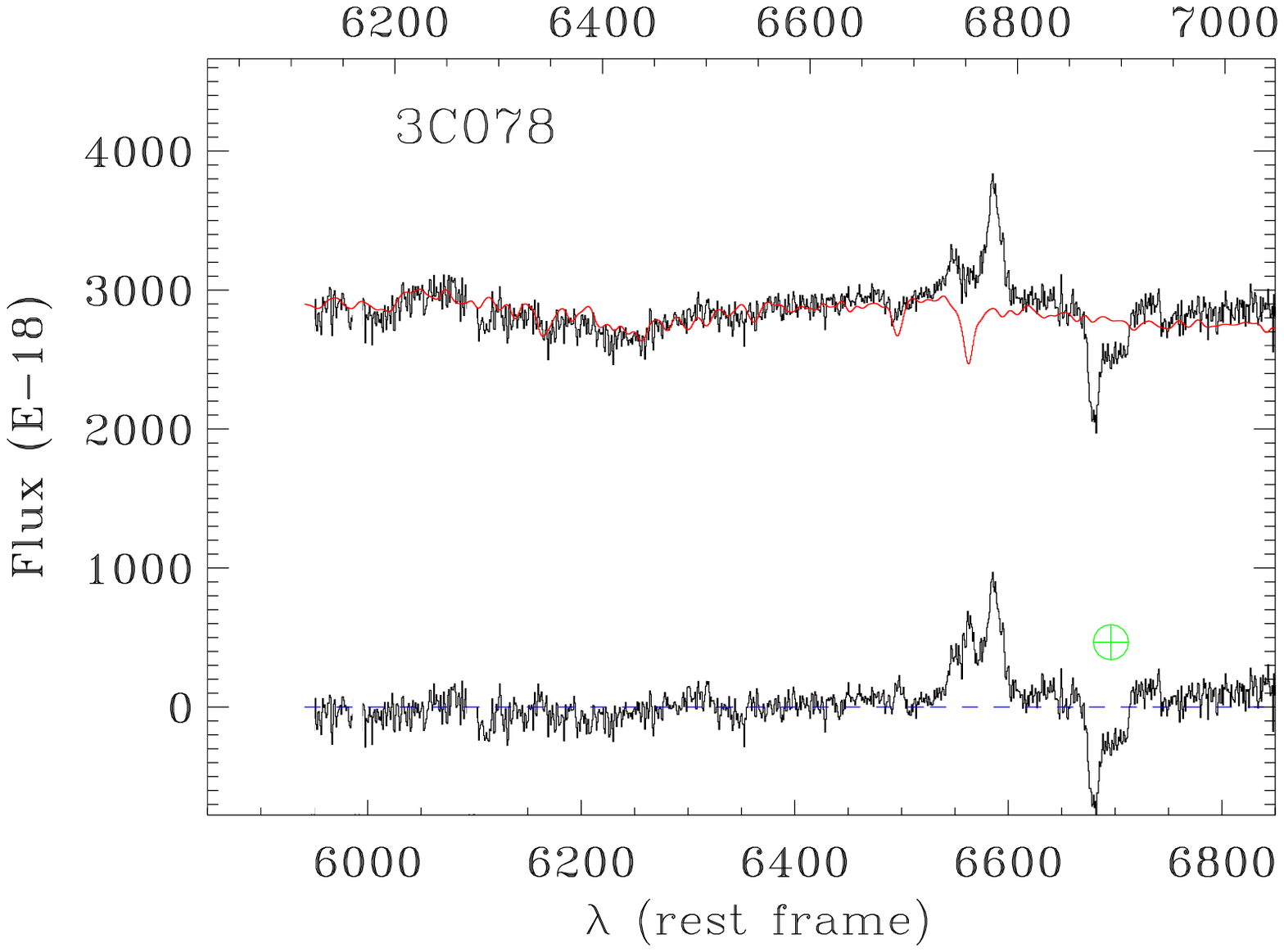,width=0.25\linewidth}} 

  \centerline{ 
    \psfig{figure=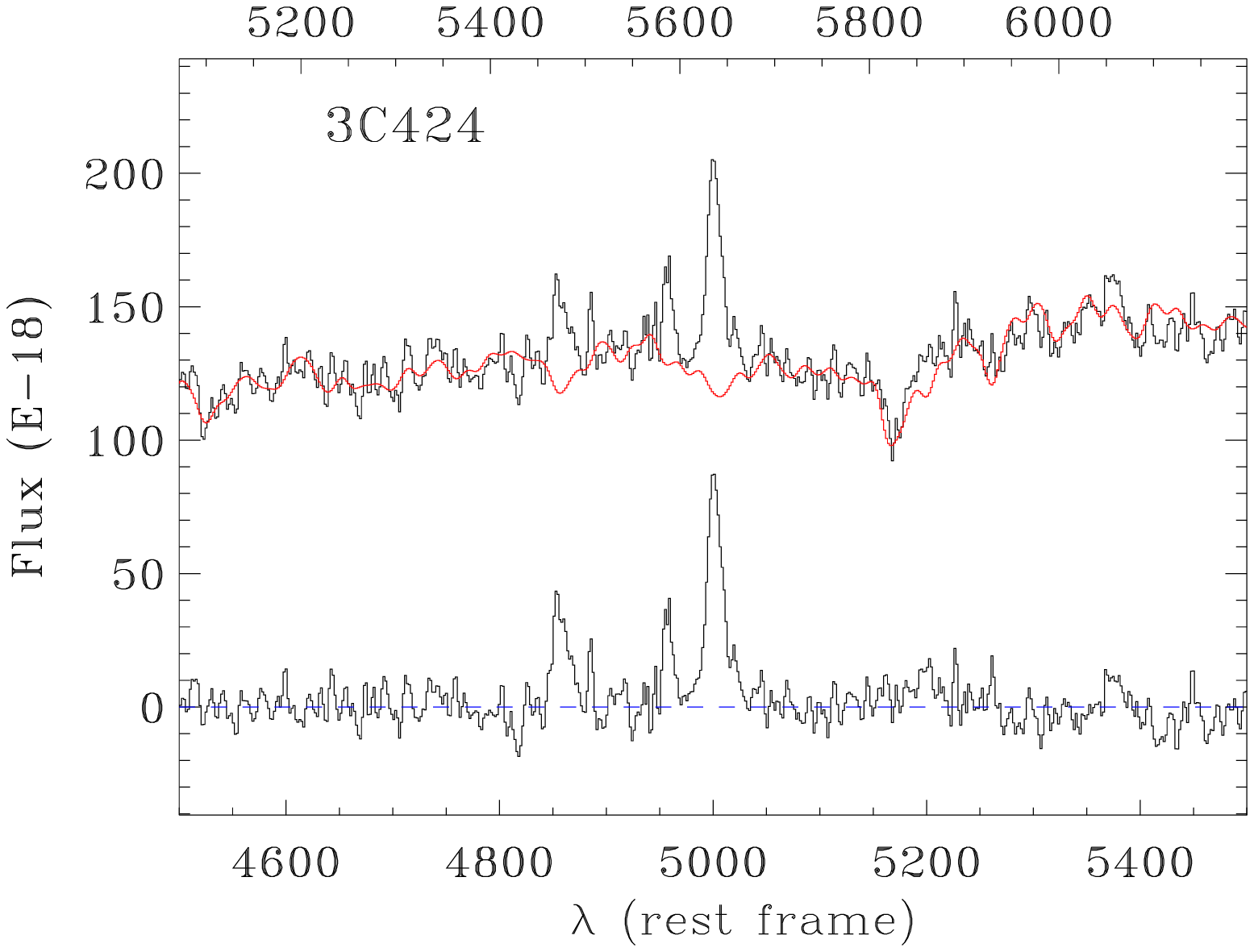,width=0.25\linewidth} 
    \psfig{figure=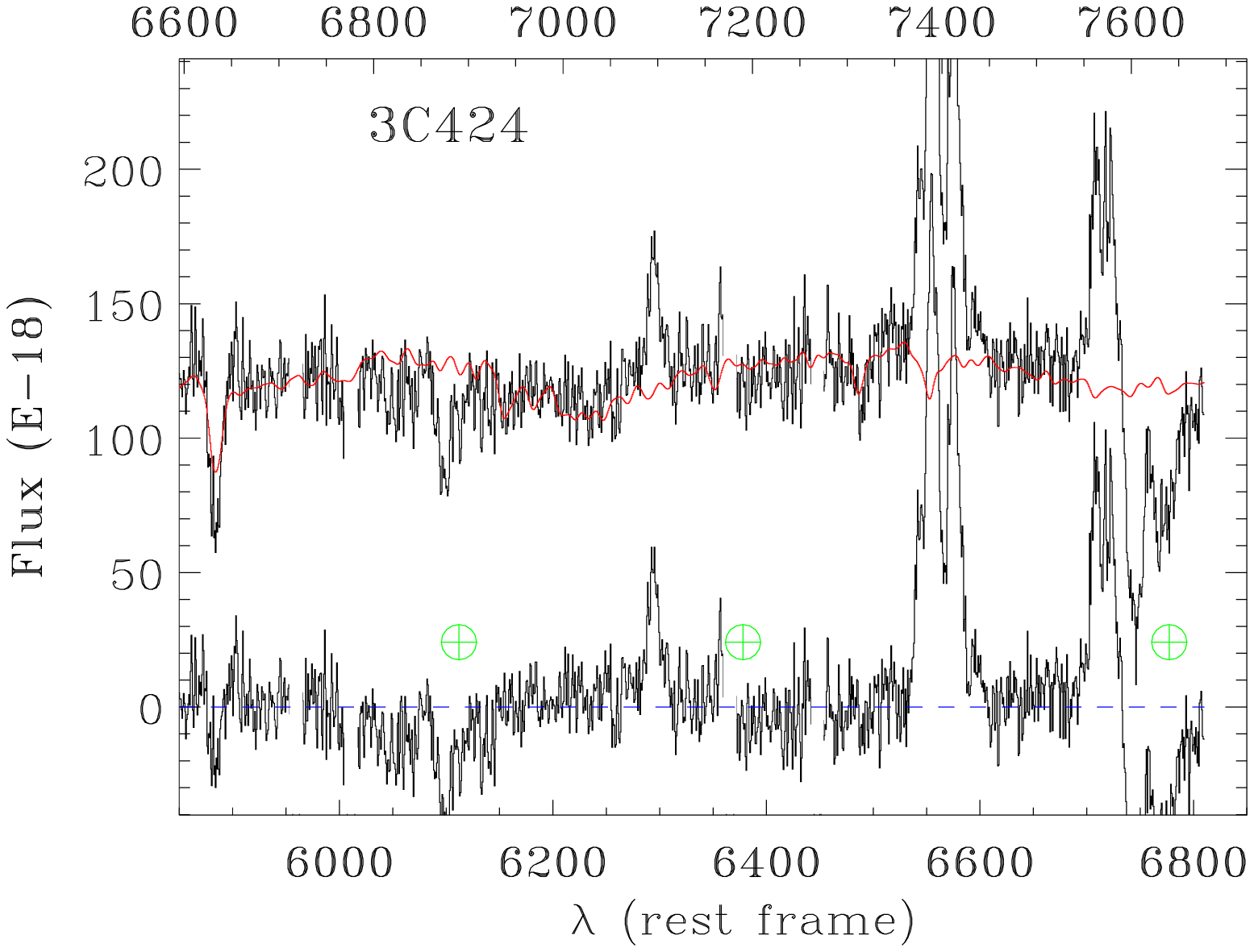,width=0.25\linewidth} 
    \psfig{figure=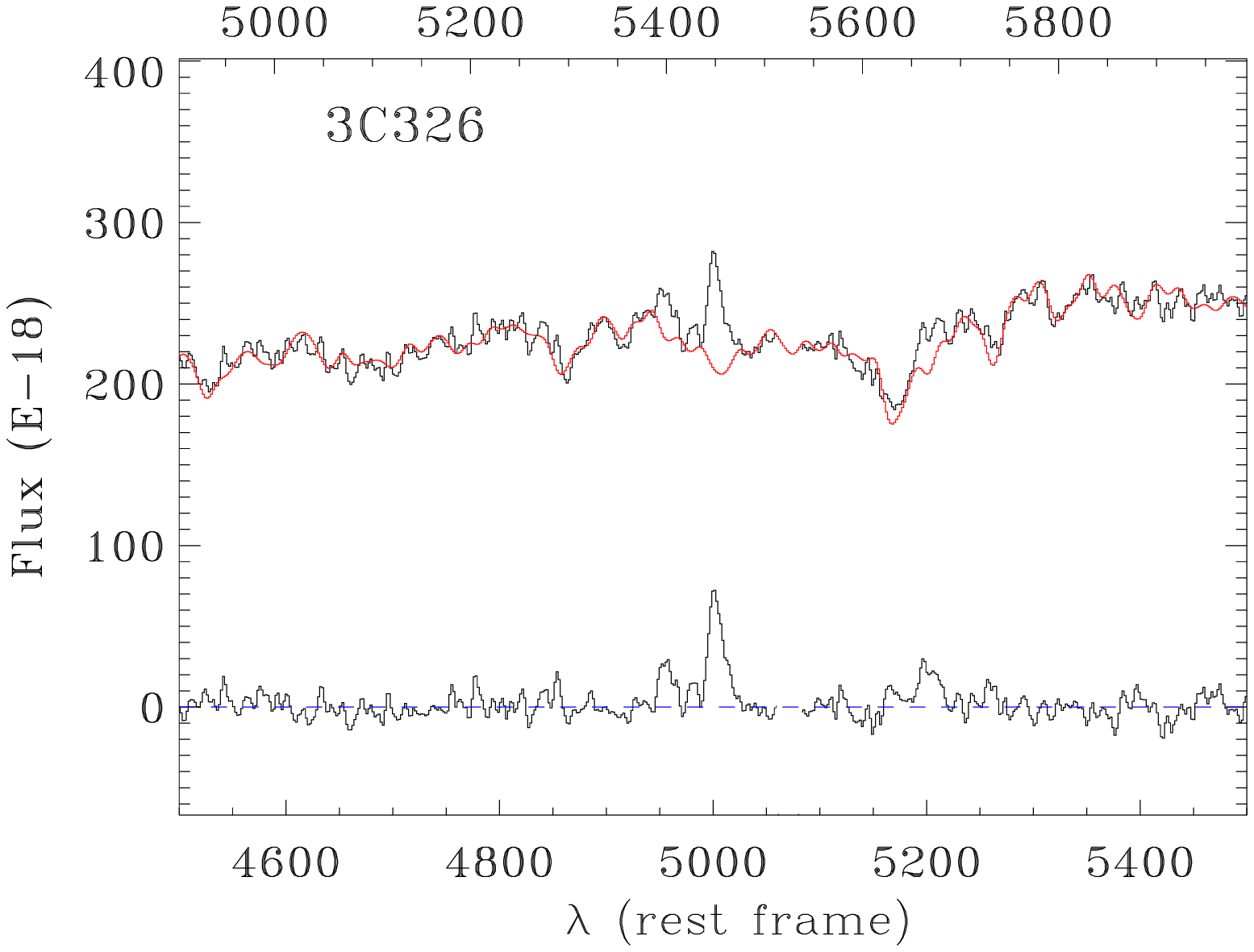,width=0.25\linewidth} 
    \psfig{figure=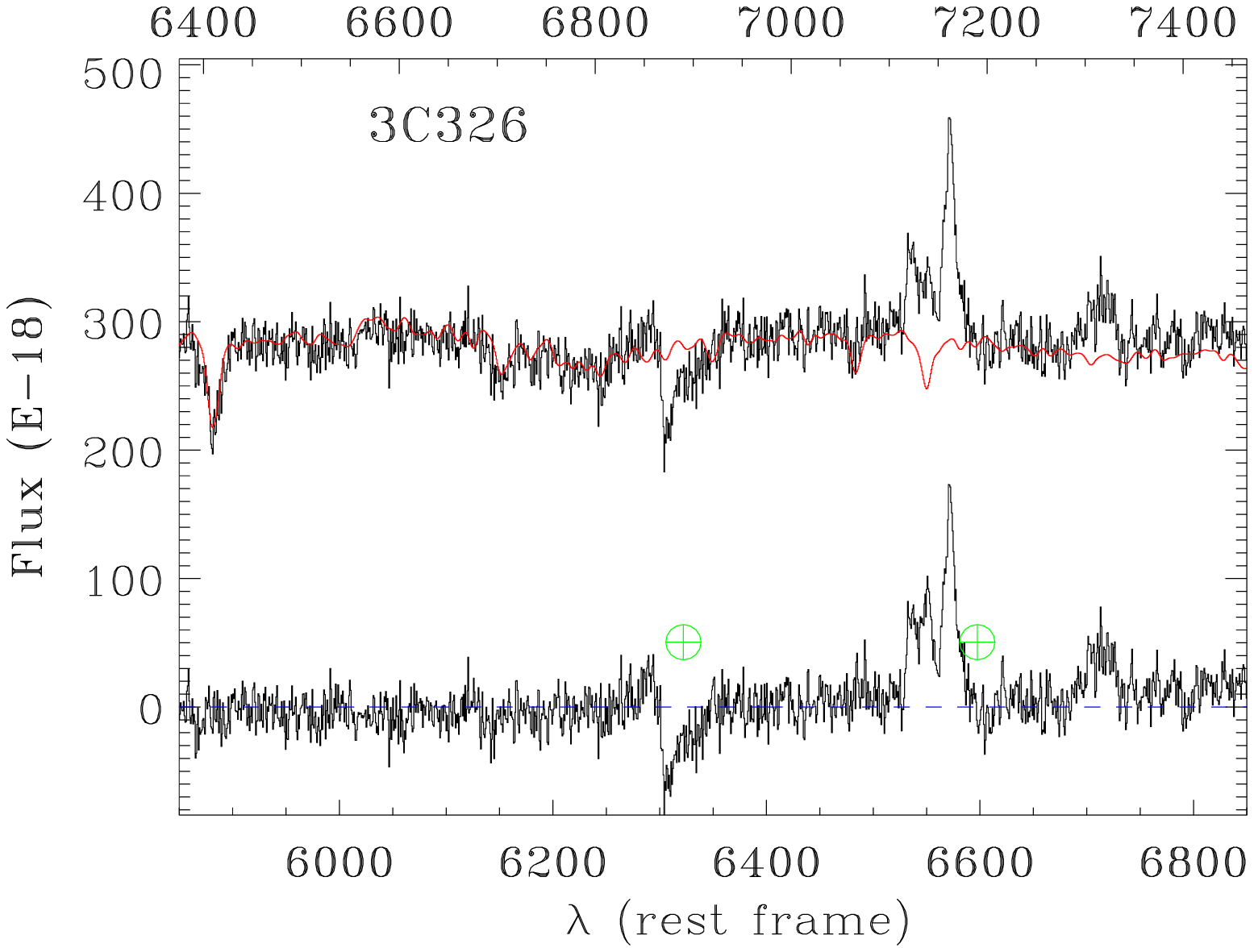,width=0.25\linewidth}}

 \centerline{ 
    \psfig{figure=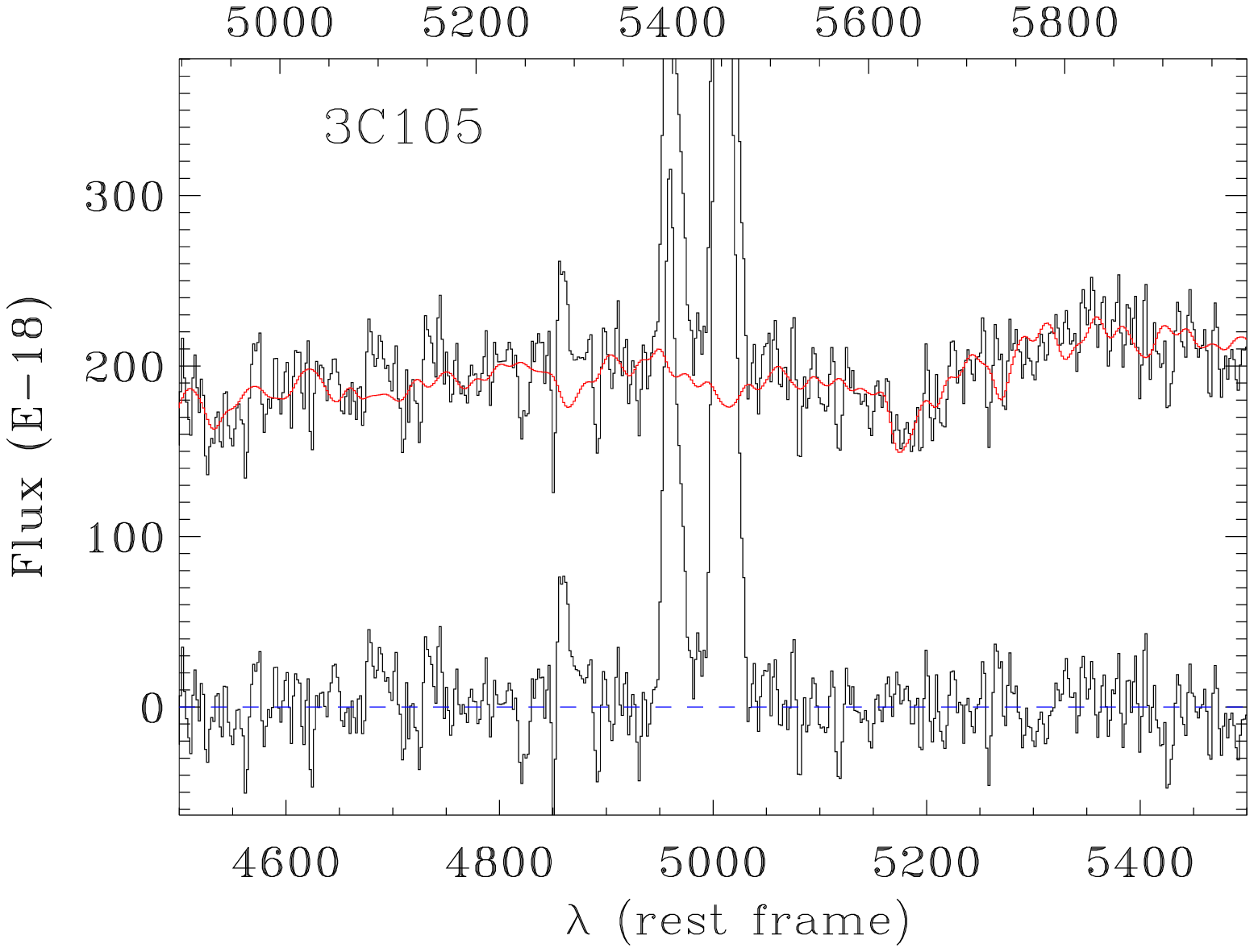,width=0.25\linewidth}
    \psfig{figure=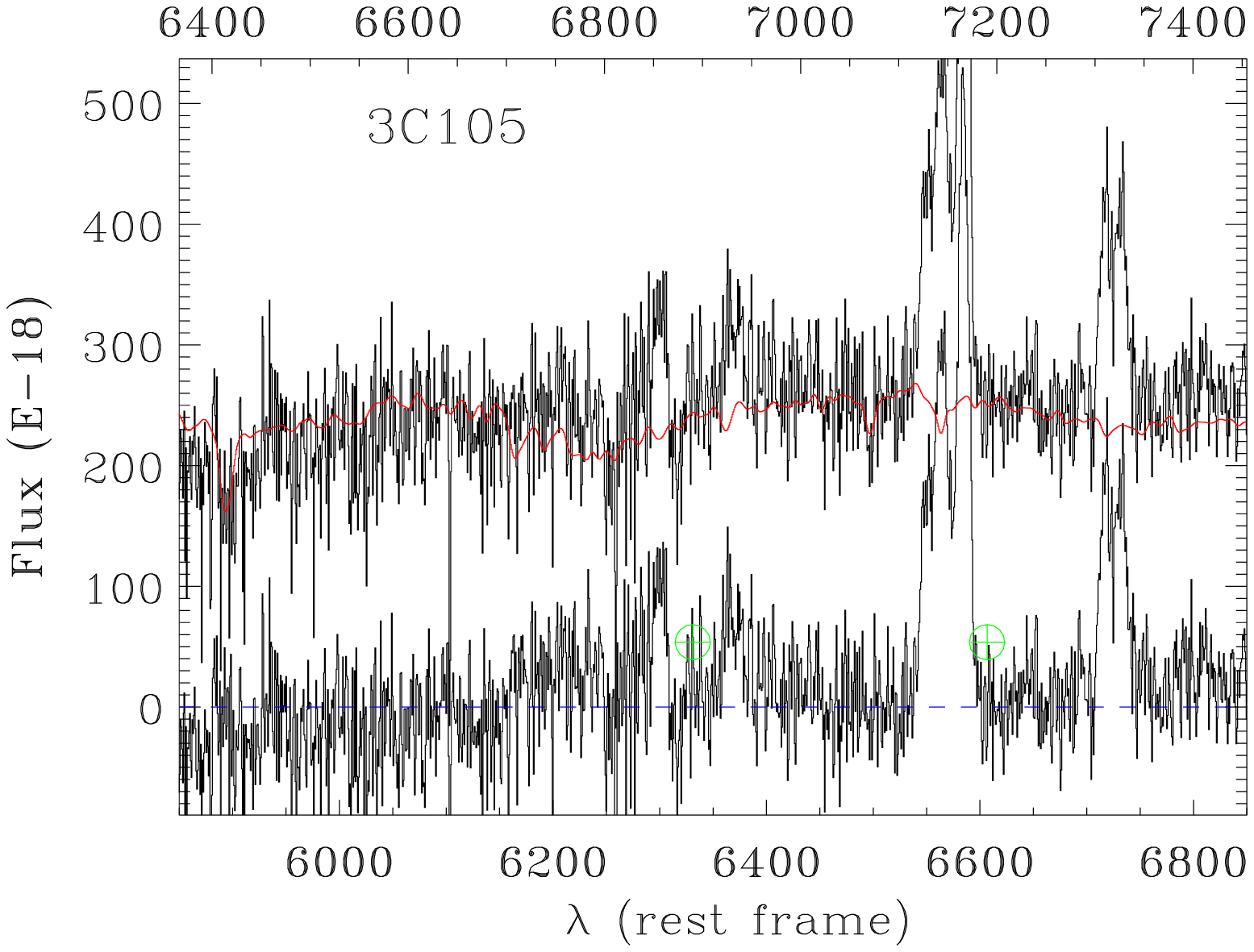,width=0.25\linewidth}
    \psfig{figure=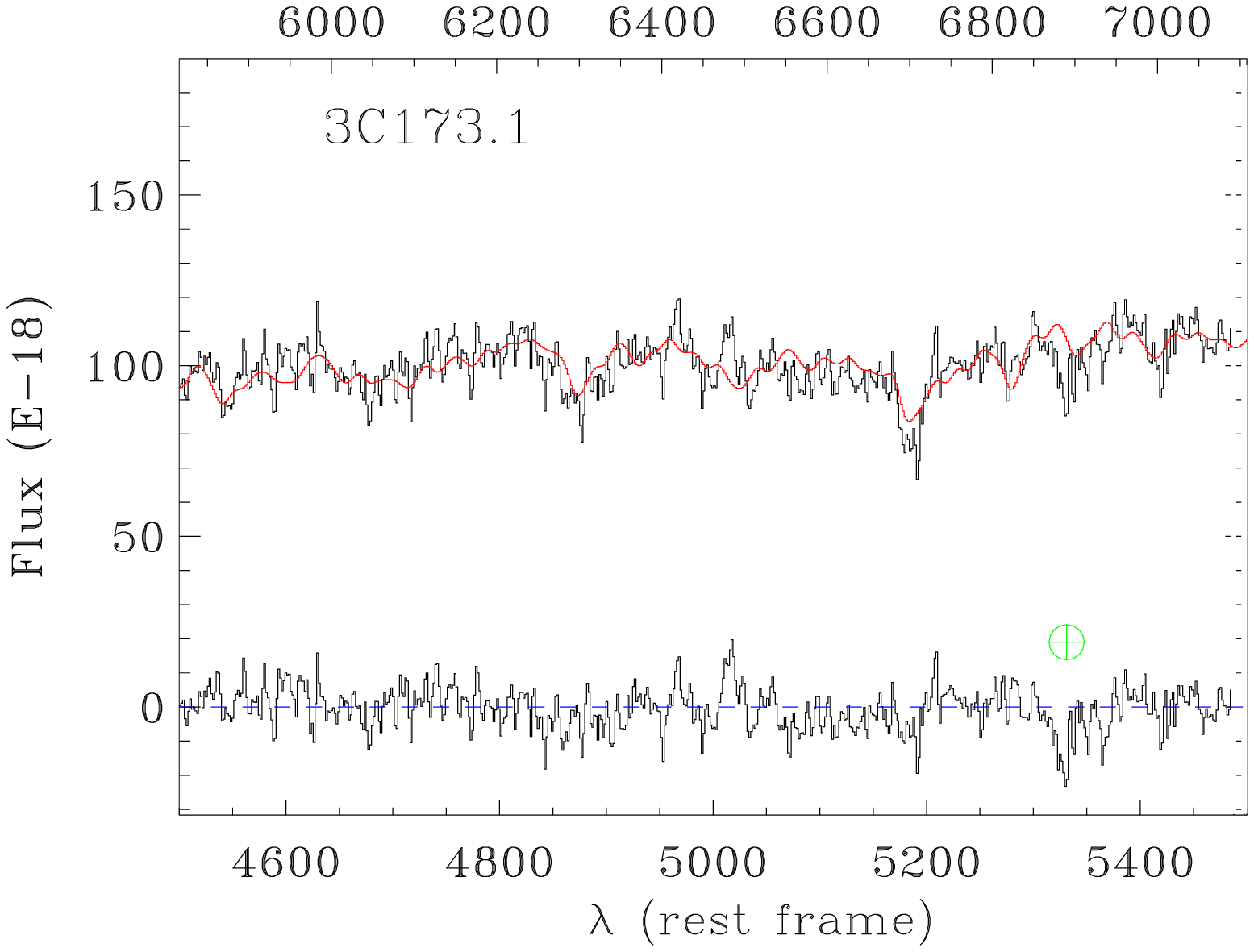,width=0.25\linewidth}
    \psfig{figure=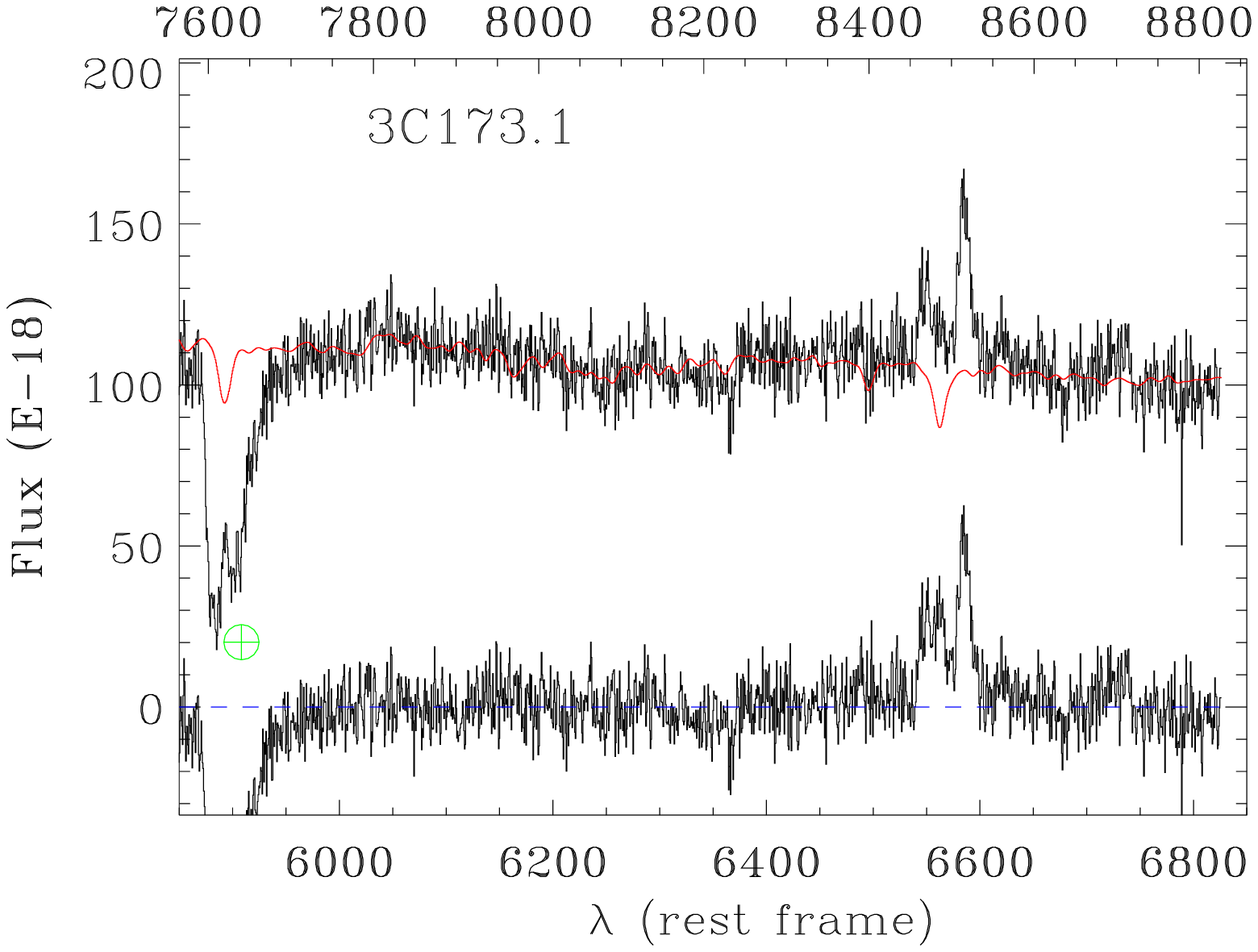,width=0.25\linewidth}}

  \caption{\label{ssp} Examples of the process of removal of the galaxies'
    starlight. The six galaxies presented are representative of the different
    quality in the sample (in order of decreasing signal-to-noise ratio from
    top to bottom) and of different levels of line equivalent widths (high EW
    in the left two columns, low EW in the right two columns).  The original
    spectrum is the upper solid black line; the best fit Single Stellar
    Population (SSP) model is in red; the lower solid black line is the
    residual spectrum after SSP subtraction; the dashed blue line is a
      reference zero level. The flux is measured in
    $10^{-18}$ erg cm$^{-2}$ s$^{-1}$ \AA$^{-1}$ while the wavelength is in
    \AA, in rest frame in the lower axes and in observed frame in the axes
    above.}
\end{figure}
\end{landscape}

The spectra modeling was done considering two wavelength ranges of $\sim1000$
\AA\ centered on the H$\alpha$ line from the HR spectrum and on the H$\beta$
line from the LR-B spectrum. For each of the two ranges we subtracted the best
fit single stellar population (SSP) model taken from the \citet{bruzual03}
library.  We selected a grid of 33 single stellar population models varying
over 11 values of stellar ages (0.1, 0.3, 0.5, 1, 2, 3, 5, 7, 9, 11, 13 Gyr)
and three values of absolute metalicity (0.008, 0.02, 0.05), see Table
\ref{ssp}. The templates assume a Salpeter Initial Mass Function (IMF) formed
in an instantaneous burst, with stars in the mass range $0.1 \leq M \leq 125$
M$_{\odot}$. The parameters free to vary independently of each other in order
to obtain the best fit are the stellar age, the metalicity, the normalization
of the model, the velocity dispersion (line-broadening function), and a small
adjustment to the published redshift\footnote{The redshifts resulting from the
  spectral fitting always agree within the uncertainties with the published
  values with the only exception of 3C~130, for which we estimate z=0.032
  instead of z=0.109, based on the position of the Na D absorption feature.}.
We allowed for different models for the population in the \Hb\ and \Ha\
regions. In fact, as discussed in Sect.  \ref{template}, the stellar content
of radio galaxies is often composed by populations of various ages and thence
the dominant population can vary in different bands.  We excluded from the
fit the spectral regions corresponding to emission lines, as well as other
regions affected by telluric absorption, cosmic rays or other impurities.  The
SSP subtracted from the spectra are listed in Table \ref{ssptab}: for each
source the best fit SSP and the $\chi^2_r$ are reported for the two wavelength
regions around H$\alpha$ and \Hb.  

\begin{table*}
  \begin{center}
    \caption{Summary of the Single Stellar Population fitting.}
    \label{ssptab}
    \begin{tabular}{l| r r| r r ||l| r r| r r}
      \hline \hline
Name & \multicolumn{2}{|c|}{H$\beta$} &   \multicolumn{2}{|c||}{H$\alpha$} & Name & \multicolumn{2}{|c|}{H$\beta$} & \multicolumn{2}{|c}{H$\alpha$} \\
     &     SSP  & $\chi^2_{\rm r}$    &   SSP     & $\chi^2_{\rm r}$       &      &     SSP  & $\chi^2_{\rm r}$    &     SSP     & $\chi^2_{\rm r}$    \\
\hline
3C~015     &      16   &      6.29  &      27   &   1.72 &  3C~264     &      29   &     12.43  &      22	&   2.29  \\  
3C~017     &       8   &      2.22  &      33	&   1.63 &  3C~270     &      31   &      2.41  &      22	&   3.41  \\  
3C~018     &      28   &      0.81  &      10	&   1.12 &  3C~272.1   &      32   &      7.88  &      22	&   2.60  \\  
3C~028     &      29   &      1.04  &      11	&   0.88 &  3C~274     &      27   &     23.84  &      27	&   1.09  \\  
3C~029     &      28   &      2.21  &      28	&   1.53 &  3C~277.3   &      31   &      1.70  &      22	&   0.90  \\  
3C~031     &      28   &      4.50  &      28	&   1.43 &  3C~284     &      26   &      1.28  &      15	&   1.50  \\  
3C~033     &       7   &      1.43  &      29	&   0.92 &  3C~285     &      22   &      1.49  &      22	&   1.69  \\  
3C~035     &      30   &      1.76  &      22	&   0.89 &  3C~287.1   &       8   &      1.21  &      10	&   1.51  \\  
3C~040     &      28   &      7.03  &      28	&   2.52 &  3C~293     &      21   &      1.26  &      33	&   1.19  \\  
3C~052     &      29   &      0.90  &      33	&   1.00 &  3C~296     &      31   &      3.84  &      22	&   2.54  \\  
3C~061.1   &      18   &      0.83  &      33	&   0.99 &  3C~300     &      11   &      0.64  &      33	&   1.60  \\  
3C~066     &      28   &      4.23  &      28	&   2.31 &  3C~303     &      16   &      1.53  &      27	&   1.02  \\  
3C~075     &      32   &      1.93  &      22	&   2.94 &  3C~303.1   &       6   &      0.83  &      27	&   1.04  \\  
3C~076.1   &      30   &      2.50  &      11	&   1.29 &  3C~305     &      10   &      2.15  &      22	&   1.39  \\  
3C~078     &      32   &      1.18  &      11	&   0.78 &  3C~310     &      29   &      1.34  &      28	&   2.09  \\  
3C~079     &       7   &      1.83  &      10	&   1.10 &  3C~314.1   &      30   &      0.90  &      22	&   0.81  \\  
3C~083.1   &      32   &      4.19  &      33	&   7.56 &  3C~315     &      11   &      1.49  &      22	&   1.13  \\  
3C~088     &      31   &      1.63  &      32	&   1.08 &  3C~317     &      31   &      2.68  &      22	&   1.68  \\  
3C~089     &      30   &      1.77  &      22	&   0.88 &  3C~318.1   &      31   &      2.47  &      22	&   1.13  \\  
3C~093.1   &      19   &      1.00  &      32	&   0.95 &  3C~319     &       8   &      1.45  &      33	&   0.96  \\  
3C~098     &      26   &      1.91  &      27	&   2.18 &  3C~321     &      29   &      1.51  &      11	&   0.91  \\  
3C~105     &      31   &      0.86  &      33	&   1.12 &  3C~326     &      31   &      1.16  &      22	&   1.08  \\  
3C~123     &      33   &      1.98  &      33	&   3.07 &  3C~327     &      30   &      1.77  &      33	&   1.74  \\  
3C~129     &      33   &      0.69  &      33	&   0.98 &  3C~338     &      31   &      5.54  &      22	&   1.35  \\  
3C~129.1   &      32   &      0.88  &      32	&   0.68 &  3C~348     &      29   &      0.92  &      21	&   0.86  \\  
3C~130     &      32   &      0.82  &      22	&   0.64 &  3C~353     &      33   &      2.06  &      22	&   1.28  \\  
3C~133     &      18   &      0.87  &      10	&   0.97 &  3C~357     &      28   &      1.38  &      28	&   1.07  \\  
3C~135     &      29   &      0.95  &      11	&   1.07 &  3C~379.1   &      22   &      1.11  &      22	&   0.96  \\  
3C~136.1   &      32   &      1.18  &      32	&   1.22 &  3C~381     &       8   &      2.92  &      29	&   0.63  \\  
3C~153     &      31   &      0.95  &      22	&   1.52 &  3C~386     &       3   &      2.04  &      25	&   1.35  \\  
3C~165     &      31   &      0.71  &      10	&   1.92 &  3C~388     &      29   &      2.79  &      31	&   3.05  \\  
3C~166     &      18   &      0.95  &      33	&   1.00 &  3C~401     &      28   &      1.31  &      16	&   0.75  \\  
3C~171     &      17   &      0.85  &      20	&   1.13 &  3C~402     &      17   &      2.13  &      28	&   1.36  \\  
3C~173.1   &      27   &      0.83  &      16	&   0.75 &  3C~403     &      29   &      1.26  &      29	&   1.99  \\  
3C~180     &       8   &      0.83  &      21	&   0.92 &  3C~424     &      30   &      1.52  &      31	&   0.63  \\  
3C~184.1   &      30   &      1.92  &      22	&   1.31 &  3C~430     &      28   &      1.02  &      18	&   1.08  \\  
3C~192     &      30   &      1.65  &      22	&   1.14 &  3C~433     &      29   &      2.04  &      31	&   1.54  \\  
3C~196.1   &      27   &      0.85  &       6	&   1.33 &  3C~436     &      30   &      0.74  &      28	&   1.32  \\  
3C~197.1   &      10   &      1.91  &      20	&   1.89 &  3C~438     &      26   &      0.84  &      24	&   1.06  \\  
3C~198     &      16   &      1.42  &       7	&   1.36 &  3C~442     &      30   &      7.43  &      22	&   1.09  \\  
3C~213.1   &      17   &      0.76  &       7	&   0.99 &  3C~449     &      27   &      9.70  &      28	&   1.29  \\  
3C~219     &       7   &      1.75  &      20	&   1.47 &  3C~452     &      31   &      2.24  &      22	&   0.98  \\  
3C~223     &      20   &      1.59  &      29	&   0.95 &  3C~456     &      15   &      1.43  &      26	&   1.07  \\  
3C~223.1   &      30   &      1.54  &      22	&   1.33 &  3C~459     &      14   &      0.81  &      27	&   1.12  \\  
3C~236     &      31   &      1.40  &      22	&   1.05 &  3C~460     &      29   &      0.89  &      33	&   0.98  \\  
3C~258     &       2   &      1.63  &       1	&   0.87 &  3C~465     &      31   &      2.20  &      22	&   1.79  \\  
\hline
    \end{tabular}
  \end{center}

  \begin{center}
Legenda for the SSP code.

    \begin{tabular}{l | c c c c c c c c c c c}
      \hline \hline
Age (Gyr)& 0.1 & 0.3 & 0.5 & 1 & 2 & 3 & 5 & 7 & 9 & 11 & 13 \\
\hline\hline
Z=0.008 &   1 &   2 &  3 &  4 &  5 &  6 &  7 &  8 &  9 & 10 & 11 \\
Z=0.02 &  12 &  13 & 14 & 15 & 16 & 17 & 18 & 19 & 20 & 21 & 22 \\
Z=0.05 &  23 &  24 & 25 & 26 & 27 & 28 & 29 & 30 & 31 & 32 & 33 \\
\hline
    \end{tabular}
  \end{center}
\end{table*}

In Fig. \ref{ssp} we show a few examples of the process of removal of the
galaxy starlight. The six spectra presented are representative of the
different quality in the sample (in order of decreasing
signal-to-noise ratio from top to bottom). We show separately 
galaxies with different levels of line
equivalent widths (high EW in the right two columns, low EW in the left 2
columns). 

\begin{figure*}[ht]
  \centerline{ \psfig{figure=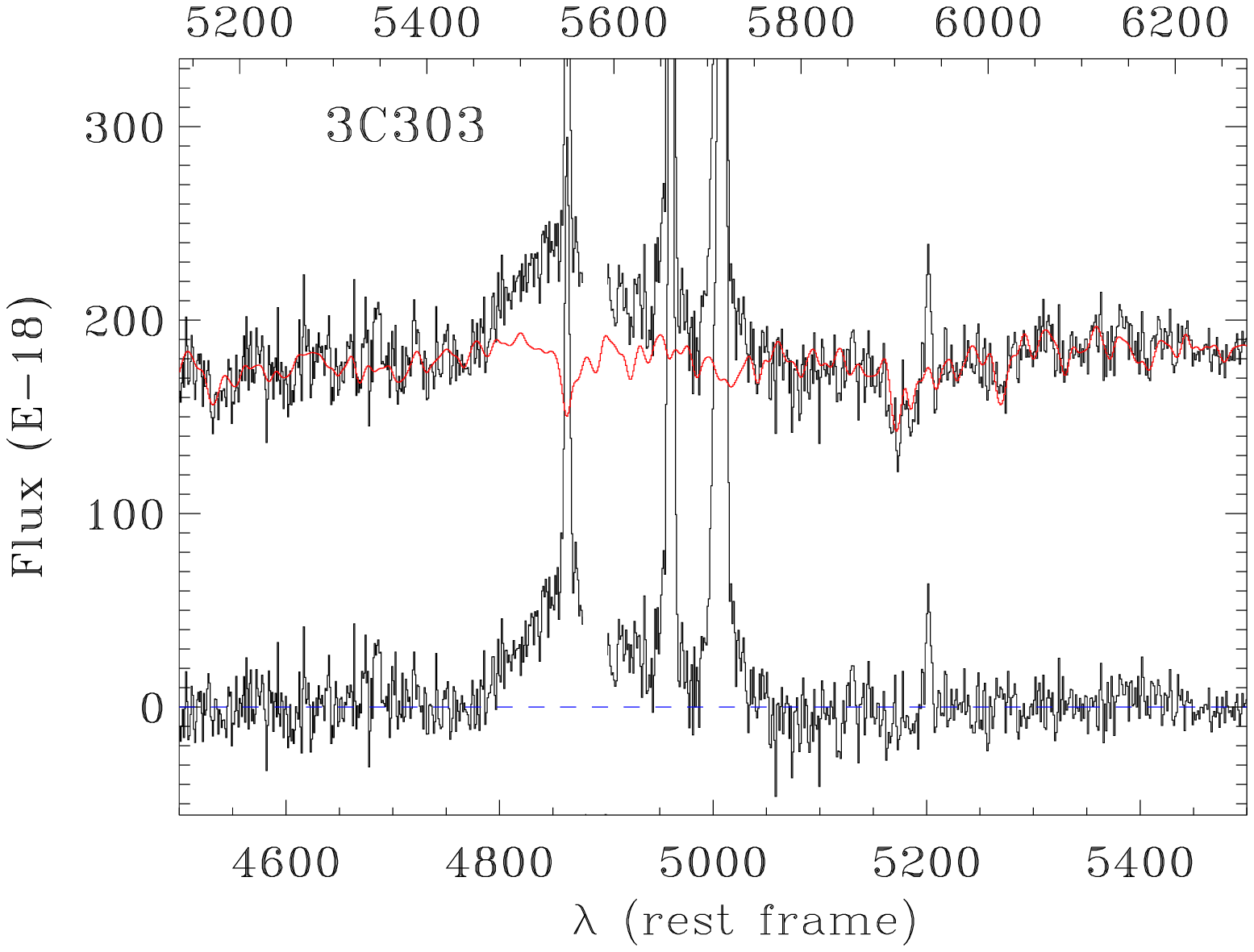,width=0.5\linewidth}
    \psfig{figure=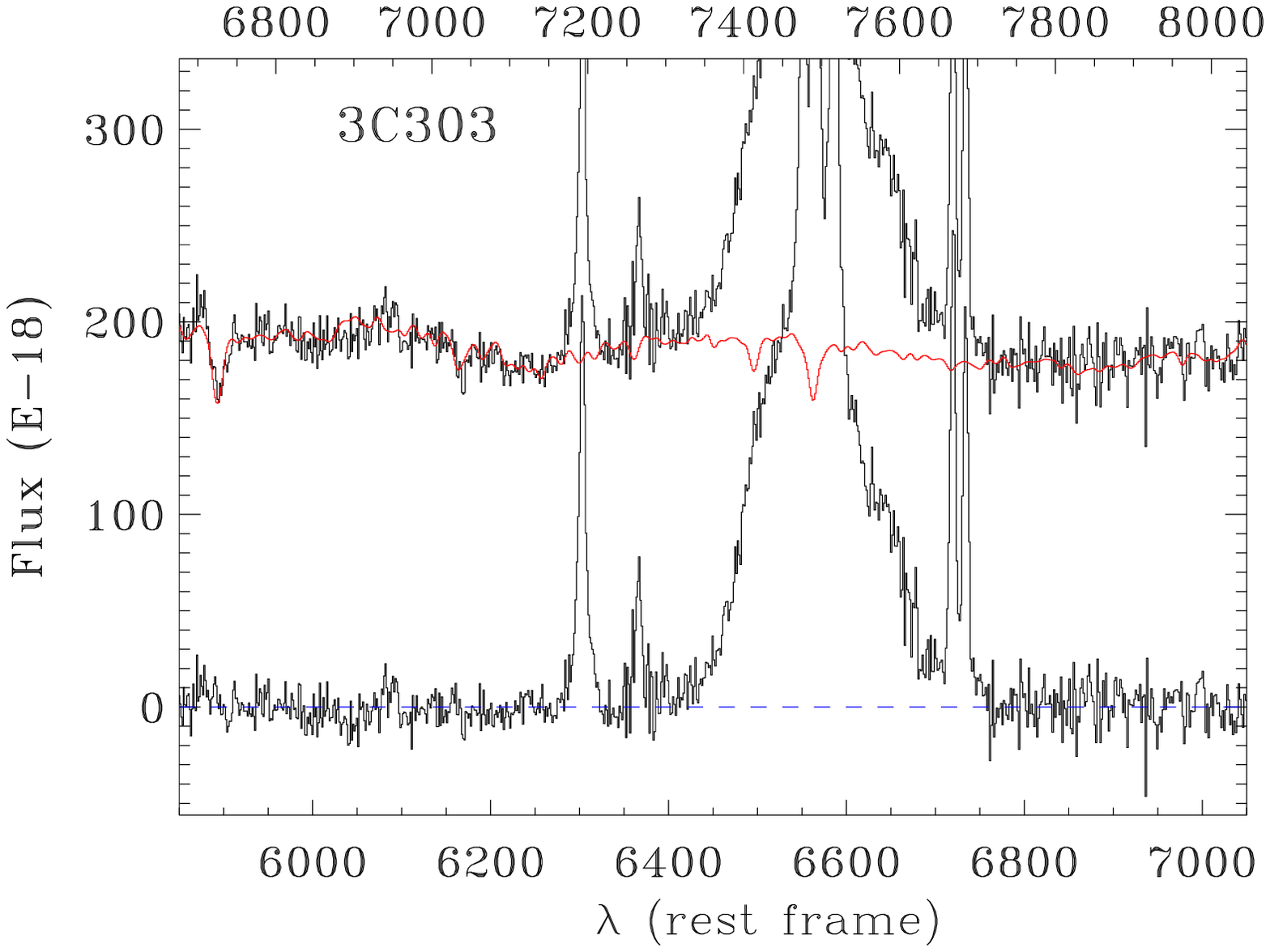,width=0.5\linewidth} }
  \caption{\label{sspbroad} 3C~303: example of starlight removal for a
    radio galaxy with prominent broad line emission. For the line legend
      see Fig.\ref{ssp}.}
\end{figure*}

In a few galaxies the continuum is essentially featureless 
(see for example, 3C~084 in Fig. \ref{spectra}) and it is likely to
be dominated by non-stellar emission.
No starlight subtraction was performed for these objects.  In several spectra a
first attempt to model the starlight left substantial positive residuals in
the region surrounding the \Hb\ and/or \Ha\ lines, indicative of the presence
of a broad line component.  In these cases we flagged a spectral region
extending over between 100 and 200 \AA\ around the broad line, extended the
spectral range considered, and repeated the fitting procedure (see Fig.
\ref{sspbroad} for an example).  

\begin{figure*}[ht]
  \centerline{ 
    \psfig{figure=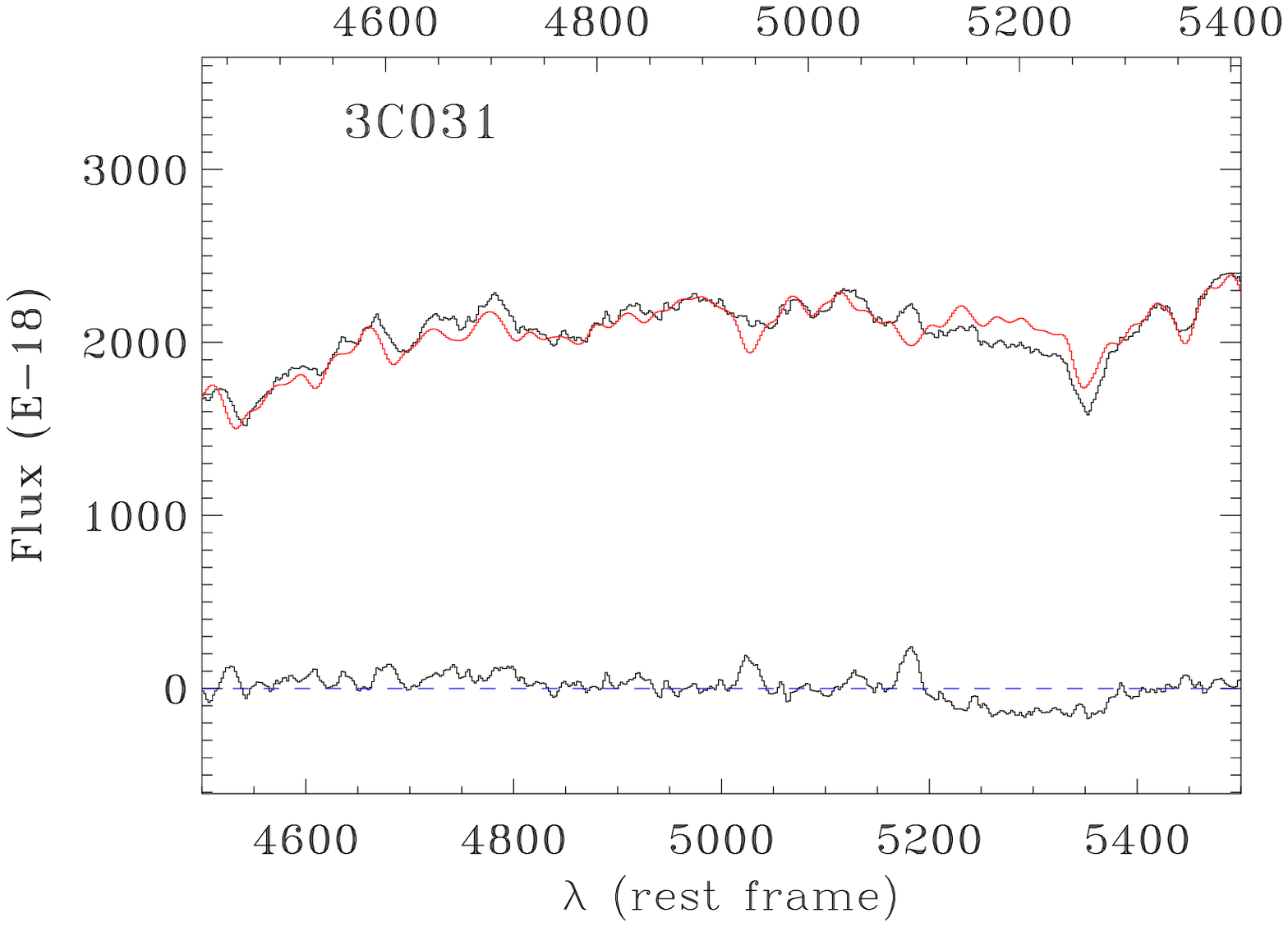,width=0.47\linewidth} 
    \psfig{figure=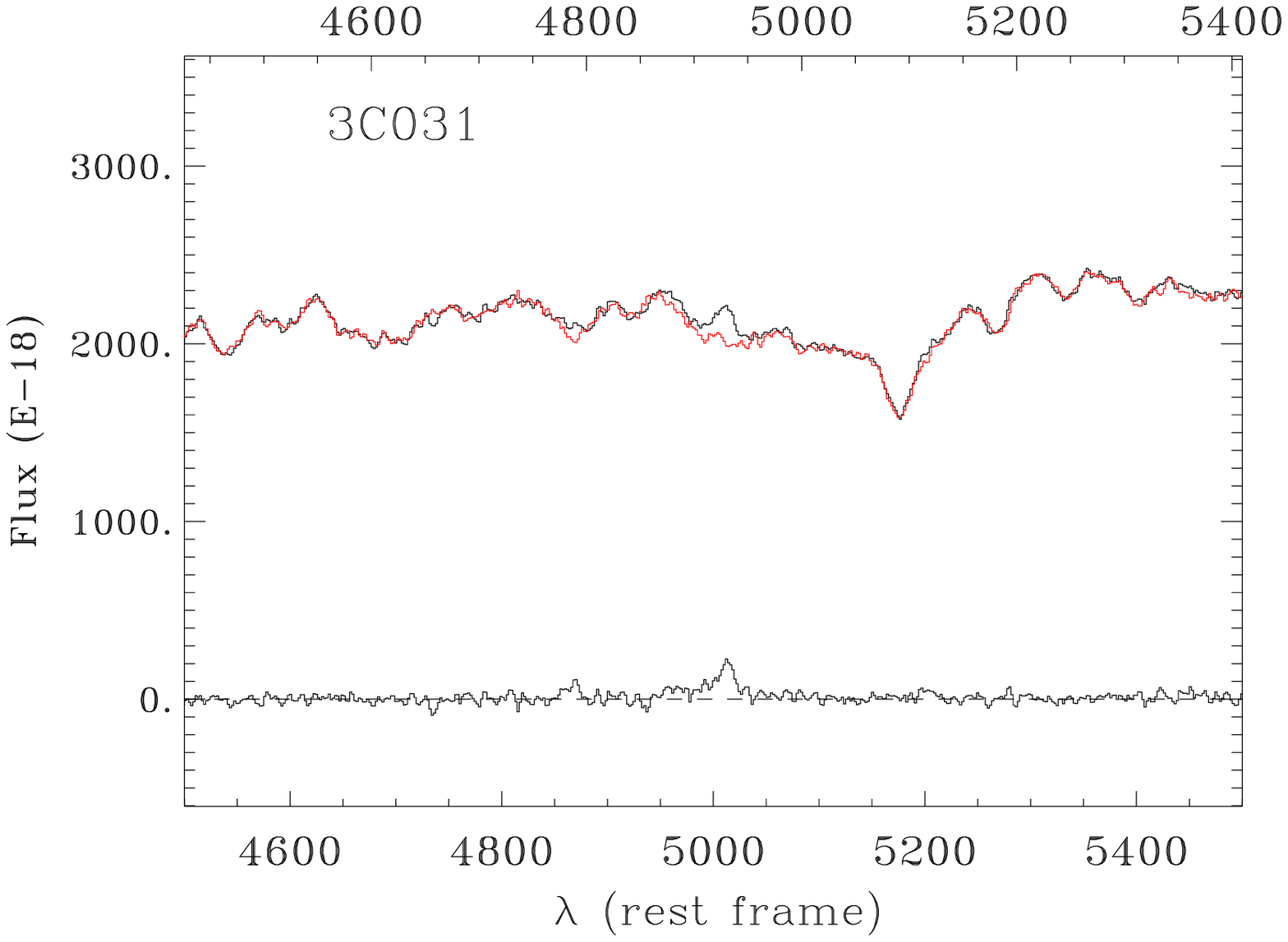,width=0.47\linewidth} }
  \caption{\label{3c031} 3C~031: example of starlight subtraction using the
    off-nuclear galactic emission as template. We compare the results obtained
    from the standard modeling technique (left panel) and from the off-nuclear
    subtraction (right panel). In the latter case, the residuals have a lower
    amplitude and show the presence of the H$\beta$ and [O III] lines. 
  For the line legend see Fig.\ref{ssp}.}
\end{figure*}

For a few objects, we used a different
technique to remove the galactic emission because we did not find a
satisfactory fit using the template models.  We then used as template the
spectra extracted from two off-nuclear regions, flanking the nuclear aperture,
summed over a spatial aperture of 2$\arcsec$ and appropriately scaled to match
the nuclear spectrum. An example of this method is given in Fig. \ref{3c031}.
The objects in which one of these different approaches was adopted are marked
appropriately in Table \ref{logoss}.

The case of 3C~386 deserves a special note.  \citet{lynds71} reports that the
peculiar observational characteristics of 3C~386 result from the chance
superposition of a F7 star on the galaxy's nucleus. This is confirmed by our
spectra that show strong Balmer absorption lines at zero 

\begin{figure}[htbp]
  \centerline{ 
    \psfig{figure=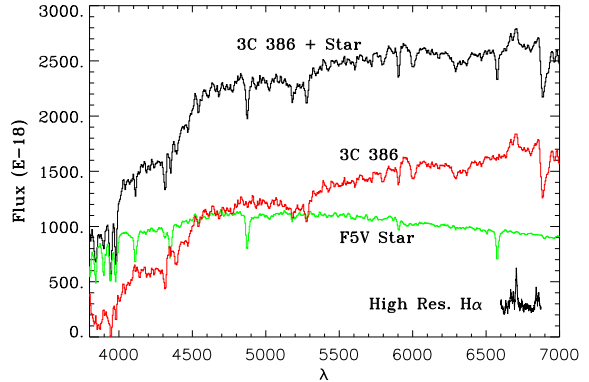,width=0.9\linewidth} }
  \caption{\label{3c386} The case of 3C~386: a Galactic star superposed on the
    radio galaxy nucleus. We show the low resolution spectrum, the F5V stellar
    template adopted, and the resulting spectrum of the galaxy. In the bottom
    right corner we show also the spectrum of 3C~386 around the \Ha\ line
    obtained from the high resolution data (vertically offset for clarity).}
\end{figure}

\noindent
redshift accompanied
by absorption and emission features at the redshift of the radio galaxy,
z=0.0177 (see Fig. \ref{3c386}).  

Before modeling the stellar population of
this galaxy, we removed the contribution of the foreground star. We obtain the
best match of the absorption lines of the star using a F5V stellar template
(from \citealt{pickles98}),
with the same reddening of the galaxy, E(B-V)= 0.335.  The low resolution
spectrum resulting from subtracting the foreground star is shown in Fig.
\ref{3c386} where we also present the H$\alpha$ spectral region from the high
resolution spectrum, with a well visible H$\alpha$+[N II] triplet and [S II]
doublet. No broad H$\alpha$ line is visible in this spectrum in contrast to the
result of \citet{simpson96}.

\subsection{Measurement of the emission lines fluxes}
\label{line}

The next step of our analysis consists of the measurement of the emission line
intensities for which we used the {\it specfit} package in IRAF.  We measured
line intensities fitting Gaussian profiles to H$\beta$,
[O~III]$\lambda\lambda$4959,5007, [O~I]$\lambda\lambda$6300,64, H$\alpha$,
[N~II]$\lambda\lambda$6548,84, and [S~II]$\lambda\lambda$6716,31.  Some
constraints were adopted to reduce the number of free parameters: we required
the FWHM and the velocity to be the same for all the lines.  The integrated
fluxes of each line were free to vary except for those with known ratios from
atomic physics: i.e. the [O~I]$\lambda\lambda$6300,64,
[O~III]$\lambda\lambda$4959,5007 and [N~II]$\lambda\lambda$6548,84
doublets.  Where required, we insert a linear continuum.

Table \ref{bigtable} summarizes the intensities of the main emission lines
(dereddened for Galactic absorption) relative to the intensity of the
narrow component of H$\alpha$, for which we give flux and luminosity. To each
line we associated its relative error, as a percentage.  We placed upper
limits at a 3$\sigma$ level to the undetected, but diagnostically important,
emission lines by measuring the noise level in the regions surrounding the
expected positions of the lines, and adopting as line width the instrumental
resolution. When no emission lines are visible, we only report the upper limit
for \Ha.  The missing values in Table \ref{bigtable} correspond to lines
outside the coverage of the spectra or severely affected by telluric bands.

For the galaxies with a broad-line component we first attempted to reproduce
the BLR emission with a gaussian profile, allowing for velocity shifts with
respect to the narrow lines and also for line asymmetry. Most line profiles
were well reproduced, but in some cases (e.g. 3C~332) two gaussians had to be
included. Nonetheless, in 3C~111 and in 3C~445 the broad line profile 
is so complex, and the prominence of the broad component with respect to
narrow lines is so large, 
it precludes any attempt to measure the [N II] doublet and the narrow component
of H$\alpha$ (or H$\beta$ in the blue region of the spectrum).  For these
objects we use the [O III] line as reference instead
of the narrow \Ha\ component. No narrow lines are visible in 3C~273 and we
only report its broad \Ha\ flux. For all 18 galaxies with a 
broad \Ha\ component we also give its flux in Table \ref{bigtable}.
These objects will be discussed in more detail in Sect.
\ref{broad}.

\subsection{On the effects of template mis-match on the emission lines fluxes}
\label{template}

\begin{figure*}[htbp]
  \centerline{ 
    \psfig{figure=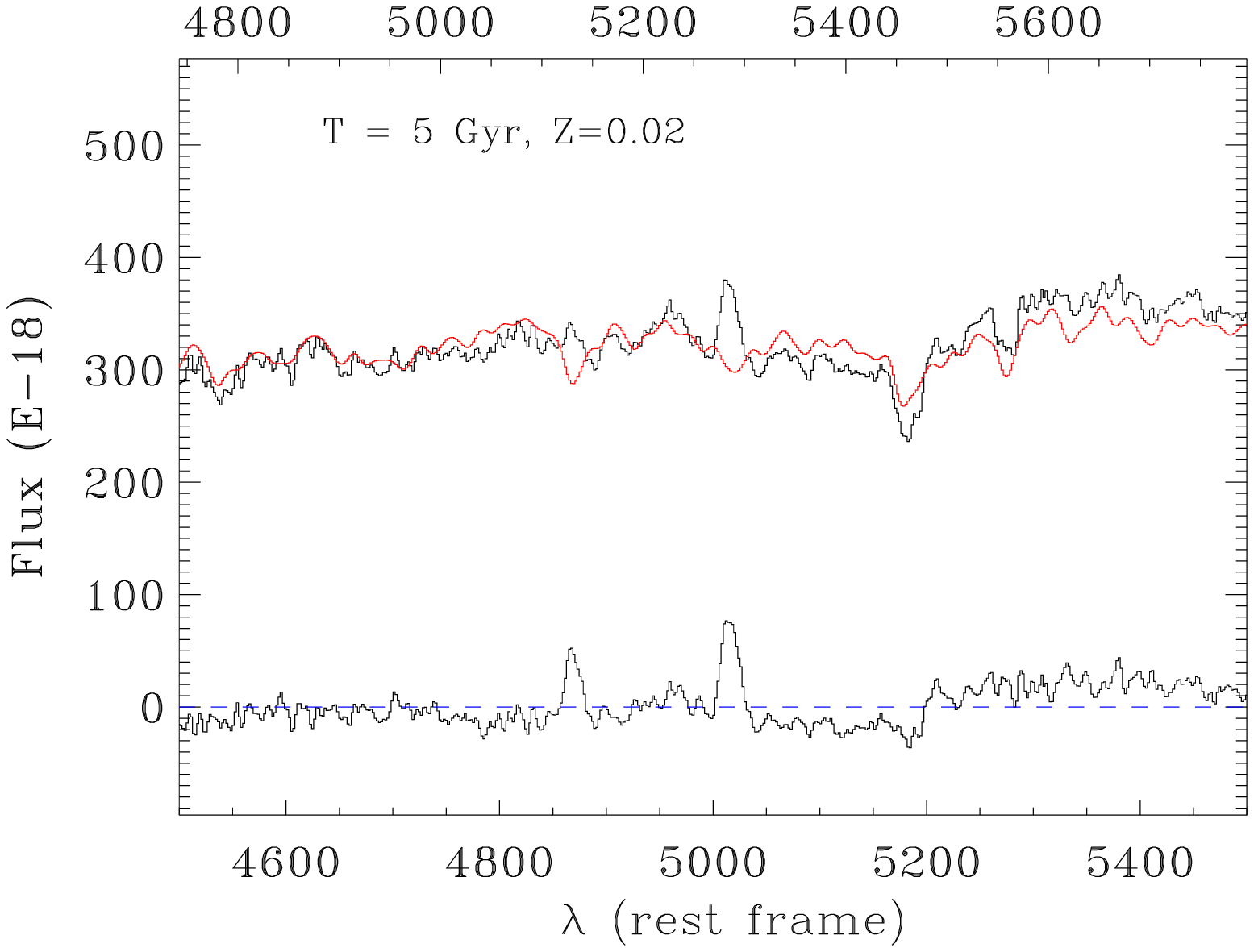,width=0.35\linewidth}
    \psfig{figure=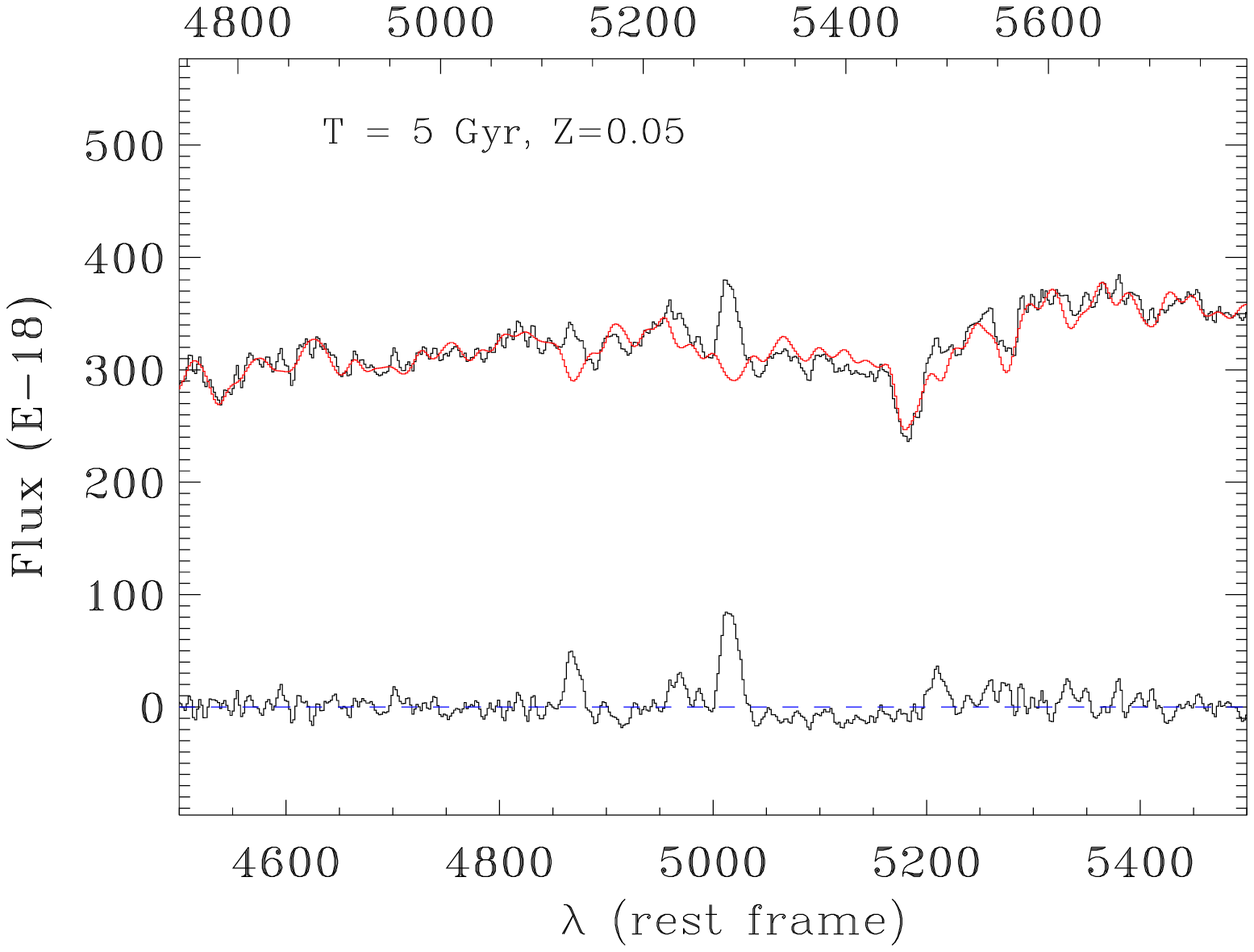,width=0.35\linewidth} 
    \psfig{figure=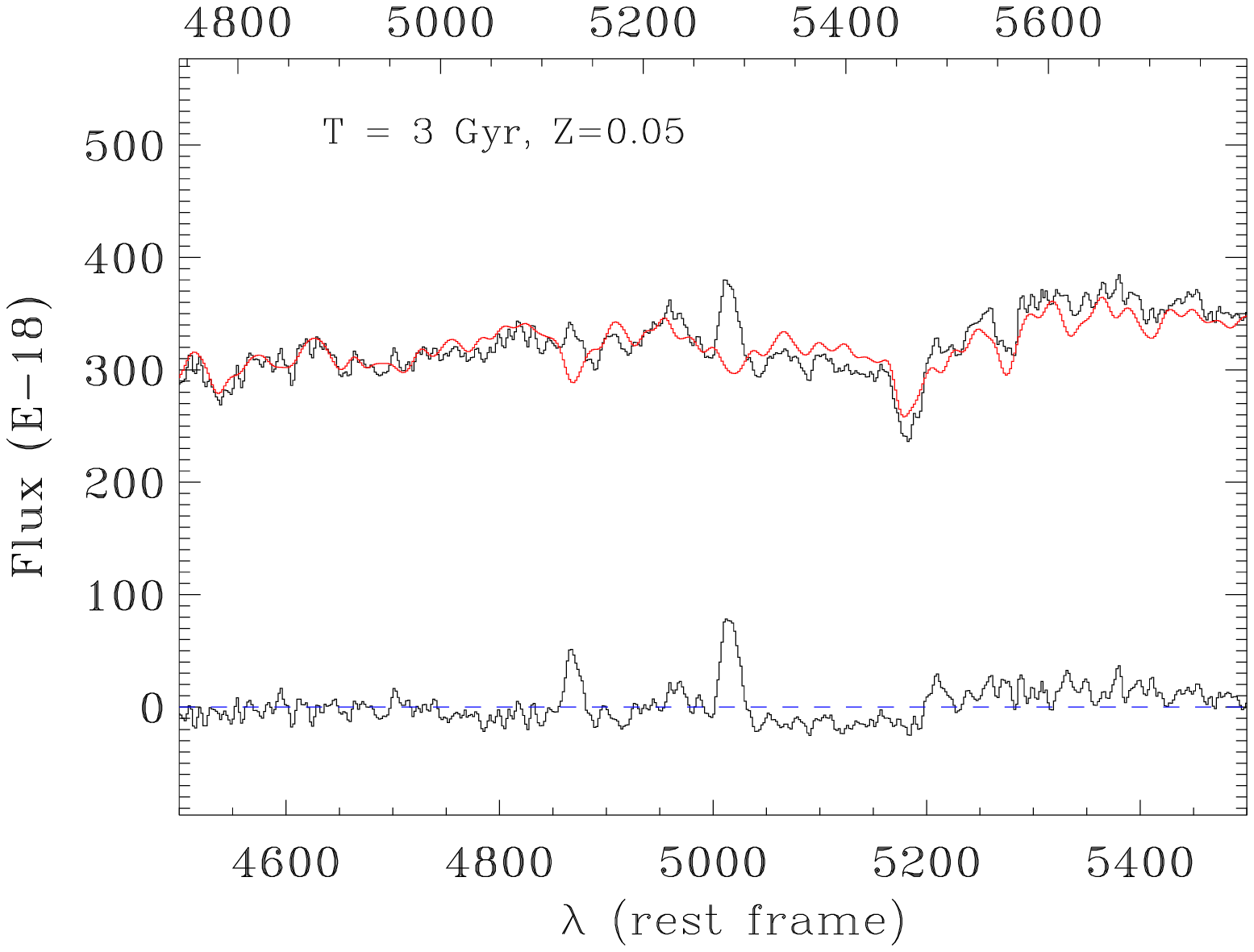,width=0.35\linewidth} }
  \caption{\label{3c310} Example of starlight removal using models differing
    by one step in stellar age or in metalicity. The center panel shows the
    best fit for 3C~310, while the two flanking figures have (left) lower
    metalicity and (right) younger stellar population.}
\end{figure*}

An important issue related to the subtraction of the stellar component is the
effect of the template mis-match for the measurements of the emission lines.
This is particularly important for the estimate of H$\beta$ in the
galaxies showing  emission lines of relatively low equivalent width. In fact
they can be strongly affected by the level of the
absorption features associated to the stellar population.

We consider as an example the case of 3C~310. Its spectrum is of average
quality in our dataset and it shows only weak emission lines, while the
H$\beta$ line is, at most, marginally detected in its spectrum.  The best
fitting stellar population has an age of 5 Gyr and a metalicity of Z=0.05 and
it is shown in red in the middle panel of Fig. \ref{3c310} superposed on the
original spectrum.  In the residual spectrum a well defined H$\beta$ line
emerges in emission, caused by the removal of a substantial absorption
associated with the stellar emission. This indicates that its intensity is
strongly influenced by the choice of the stellar template.  To associate a
proper error to this procedure it is necessary to establish the range of
acceptable stellar models (in terms of age and metalicity) and the resulting
uncertainty in the line measurement.

As a formal error propagation across all steps of the data reduction is clearly
unfeasible, we estimated the typical error of our spectra by measuring
the rms flux in continuum dominated spectral regions.  Even with this approach,
the value of minimum reduced $\chi^2_{\rm r}$ obtained from the best fitting
stellar population is often larger than the value indicative of a good fit
($\chi^2_{\rm r} \sim 1$) and this is in contrast with the fact that the
stellar models seem to trace in general the spectra rather well. There are
several reasons for this discrepancy: i) the $\chi^2_{\rm r}$ is not properly
normalized (e.g. because not all data points are independent) ii) our estimate
of the signal-to-noise ratio does not include the uncertainties in e.g. the
wavelength and flux calibration, iii) there are real mismatches between data
and models, in part caused by the use of a discrete grid of stellar models. We
then decided, when necessary, to rescale our error bars such that the overall best fitting
model provides $\chi^2$/d.o.f. = 1, following the approach proposed by
\citet{barth01} in a different context. This is a
conservative approach since it has the effect of increasing the range of the
acceptable templates with respect to what would 
have been obtained only using
the measured rms of each spectrum.

\onecolumn
\begin{landscape}
\begin{longtable}{l c c c c c c c c c c c c}

\caption[Emission line measurements.]{Emission line measurements.} 
\label{bigtable} \\

\hline \hline 
Name     & Redshift   & E(B-V) & L(H$\alpha$) & F(H$\alpha$) & H$\beta$  & [O III]$\lambda$5007 & [O I]$\lambda$6364 & [N II]$\lambda$6584 & [S II]$\lambda$6716 & [S II]$\lambda$6731 & F(H$\alpha$) broad\\
\hline	
\endfirsthead

\multicolumn{3}{c}{{\tablename} \thetable{} -- Continued} \\[0.5ex]
\hline \hline 
Name     & Redshift   & E(B-V) & L(H$\alpha$) & F(H$\alpha$) & H$\beta$  & [O III]$\lambda$5007 & [O I]$\lambda$6364 & [N II]$\lambda$6584 & [S II]$\lambda$6716 & [S II]$\lambda$6731 & F(H$\alpha$) broad\\
\hline
\endhead

\hline
  \multicolumn{10}{c}{{Continued on Next Page}} \\
\endfoot

  \\[-1.8ex] 
\endlastfoot

3C~015.0 & 0.073  &  0.022 &  40.40 &    -14.70 ( 2) &   0.32 (20) &   1.58 ( 4) &   0.29 ( 8) &   2.06 ( 1) &   0.32 ( 8) &   0.41 ( 8)  & \\
3C~017.0 & 0.220  &  0.023 &  41.88 &    -14.27 ( 1) &   0.24 ( 8) &   1.28 ( 1) &   0.44 ( 1) &   0.70 ( 1) &   0.23 ( 1) &   0.21 ( 2)  & -13.87 \\
3C~018.0 & 0.188  &  0.158 &  41.93 &    -14.06 ( 1) &   0.33 ( 1) &   4.17 ( 1) &   0.43 ( 3) &   1.13 ( 1) &   0.38 ( 2) &   0.42 ( 1)  & -13.03 \\
3C~028.0 & 0.195  &  0.058 &  41.51 &    -14.52 ( 2) &   0.53 (15) &   0.28 (20) &   0.17 (10) &   0.92 ( 3) &   0.55 (10) &          --  & \\
3C~029.0 & 0.045  &  0.036 &  40.06 &    -14.60 ( 6) &   0.24 (25) &   1.07 ( 2) &   0.19 (16) &   1.85 ( 1) &   0.50 ( 6) &   0.52 ( 4)  & \\
3C~031.0 & 0.017  &  0.001 &  39.83 &    -13.96 ( 1) &   0.15 ( 8) &   0.43 ( 2) &   0.14 (12) &   0.99 ( 1) &   0.37 ( 1) &   0.32 ( 1)  & \\
3C~033.0 & 0.060  &  0.028 &  41.63 &    -13.29 ( 1) &   0.31 ( 1) &   3.55 ( 1) &   0.25 ( 1) &   0.63 ( 1) &   0.39 ( 1) &   0.33 ( 1)  & \\
3C~033.1 & 0.181  &  0.633 &  41.85 &    -14.11 ( 1) &   0.22 ( 5) &   2.80 ( 1) &   0.25 ( 1) &   0.57 ( 1) &   0.27 ( 1) &   0.21 ( 1)  & -13.26  \\
3C~035.0 & 0.067  &  0.141 &  40.22 &    -14.81 ( 1) & $<$  1.27   &   0.62 (23) &   0.46 ( 5) &   0.77 ( 2) &   0.62 (15) &          --  & \\
3C~040.0 & 0.019  &  0.041 &  39.08 &    -14.79 (11) &   0.32 (32) &   1.38 ( 6) &   0.24 (28) &   2.32 ( 1) &   0.81 (10) & $<$  0.52    & \\
3C~052.0 & 0.285  & 0.232  &$<$ 40.64 & $<$ -15.76   &     --      &         --  &         --  &        --   &        --   &          --  & \\
3C~061.1 & 0.184  &  0.176 &  42.05 &    -13.92 ( 1) &   0.25 ( 4) &   2.63 ( 1) &   0.08 (10) &   0.31 ( 1) &   0.16 ( 1) &   0.15 ( 3)  & \\
3C~066B  & 0.022  &  0.080 &  40.11 &    -13.90 ( 4) &   0.22 ( 7) &   0.87 ( 1) &   0.26 (16) &   2.45 ( 1) &   0.56 ( 9) &       --     & \\
3C~075N  & 0.023  &  0.180 &  39.58 &    -14.50 ( 1) & $<$  2.20   & $<$  2.20   &   0.42 ( 2) &   2.48( 1)  & $<$  0.69   &   0.37 ( 1)  & \\
3C~076.1 & 0.033  &  0.138 &  39.89 &    -14.50 ( 2) & $<$  0.85   & $<$  0.92   & $<$  0.18   &   1.57 ( 1) &   0.33 ( 4) &   0.54 ( 1)  & \\
3C~078.0 & 0.029  &  0.173 &  39.73 &    -14.53 ( 3) &   0.18 (30) &   0.48 (16) &   0.18 (13) &   1.88 ( 2) &       --    &       --     & \\
3C~079.0 & 0.256  &  0.127 &  42.39 &    -13.91 ( 1) &   0.29 ( 3) &   2.97 ( 1) &   0.05 ( 2) &   0.32 ( 1) &   0.12 ( 3) &   0.11 ( 7)  & \\
3C~083.1 & 0.027  &  0.164 &  39.40 &    -14.83 (14) & $<$  0.72   & $<$  1.25   &       --    &   1.35 ( 3) &       --    &       --     & \\
3C~084.0 & 0.018  &  0.163 &  41.28 &    -12.55 ( 1) &   0.42 ( 1) &   2.09 ( 1) &   0.64 ( 1) &   1.12 ( 1) &   0.54 ( 1) &   0.51 ( 1)  & \\
3C~088.0 & 0.030  &  0.126 &  39.98 &    -14.33 ( 1) &   0.29 (11) &   1.44 ( 2) &   0.50 ( 3) &   2.39 ( 1) &   0.97 ( 1) &   0.79 ( 3)  & \\
3C~089.0 & 0.139  & 0.134  &  40.28 &    -15.42 (11) & $<$  1.86   & $<$  1.69   & $<$  1.26   &   1.43 ( 7) &       --    &       --     & \\
3C~093.1 & 0.243  &  0.389 &  42.35 &    -13.89 ( 1) &   0.28 ( 4) &   2.08 ( 1) &   0.28 ( 3) &   1.36 ( 1) &   0.54 ( 1) &   0.55 ( 1)  & \\
3C~098.0 & 0.030  &  0.229 &  40.52 &    -13.79 ( 1) &   0.25 ( 3) &   3.01 ( 1) &   0.15 ( 3) &   0.76 ( 1) &   0.34 (10) &   0.23 ( 7)  & \\
3C~105.0 & 0.089  &  0.480 &  40.89 &    -14.39 ( 3) &   0.26 (28) &   3.59 ( 1) &   0.38 ( 5) &   1.59 ( 1) &   0.55 ( 4) &   0.55 ( 1)  & \\
3C~111.0 & 0.049  &  1.647 &42.44$^a$& -12.28$^a$ (1)   &   --     & 1.00$^a$( 1)&   0.04$^a$ (10)  &  -- &  0.03$^a$ ( 7) & 0.03$^a$ ( 9 )& -11.64 \\ 
3C~123.0 & 0.218  &  0.981 &  41.96 &    -14.18 ( 1) &   0.61 (32) &   1.09 (18) &   0.25 (32) &   2.34 ( 1) &   0.48 ( 1) &   0.35 ( 1)  & \\
3C~129.0 & 0.022  &  1.058 &  39.81 &    -14.20 ( 1) & $<$  0.99   & $<$  1.10   &   0.61 ( 1) &   1.51 ( 1) &   0.50 ( 1) &   0.50 (20)  & \\
3C~129.1 & 0.022  &  1.131 & $<$ 39.83 & $<$ -14.21  &      --     &         --  &         --  &        --   &        --   &          --  & \\
3C~130.0 & 0.032  &  1.309 & $<$ 40.17 & $<$ -14.19  &      --     &         --  &         --  &        --   &        --   &          --  & \\
3C~133.0 & 0.278  &  0.949 &  42.41 &    -13.97 ( 1) &   0.32 ( 1) &   2.26 ( 1) &   0.25 ( 9) &   0.70 ( 1) &   0.17 ( 1) &   0.14 ( 1)  & \\
3C~135.0 & 0.125  &  0.115 &  41.52 &    -14.09 ( 1) &   0.33 ( 3) &   3.40 ( 1) &   0.18 ( 2) &   0.80 ( 1) &   0.30 ( 1) &   0.30 ( 3)  & \\
3C~136.1 & 0.064  &  0.762 &  41.41 &    -13.57 ( 1) &   0.14 ( 1) &   1.08 ( 1) &   0.05 ( 1) &   0.59 ( 1) &   0.20 ( 1) &   0.20 ( 3)  & \\
3C~153.0 & 0.277  &  0.162 &  41.60 &    -14.77 ( 3) &   0.23 (15) &   1.07 ( 3) &   0.36 ( 7) &   1.21 ( 3) &   0.54 ( 8) &   0.33 (13)  & \\
3C~165.0 & 0.296  &  0.174 &  41.44 &    -15.00 (15) &   0.44 (15) &   1.68 ( 4) &   0.31 (32) &   1.14 (27) &   0.32 (27) &   0.33 (27)  & \\
3C~166.0 & 0.245  &  0.211 &  41.51 &    -14.75 ( 3) &   0.42 ( 6) &   1.40 ( 2) &   0.38 ( 7) &   0.73 ( 4) &   0.41 (16) &   0.34 (24)  & \\
3C~171.0 & 0.238  &  0.054 &  42.45 &    -13.78 ( 1) &   0.36 ( 1) &   2.73 ( 1) &   0.24 ( 2) &   0.57 ( 1) &   0.38 ( 6) &   0.29 ( 2)  & \\
3C~173.1 & 0.292  &  0.044 &  41.05 &    -15.38 ( 7) &  $<$   0.23 &   0.63 (12) &   0.24 (28) &   2.04 ( 3) &   0.24 (29) &   0.31 (24)  & \\
3C~180.0 & 0.220  &  0.098 &  41.79 &    -14.36 ( 1) &   0.25 ( 4) &   3.53 ( 1) &   0.10 (28) &   0.69 ( 1) &   0.28 ( 3) &   0.19 ( 5)  & \\
3C~184.1 & 0.118  &  0.032 &  41.79 &    -13.77 ( 3) &   0.29 ( 4) &   2.71 ( 1) &   0.08 ( 7) &   0.27 ( 8) &   0.11 ( 1) &   0.09 ( 3)  & -13.99 \\
3C~192.0 & 0.060  &  0.054 &  40.95 &    -13.97 ( 1) &   0.30 ( 1) &   2.48 ( 1) &   0.11 ( 3) &   0.71 ( 1) &   0.35 ( 1) &   0.26 ( 1)  & \\
3C~196.1 & 0.198  &  0.065 &  41.56 &    -14.48 ( 2) &   0.22 (10) &   0.91 ( 3) &   0.20 (15) &   1.19 ( 1) &   0.52 ( 1) &   0.49 ( 1)  & \\
3C~197.1 & 0.128  &  0.041 &  40.69 &    -14.93 ( 3) &   0.37 (11) &   1.69 ( 2) &   0.33 ( 8) &   0.76 ( 4) &   0.29 (11) &   0.31 ( 9)  & -13.95 \\
3C~198.0 & 0.082  &  0.026 &  41.31 &    -13.89 ( 1) &   0.27 ( 1) &   0.46 ( 1) &   0.05 ( 9) &   0.28 ( 1) &       --    &       --     & \\
3C~213.1 & 0.194  &  0.028 &  41.01 &    -15.02 ( 3) &   0.21 (14) &   1.13 ( 3) &   0.46 ( 6) &   1.41 ( 2) &   0.38 (10) &   0.28 (12)  & \\
3C~219.0 & 0.175  &  0.018 &  41.55 &    -14.38 ( 2) &   0.25 (10) &   1.67 ( 1) &   0.44 ( 3) &   0.90 ( 1) &   0.24 ( 6) &   0.24 ( 7)  & -13.87 \\
3C~223.0 & 0.137  &  0.012 &  41.68 &    -14.01 ( 1) &   0.23 ( 5) &   3.09 ( 1) &   0.19 ( 3) &   0.63 ( 1) &   0.25 ( 3) &   0.21 ( 3)  & \\
3C~223.1 & 0.107  &  0.017 &  41.16 &    -14.30 ( 1) &   0.28 ( 7) &   2.63 ( 1) &   0.06 (15) &   0.81 ( 1) &   0.22 ( 4) &   0.19 ( 5)  & \\
3C~227.0 & 0.086  &  0.026 &  41.08 &    -14.17 ( 2) &   0.44 ( 1) &   4.73 ( 1) &   0.08 (18) &   0.19 ( 4) &   0.16 ( 5) &   0.16 (11)  & -12.52 \\
3C~234.0 & 0.185  &  0.019 &  42.64 &    -13.33 ( 1) &   0.25 ( 2) &   2.96 ( 1) &   0.04 ( 8) &   0.28 ( 1) &   0.08 ( 3) &   0.07 ( 1)  & -13.29 \\
3C~236.0 & 0.099  &  0.011 &  41.13 &    -14.25 ( 1) &   0.22 ( 4) &   0.57 ( 2) &   0.30 ( 3) &   0.69 ( 1) &   0.49 ( 2) &   0.35 ( 3)  & \\
3C~258.0 & 0.165  &  0.020 &  40.96 &    -14.90 (23) &   0.11 (24) &   0.17 (15) & $<$  0.47   & $<$  2.14   &       --    &       --     & \\
3C~264.0 & 0.022  &  0.023 &  39.68 &    -14.35 ( 1) &   0.27 ( 7) &   0.33 (10) &   0.22 ( 9) &   1.45 ( 1) &   0.33 ( 8) &   0.33 (22)  & \\
3C~270.0 & 0.007  &  0.018 &  38.13 &    -14.93 ( 7) & $<$  1.09   &   0.63 (25) &   0.52 (16) &   0.72 (14) & $<$  0.37   & $<$  0.53    & \\
3C~272.1 & 0.004  &  0.040 &  38.92 &    -13.57 ( 1) &   0.10 ( 3) &   0.19 ( 4) &   0.23 ( 6) &   1.28 ( 1) &   0.52 ( 1) &   0.34 ( 1)  & \\
3C~273.0 & 0.158  &  0.021 &   --    &    --           &   --         &    --        &        --    &  --          &   --         &   --  & -11.51 \\ 
3C~274.0 & 0.004  &  0.022 &  39.50 &    -13.11 ( 1) &   0.17 ( 1) &   0.31 ( 1) &   0.36 ( 1) &   2.32 ( 1) &   0.68 ( 1) &   0.77 ( 1)  & \\
3C~277.3 & 0.086  &  0.012 &  40.83 &    -14.43 ( 1) &   0.19 ( 9) &   1.29 ( 1) &   0.29 ( 5) &   0.79 ( 3) &   0.32 ( 1) &   0.28 ( 5)  & \\
3C~284.0 & 0.239  &  0.016 &  41.41 &    -14.82 ( 7) &   0.20 (12) &   1.52 ( 1) & $<$  0.24   &   0.61 (11) & $<$  0.45   & $<$  0.41    & \\
3C~285.0 & 0.079  &  0.017 &  40.66 &    -14.52 ( 1) &   0.19 ( 6) &   0.78 ( 1) &   0.10 (10) &   0.54 ( 2) &   0.27 ( 4) &   0.19 ( 6)  & \\
3C~287.1 & 0.216  &  0.025 &  41.50 &    -14.62 ( 3) &   0.27 ( 9) &   1.71 ( 1) &   0.48 ( 4) &   0.68 ( 4) &   0.29 ( 7) &   0.29 ( 7)  & -13.85 \\
3C~293.0 & 0.045  &  0.017 &  40.18 &    -14.49 ( 3) &   0.19 (17) &   0.42 (10) &   0.26 ( 3) &   0.88 ( 1) &   0.66 ( 2) &   0.66 (10)  & \\
3C~296.0 & 0.025  &  0.025 &  39.87 &    -14.28 ( 1) &   0.30 (12) &   0.81 ( 2) &   0.22 (23) &   1.84 ( 1) &   0.43 ( 2) &   0.38 (10)  & \\
3C~300.0 & 0.272  &  0.035 &  41.78 &    -14.58 ( 2) &   0.25 ( 9) &   1.71 ( 1) &   0.15 (13) &   0.48 ( 4) &   0.33 ( 4) &   0.23 ( 4)  & \\
3C~303.0 & 0.141  &  0.019 &  41.33 &    -14.39 ( 1) &   0.35 ( 4) &   2.55 ( 1) &   0.41 ( 2) &   1.06 ( 1) &   0.46 ( 2) &   0.39 ( 3)  & -13.46 \\
3C~303.1 & 0.269  &  0.036 &  42.10 &    -14.24 ( 1) &   0.26 ( 2) &   2.07 ( 1) &   0.24 ( 3) &   1.03 ( 1) &   0.38 ( 2) &   0.48 ( 1)  & \\
3C~305.0 & 0.042  &  0.026 &  40.92 &    -13.68 ( 1) &   0.12 ( 4) &   1.30 ( 1) &   0.17 ( 7) &   1.77 ( 1) &   0.48 ( 1) &   0.40 ( 1)  & \\
3C~310.0 & 0.054  &  0.042 &  40.32 &    -14.50 ( 1) &   0.23 ( 7) &   0.54 ( 3) &   0.30 ( 4) &   1.74 ( 1) &   0.84 ( 1) &   0.74 ( 2)  & \\
3C~314.1 & 0.120  &  0.020 &  40.31 &    -15.25 ( 4) &   0.37 (16) &   0.24 (24) &   0.18 (23) &   0.53 ( 7) &   0.52 ( 8) &   0.33 (13)  & \\
3C~315.0 & 0.108  &  0.062 &  41.15 &    -14.32 ( 1) &   0.20 ( 4) &   0.53 ( 1) &   0.25 ( 2) &   0.72 ( 1) &   0.51 ( 1) &   0.37 ( 2)  & \\
3C~317.0 & 0.034  &  0.037 &  40.35 &    -14.08 ( 1) &   0.30 ( 5) &   1.00 ( 2) &   0.25 ( 5) &   1.92 ( 1) &   0.58 ( 1) &   0.49 ( 1)  & \\
3C~318.1 & 0.044  &  0.035 &  39.95 &    -14.70 ( 4) &  $<$ 0.58   &   0.26 (32) &   0.21 (19) &   1.02 ( 3) &   0.29 ( 9) &   0.16 (13)  & \\
3C~319.0 & 0.189  & 0.012  &  41.16 &    -14.84 ( 7) & $<$  0.13   & $<$  0.10   & $<$  0.27   &   0.30 ( 6) &   0.15 (25) & $<$  0.14    & \\
3C~321.0 & 0.097  &  0.044 &  40.50 &    -14.87 ( 2) &   0.30 (11) &   2.58 ( 1) &   0.06 (28) &   0.50 ( 3) &   0.18 ( 8) &   0.15 (11)  & \\
3C~323.1 & 0.264  &  0.042 &  42.21 &    -14.12 ( 1) &   0.26 (11) &   3.93 ( 1) &   0.20 (11) &   0.29 (32) &   0.10 (11) &   0.08 (21)  & -12.37 \\
3C~326.0 & 0.090  &  0.053 &  40.28 &    -15.02 ( 5) & $<$  0.23   &   1.31 ( 4) &   0.37 (13) &   1.93 ( 3) &   0.41 (12) &   0.50 (11)  & \\
3C~327.0 & 0.104  &  0.089 &  41.73 &    -13.70 ( 1) &   0.27 ( 4) &   3.20 ( 1) &   0.14 ( 1) &   0.73 ( 1) &   0.30 ( 1) &   0.24 ( 1)  & \\
3C~332.0 & 0.151  &  0.024 &  41.31 &    -14.47 ( 2) &   0.28 ( 8) &   3.14 ( 1) &   0.21 ( 9) &   0.97 ( 1) &   0.37 ( 6) &   0.36 ( 7)  & -12.77 \\
3C~338.0 & 0.032  &  0.012 &  40.25 &    -14.11 ( 1) &   0.18 (19) &   0.21 ( 6) &   0.18 ( 2) &   1.63 ( 1) &   0.41 ( 1) &   0.33 ( 1)  & \\
3C~348.0 & 0.154  &  0.094 &  41.29 &    -14.51 ( 1) &   0.25 ( 5) &   0.13 ( 9) &   0.27 ( 5) &   1.27 ( 1) &       --    &       --     & \\
3C~353.0 & 0.030  &  0.439 &  40.42 &    -13.90 ( 1) &   0.20 (17) &   0.53 ( 7) &   0.30 ( 2) &   1.09 ( 1) &   0.56 ( 1) &   0.43 ( 1)  & \\
3C~357.0 & 0.166  &  0.045 &  40.92 &    -14.96 ( 3) &   0.23 (16) &   1.08 ( 4) &   0.46 ( 7) &   1.37 ( 4) &       --    &       --     & \\
3C~371.0 & 0.050  &  0.036 &  40.94 &    -13.82 ( 1) &   0.29 (10) &   1.01 ( 4) &   0.51 ( 2) &   1.14 ( 1) &   0.32 ( 3) &   0.29 ( 1)  & \\
3C~379.1 & 0.256  &  0.062 &  41.41 &    -14.89 ( 6) &   0.31 ( 6) &   2.80 ( 1) &   0.21 (19) &   1.54 ( 5) &   0.30 (13) &   0.28 (14)  & \\
3C~381.0 & 0.161  &  0.053 &  41.79 &    -14.05 ( 1) &   0.31 ( 2) &   3.83 ( 1) &   0.11 ( 9) &   0.53 ( 1) &       --    &       --     & \\
3C~382.0 & 0.058  &  0.070 &  41.39 &    -13.51 ( 1) &   0.31 ( 1) &   2.45 ( 1) &   0.29 ( 1) &   1.49 ( 1) &   0.12 ( 8) &   0.13 ( 9)  & -11.61 \\
3C~386.0 & 0.017  &  0.335 &  40.17 &    -13.63 ( 1) & $<$  1.08   & $<$  1.08   &   0.10 (17) &   0.57 ( 1) &   0.11 ( 7) &   0.08 (10)  & \\
3C~388.0 & 0.091  &  0.080 &  40.83 &    -14.47 ( 2) &   0.23 (14) &   0.74 ( 3) &   0.26 (13) &   2.33 ( 1) &   0.41 ( 1) &   0.37 ( 1)  & \\
3C~390.3 & 0.056  &  0.071 &  41.57 &    -13.29 ( 1) &   0.32 ( 1) &   3.24 ( 1) &   0.27 ( 1) &   0.47 ( 1) &   0.10 ( 1) &   0.10 ( 1)  & -11.60 \\
3C~401.0 & 0.201  &  0.059 &  41.01 &    -15.05 ( 3) &   0.30 (15) &   1.10 ( 5) &   0.24 (16) &   1.77 ( 2) &   0.46 ( 9) &   0.34 (14)  & \\
3C~402.0 & 0.024  &  0.121 &  39.08 &    -15.03 ( 3) & $<$  1.95   & $<$  2.19   &   0.37 (13) &   2.97 ( 1) & $<$  0.44   & $<$  0.97    & \\
3C~403.0 & 0.059  &  0.187 &  41.20 &    -13.71 ( 1) &   0.25 ( 3) &   3.54 ( 1) &   0.13 ( 3) &   0.84 ( 1) &   0.25 ( 1) &   0.24 ( 2)  & \\
3C~424.0 & 0.127  &  0.096 &  41.07 &    -14.55 ( 1) &   0.24 ( 6) &   0.54 ( 3) &   0.27 ( 5) &   0.79 ( 1) &   0.44 ( 3) &   0.40 (12)  & \\
3C~430.0 & 0.054  &  0.630 &  40.12 &    -14.72 ( 3) & $<$ 1.09    &   1.61 ( 9) &   0.33 (20) &   1.43 ( 1) &   0.37 ( 8) &   0.40 ( 4)  & \\
3C~433.0 & 0.102  &  0.145 &  41.40 &    -14.01 ( 1) &   0.19 ( 4) &   1.88 ( 1) &   0.22 ( 3) &   1.09 ( 1) &   0.46 ( 2) &   0.30 ( 1)  & \\
3C~436.0 & 0.214  &  0.089 &  41.07 &    -15.06 (10) &   0.20 (22) &   3.09 ( 1) &       --    &   2.08 ( 9) &   0.63 (19) &   0.49 (18)  & \\
3C~438.0 & 0.290  &  0.358 &  41.55 &    -14.87 ( 1) & $<$  0.54   & $<$  0.82   & $<$  0.67   &   1.61 ( 1) &  $<$  0.66  &   0.46 (28)  & \\
3C~442.0 & 0.026  &  0.065 &  39.78 &    -14.40 ( 1) &   0.08 (24) &   0.27 ( 7) &   0.19 ( 8) &   1.84 ( 1) &   0.35 ( 5) &   0.45 ( 1)  & \\
3C~445.0 & 0.056  &  0.083 &42.50$^a$&   -12.37$^a$ ( 1)  & -- & 1.00$^a$ ( 1) & 0.04$^a$ ( 6)   &  --  &   0.02$^a$ (10) &  0.02$^a$ (10) & -12.03\\ 
3C~449.0 & 0.017  &  0.167 &  39.71 &    -14.09 ( 1) &   0.10 (23) &   0.30 (24) &   0.13 ( 9) &   1.38 ( 1) &   0.29 ( 7) &   0.22 ( 1)  & \\
3C~452.0 & 0.081  &  0.137 &  41.16 &    -14.05 ( 1) &   0.23 ( 5) &   1.53 ( 1) &   0.27 ( 2) &   1.08 ( 1) &   0.36 ( 1) &   0.29 ( 1)  & \\
3C~456.0 & 0.233  &  0.038 &  42.48 &    -13.72 ( 1) &   0.32 ( 1) &   2.15 ( 1) &   0.15 ( 2) &   0.78 ( 1) &   0.22 ( 1) &   0.24 ( 1)  & \\
3C~459.0 & 0.220  &  0.066 &  42.17 &    -13.97 ( 1) &   0.16 ( 6) &   0.73 ( 1) &   0.12 ( 1) &   1.77 ( 1) &   0.35 ( 1) &   0.33 ( 1)  & \\
3C~460.0 & 0.269  &  0.092 &  42.09 &    -14.25 ( 3) &   0.25 ( 3) &   0.49 ( 7) &   0.39 ( 4) &   1.23 ( 2) &   0.60 ( 1) &   0.54 ( 4)  & \\
3C~465.0 & 0.030  &  0.069 &  40.15 &    -14.17 ( 1) &   0.17 (27) &   0.46 ( 6) &   0.26 ( 9) &   2.77 ( 1) &   0.47 ( 1) &   0.32 ( 1)  & \\
\hline
\hline
\end{longtable}
Column description: (1) source name; (2) redshift; (3) Galactic absorption;
(4) logarithm of the luminosity of the H$\alpha$ narrow line, in erg s$^{-1}$;
(5) logarithm of the \Ha\ flux in erg cm$^{-2}$ s$^{-1}$; (6 through 11)
de-reddened flux ratios of the key diagnostic lines with respect to H$\alpha$.
The values in parentheses report the errors (in percentage) of each line.
Missing values marked with `--' correspond to lines outside the coverage of
the spectra or severely affected by telluric bands. When no lines are visible
we only give the upper limit for \Ha; (12) logarithm of the flux of the \Ha\
broad component, when visible. \\ Notes: (a) for 3C~111 and 3C~445, no narrow
\Ha\ measurement is possible and we give instead the [O III] luminosity,
referring the flux ratios to this line. No narrow lines are visible in 3C~273
and we only report its broad \Ha\ flux.
\end{landscape}
\twocolumn

We consider as acceptable templates all stellar populations for which the
resulting $\chi^2$ is increased by less than $\Delta\chi^2$ = 9.21,
corresponding to a confidence level of 99\% ($\sim$ 2.6 $\sigma$ level) for 2
parameters.  In the case of 3C~310, the error derived setting $\chi^2_{\rm r}
= 1$ is 9.5 $10^{-18}$ erg cm$^{-2}$ s$^{-1}$ \AA$^{-1}$. This value is
similar to the rms of the stellar subtracted spectrum, which is 8.2
$10^{-18}$ erg cm$^{-2}$ s$^{-1}$ \AA$^{-1}$. Adopting the procedure outlined
above, the best fit is the only statistically acceptable model. Nonetheless,
in Fig.  \ref{3c310} we show the two models with the lower values of
$\chi^2_{\rm r}$, differing by one step in metalicity or age.  In both cases
significant residuals already emerge in the continuum subtracted spectrum,
confirming visually the poorer quality of the fit. We re-estimate the H$\beta$
flux resulting from these two templates. They differ by $\sim$ 4\% with
respect to our original measurement using the best fit model. This is lower
than the statistical error associated by {\it specfit} to this measurement,
which is 7 \%.  We conclude that, in the case of 3C~310, a possible template
mis-match does not have a significant effect on the H$\beta$ flux measurement.

The influence of the stellar templates on the other emission lines 
(and in particular on the H$\alpha$ line that is 
$\gtrsim$ 3 brighter that H$\beta$ while they have the same EW in absorption) 
is negligible because the stellar population has not significant 
absorption features at the line wavelengths.

For the majority of the sources, only models differing by at most one step in
age or in metalicity are consistent with the data, without a significant
impact on the H$\beta$ flux. Only for the lower quality spectra can this range
be broader.  For low SNR spectra, coupled with emission lines of low
equivalent widths, the resulting measurement error is dominated by the choice
of the stellar template. In these cases we adopted the following strategy: we
subtracted the acceptable template with the lowest age (i.e.  with the
strongest H$\beta$ absorption) and considered the \Hb\ measurement as an upper
limit. One of such galaxy is 3C~173.1, already shown in
the bottom right panel of Fig.  \ref{ssp}.

Finally we note that our approach to the starlight subtraction is limited to
the choice of the best fitting single population model, similarly to the
method adopted by \citet{kauffmann03} for the SDSS sample of emission line
galaxies and by \citet{tadhunter93} for the 2Jy sample of radio galaxies.
Nonetheless, this procedure does not grasp the complexity of the emission
processes in radio galaxies. As clearly shown by several studies in the
literature the stellar content of radio galaxies is often composed by
populations of different ages, and there are also evidence for internal
absorption, particularly associated to the young stars component (e.g.
\citealt{wills04}; \citealt{raimann05}; \citealt{tadhunter05};
\citealt{holt07}). The presence of stars of various ages is indirectly
confirmed by the fact that the best fit stellar population found in the \Hb\
regions is on average $\sim$ one step in age younger than for the \Ha\ band.
This can be explained by the higher relative contribution of young stars at
shorter wavelengths that drives the age of the SSP to lower values with
respect to the redder part of of the spectrum.  Furthermore, other mechanisms
contribute to the continuum emission, e.g. nebular continuum and light from
the accretion disk (direct or via scattering). However, a full separation of
these contributions requires a detailed analysis on a object-by-object basis
in order to properly measure the rather large set of free parameters
describing the various components.  Moreover this fitting requires spectra
with a high signal to noise, and this is not always achieved in our data.  We
defer such a study to a forthcoming paper.

There is however a general consideration that can be drawn already at this
stage. Nuclear emission and nebular continuum are expected to be in general
bluer that the starlight. This biases the derived age of the single stellar
population that appears younger that it actually is. It also dilutes the
stellar absorption features, thus biasing the metalicity to lower values.  The
main resulting effect is an overestimate of the correction of the \Hb\ line
and, consequently, an over-rating of this line, particularly in the objects
with brightest optical nuclei.

\section{Results}
\label{sect4}

\subsection{Emission line data quality}
\label{per}
We summarize here the data quality of our spectra from the point of view of
the measurements of the emission line intensities. Only in three radio galaxies
(namely 3C~052, 3C~129.1 and 3C~130) did we fail to detect any emission line;
in all cases an accurate redshift can be measured by fitting the stellar
population models based on stellar absorption features.

Leaving aside these objects, the \Ha\ line can be measured with a statistical
accuracy of better than 10 \% with only three exceptions. In the red part of
the spectrum, the [O I]$\lambda$6300 line is detected in 95 \% of the 3C
galaxies.  The completeness of the [S II] doublet measurements is slightly
lower (87 \%) due to our choice to favor the inclusion of the [O I] in the
high resolution spectra but, when covered by the data, the [S II] line is also
measurable in 95 \% of the objects. In the blue part of the spectrum there are
$\sim$ 15 \% of the sources in which \Hb\ and/or [O III] are not detected.

All together, these measurements will enable us to locate the vast majority of
the sources in the diagnostic planes that compare pairs of emission line
ratios and also to perform a detailed analysis of the relationship between the
multiwavelength characteristics of the 3C sources with their emission line
properties.

In addition, a broad \Ha\ line is seen in 18 galaxies and these are discussed
in more detail in the Sect. \ref{broad}.

\begin{figure*}[htbp]
  \centerline{ 
    \psfig{figure=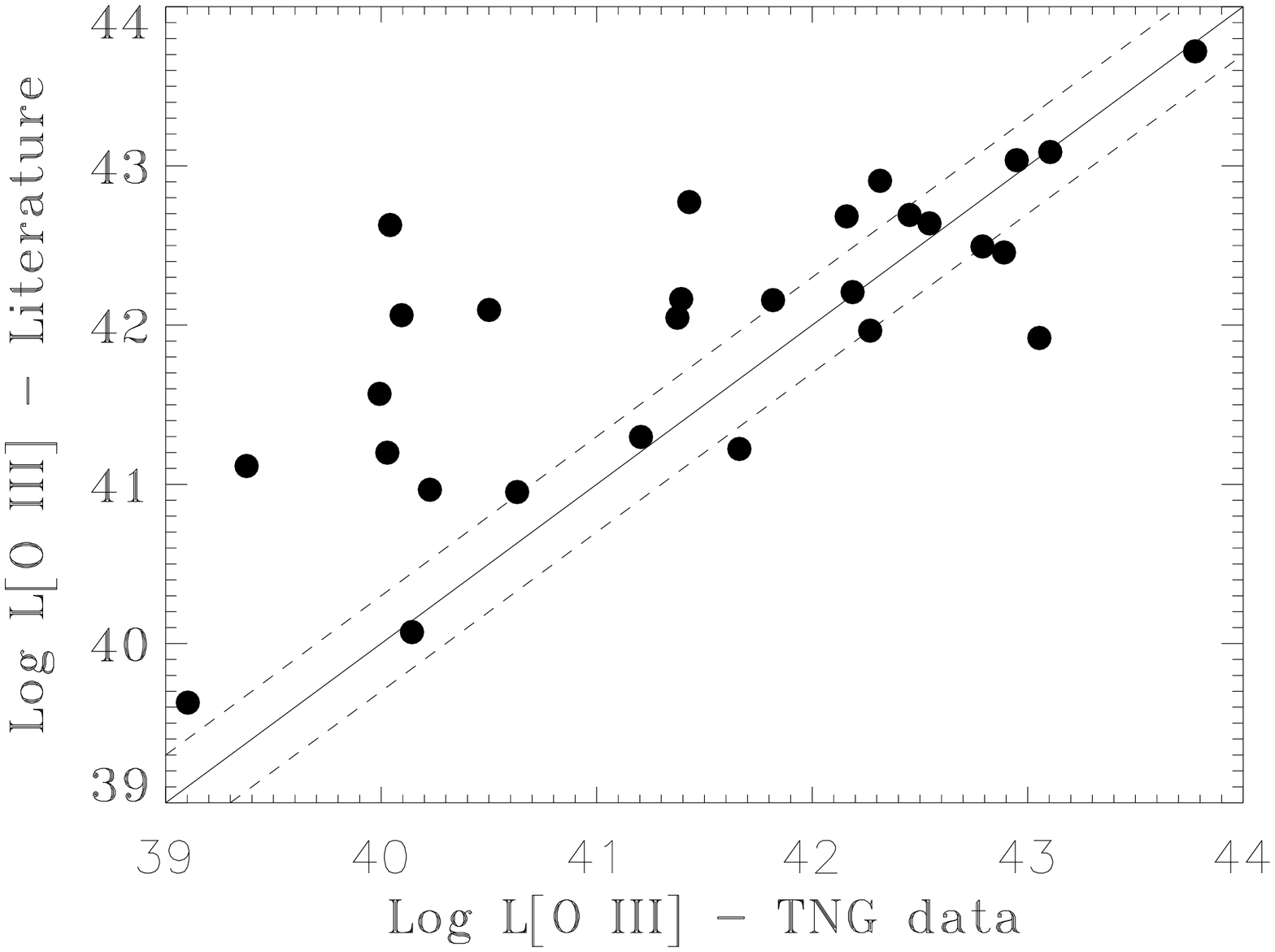,angle=90,width=0.35\linewidth} 
    \psfig{figure=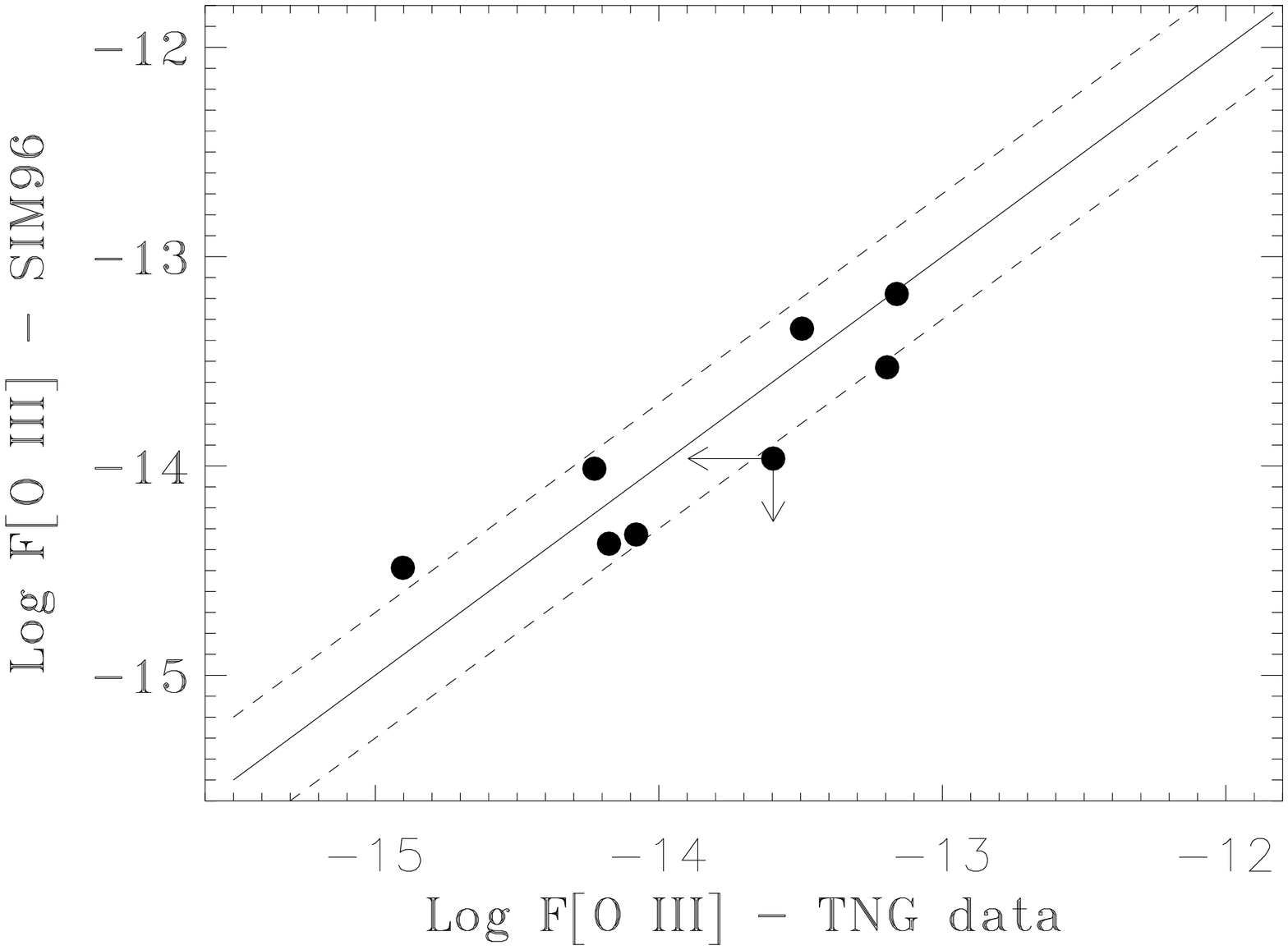,angle=90,width=0.35\linewidth} 
    \psfig{figure=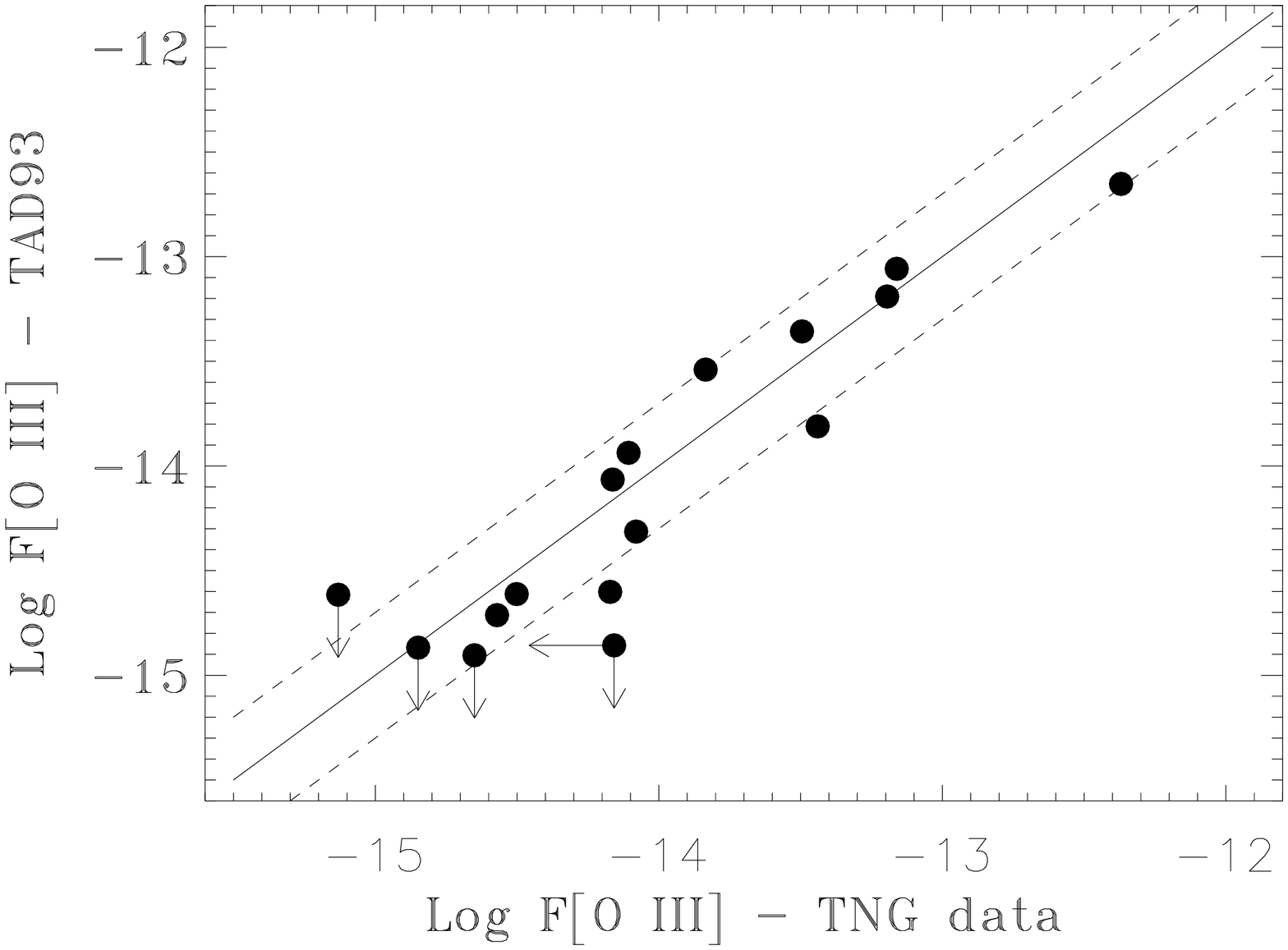,angle=90,width=0.35\linewidth} }
  \caption{\label{compwst} Comparison of our [O~III] luminosity
      measurements with the values collected by \citet{willott99} from
      dishomogeneous works in literature (left panel), by \citet{simpson96}
      (center panel) and by \citet{tadhunter93} (right panel). The
      luminosities are in units of erg s$^{-1}$ while the fluxes are in units
      of erg s$^{-1}$ cm$^{-2}$. The solid line traces where the compared
      objects have the same fluxes, while the two dotted lines above and below
      the solid line limit the region in which the fluxes differ by less than
      a factor of 2. The arrows indicate the upper limits. }
\end{figure*}

\subsection{Comparison with previous results}
\label{comp} 

There is a extremely vast amount of spectroscopic data accumulated for
  the 3CR sample in the
  literature over the last 40 years and it is therefore, on the one hand, very
  interesting to contrast our measurements with those available, but, on the
  other hand, difficult to perform a full comparison on an object-by-object
  basis. We therefore decided to focus on three papers, those having
  the largest overlap with our sample.

  \citet{willott99} collected [O III] measurements from optical spectra and
  images of 3CR objects from the literature (these data are available at the
  web-page: {\it http://www.science.uottawa.ca/$\sim$cwillott/3crr/3crr.html}).
  There are 28 sources in common with our sample.  Fig. \ref{compwst} (left
  panel) shows the comparison between their data collection and our
  measurements, performed after scaling their reported luminosities to our
  cosmological parameters. These data show large discrepancies, even in
  excess of a factor of 100, particularly for the sources with the lowest
  [O~III] fluxes. Sources with stronger fluxes are in better agreement with
  our results, but there are still large differences.  

  In the central and right panels we compare our [O~III] fluxes with the
  values found by \citet{simpson96} and \citet{tadhunter93}, derived from
  long-slit spectra of two complete samples, with respectively 8 and 16
  sources in common with our 3CR subsample. Their data have been corrected
  for Galactic absorption before comparison. Quite reassuringly, essentially
  all sources have [O III] fluxes differing by less than a factor of 2.

\subsection{Radio galaxies with broad emission lines}
\label{broad} 
\begin{figure*}[htbp]
  \centerline{ 
    \psfig{figure=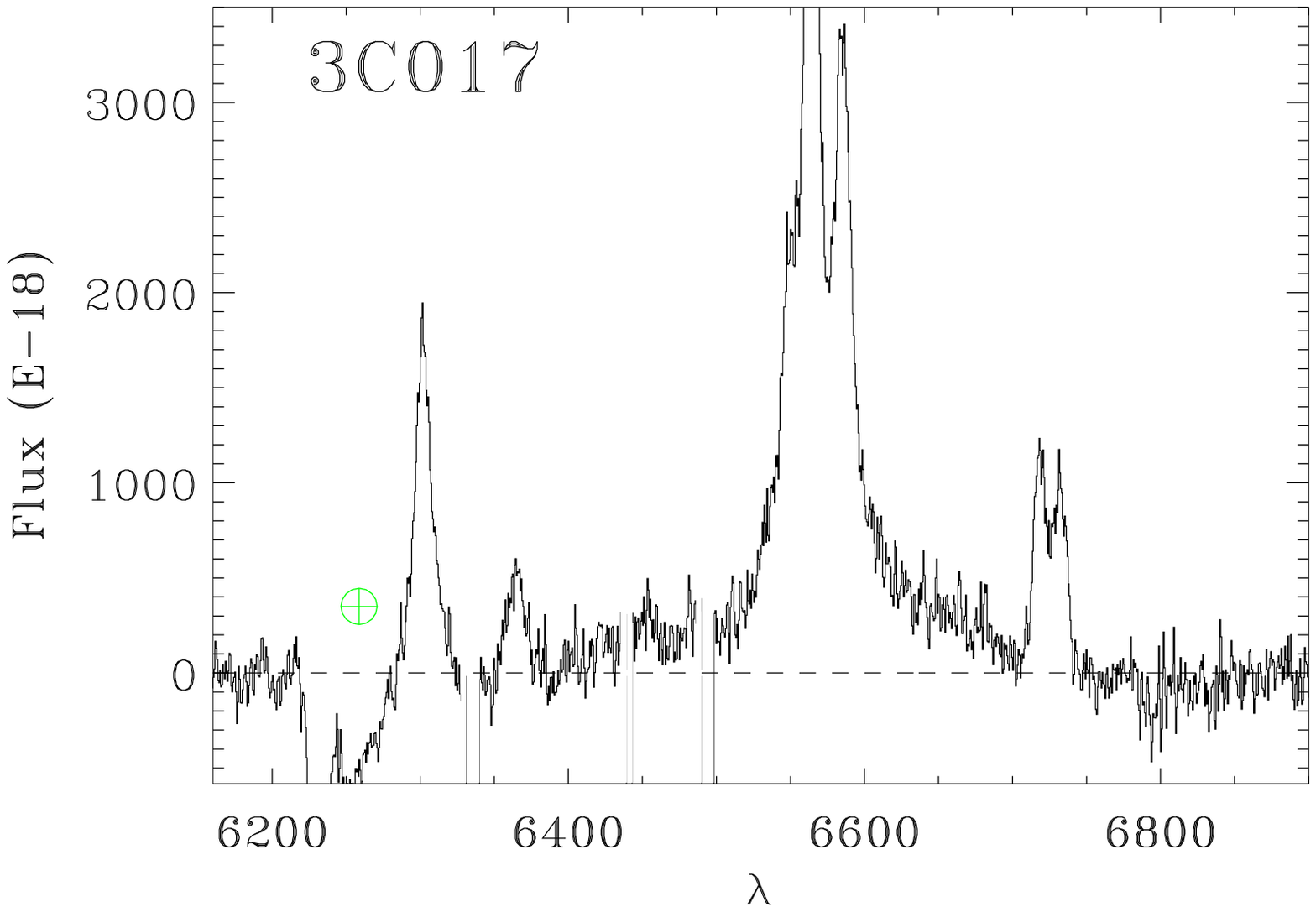,width=0.30\linewidth}
    \psfig{figure=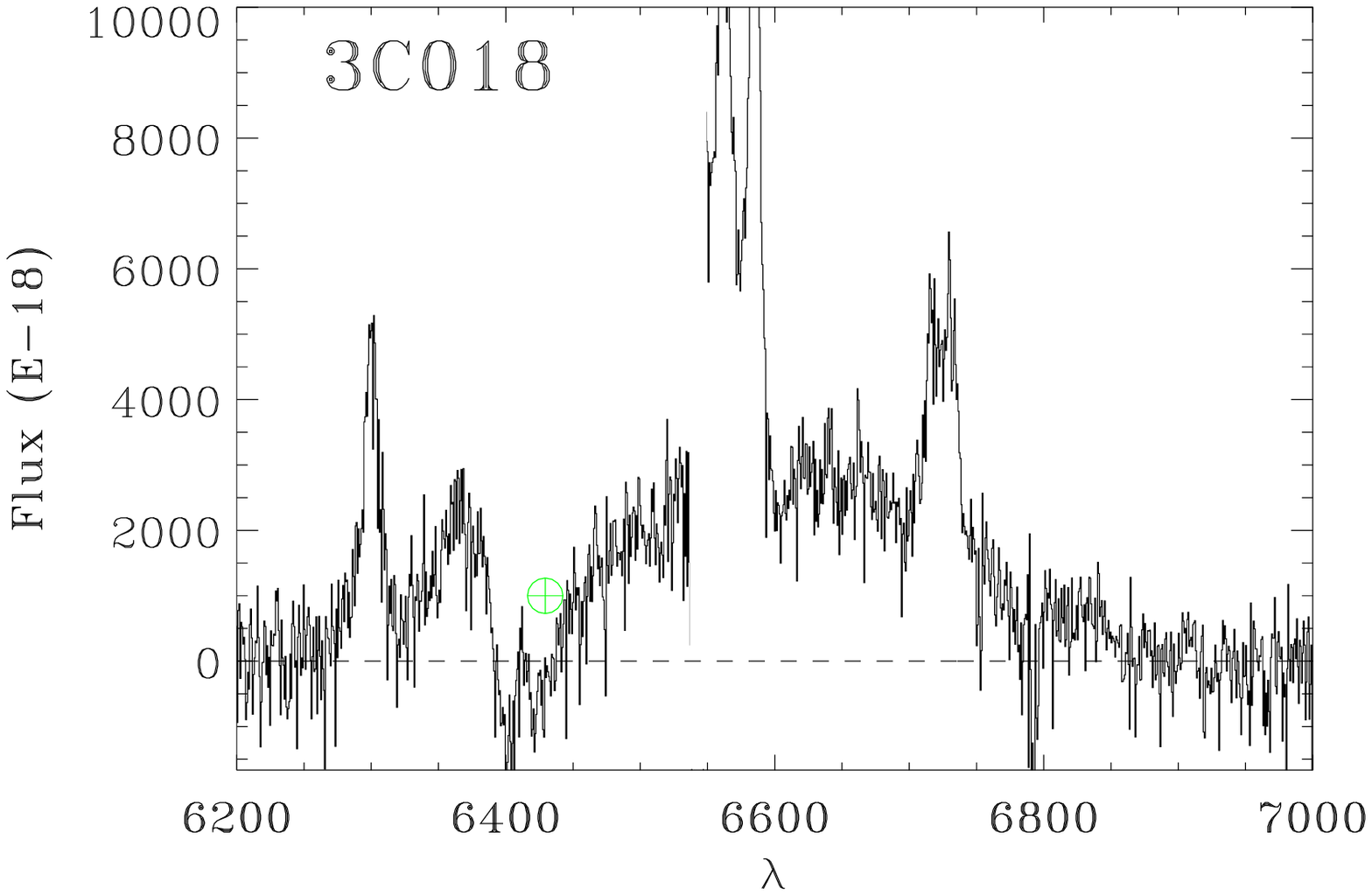,width=0.30\linewidth}
    \psfig{figure=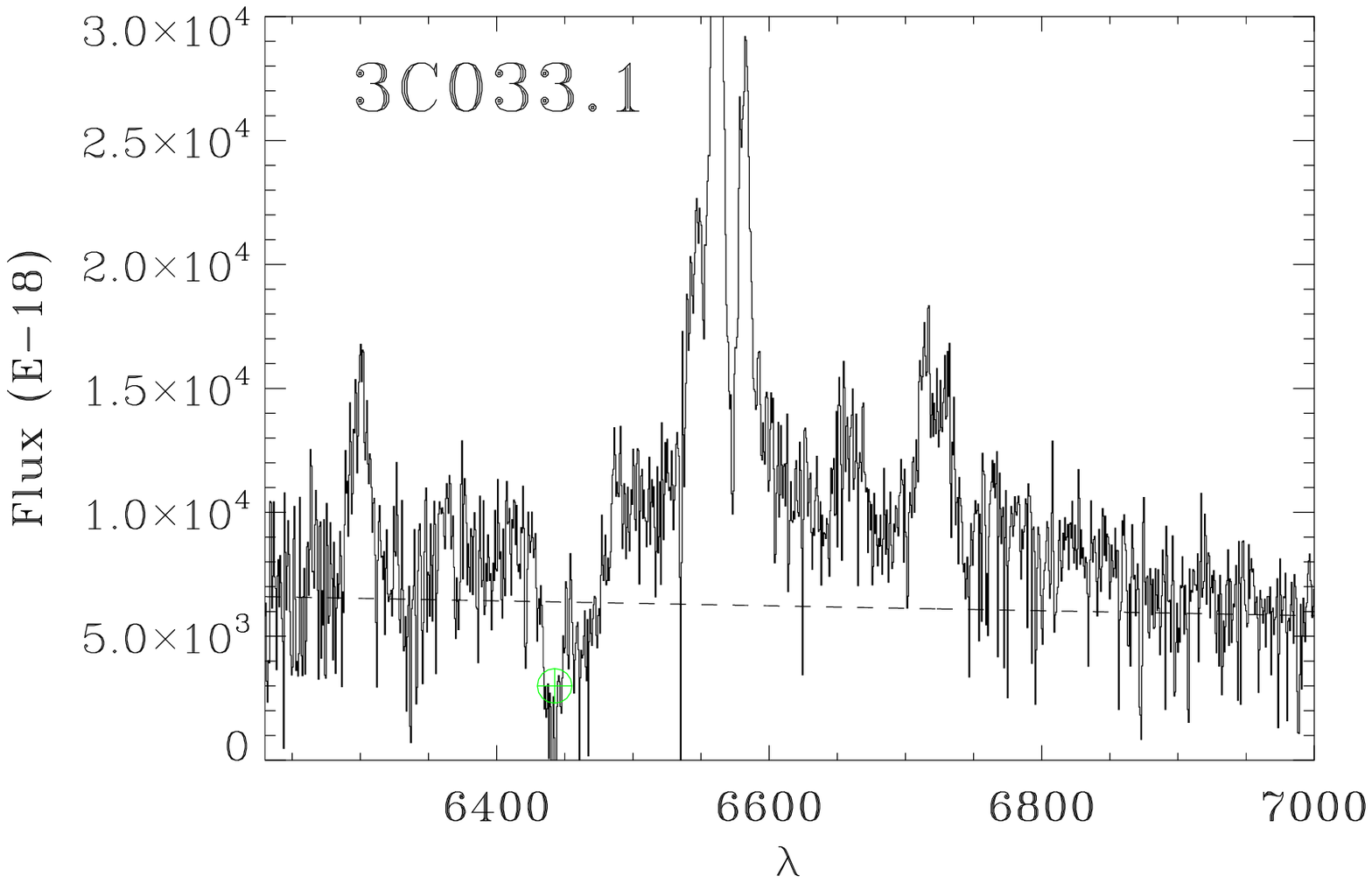,width=0.30\linewidth}}

  \centerline{ 
    \psfig{figure=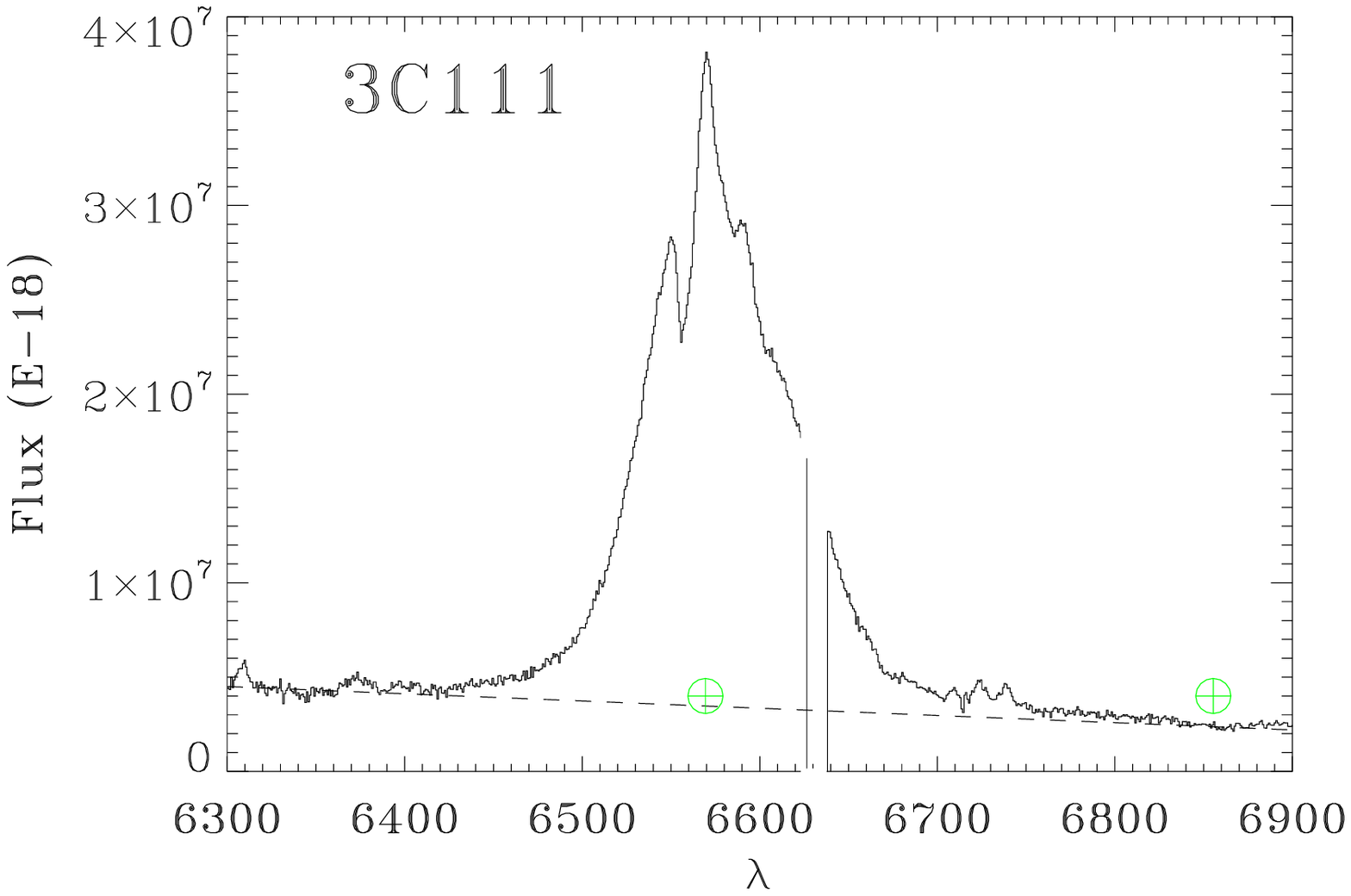,width=0.30\linewidth}
    \psfig{figure=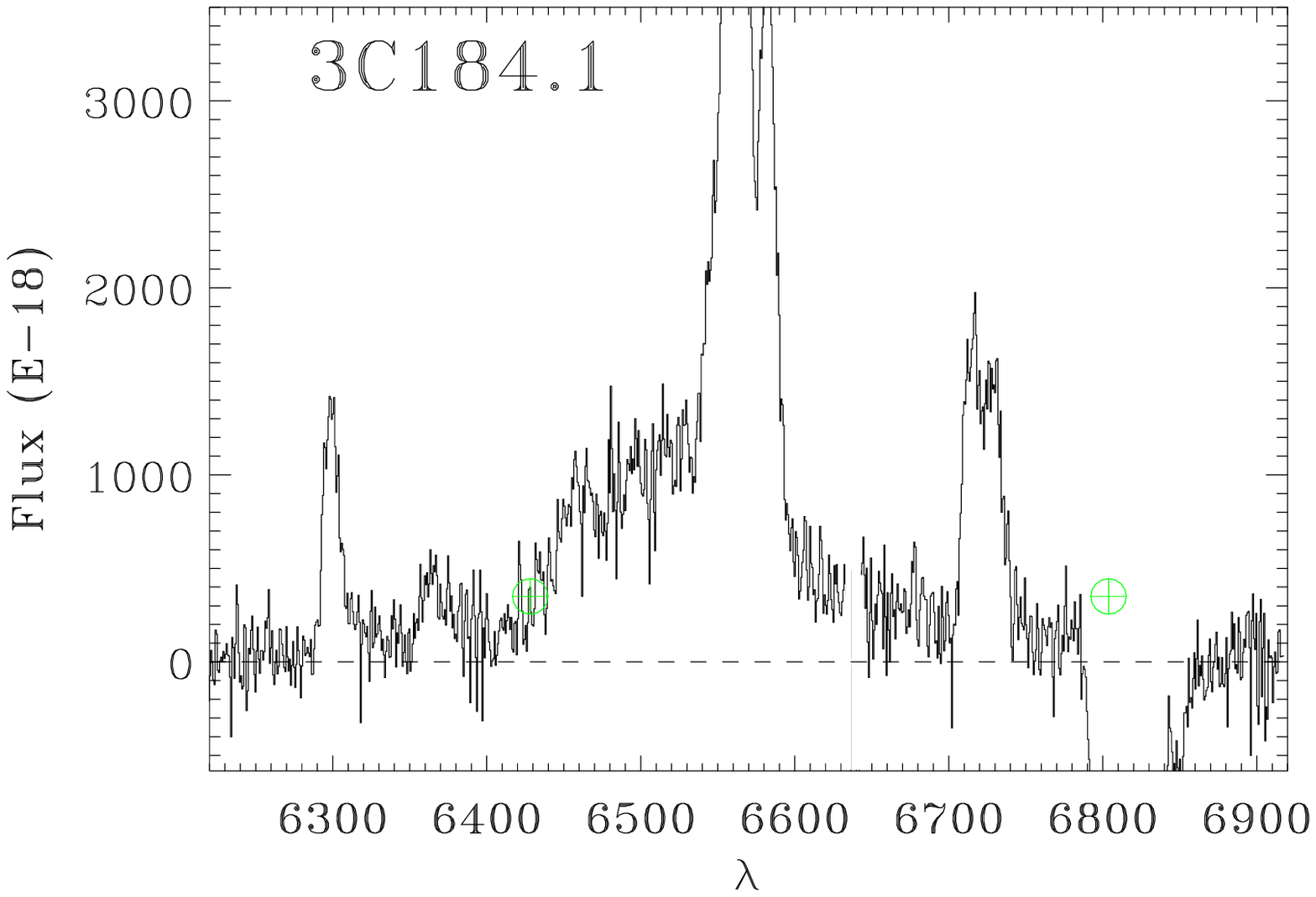,width=0.30\linewidth}
    \psfig{figure=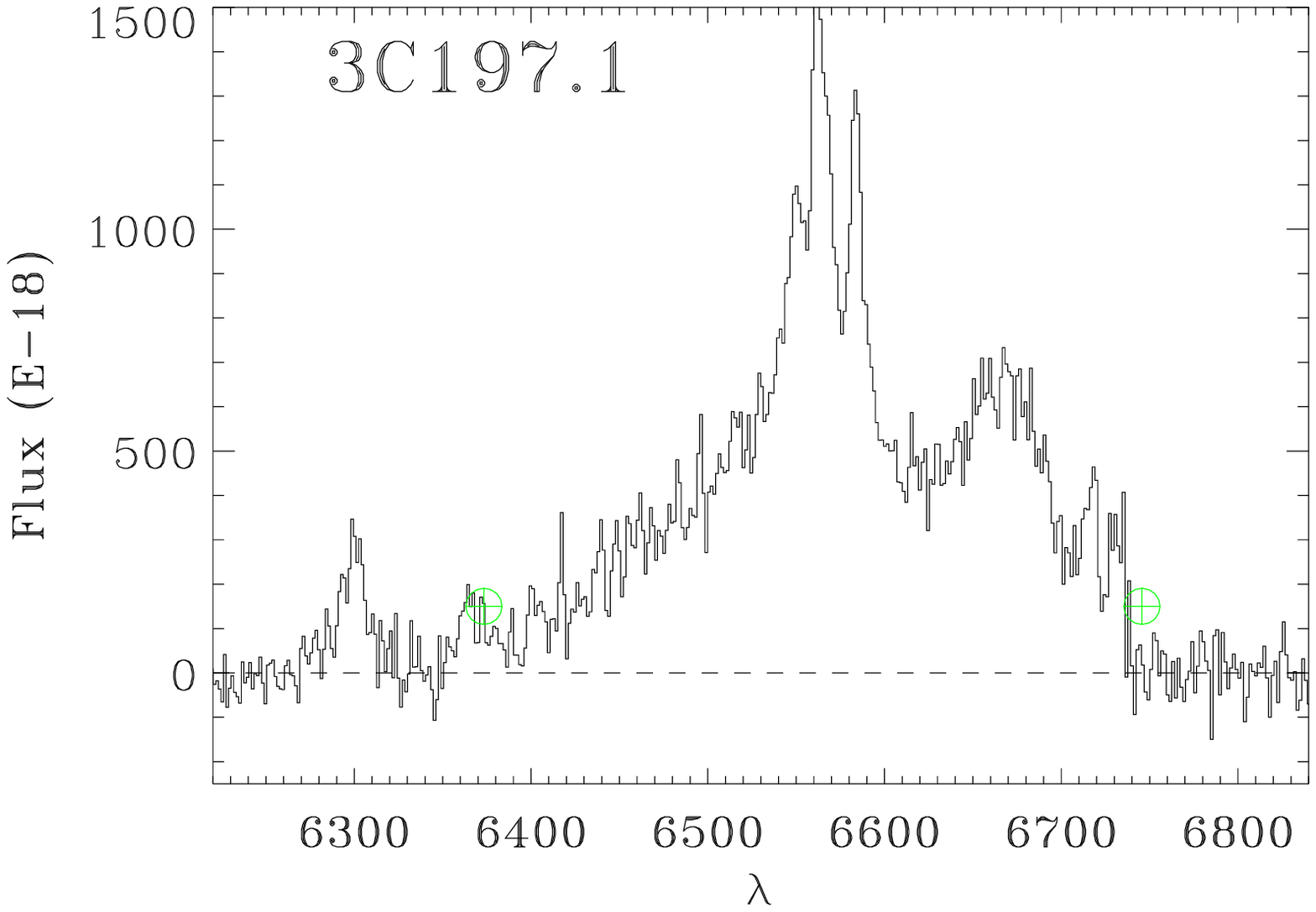,width=0.30\linewidth}}

  \centerline{ 
    \psfig{figure=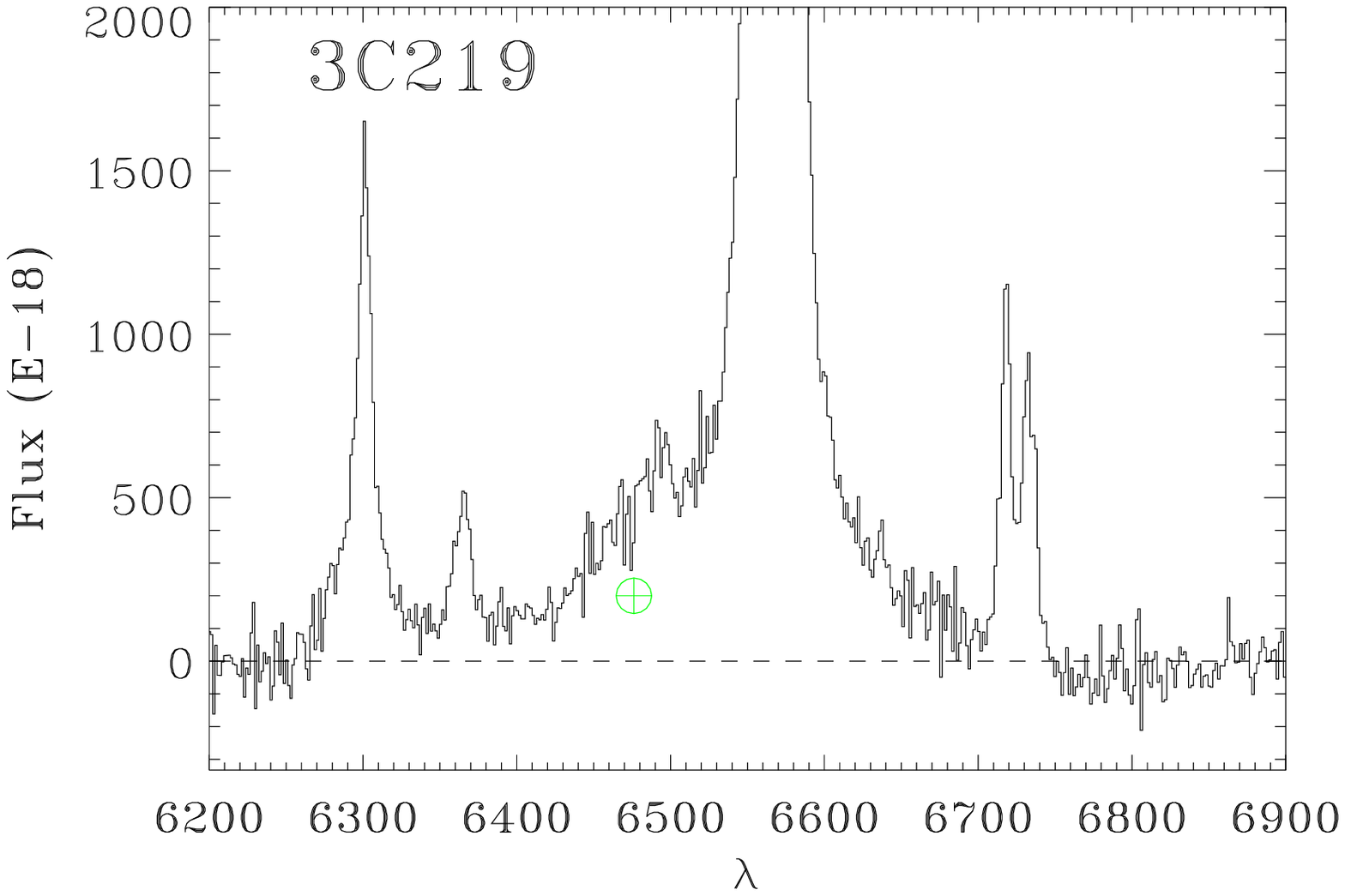,width=0.30\linewidth}
    \psfig{figure=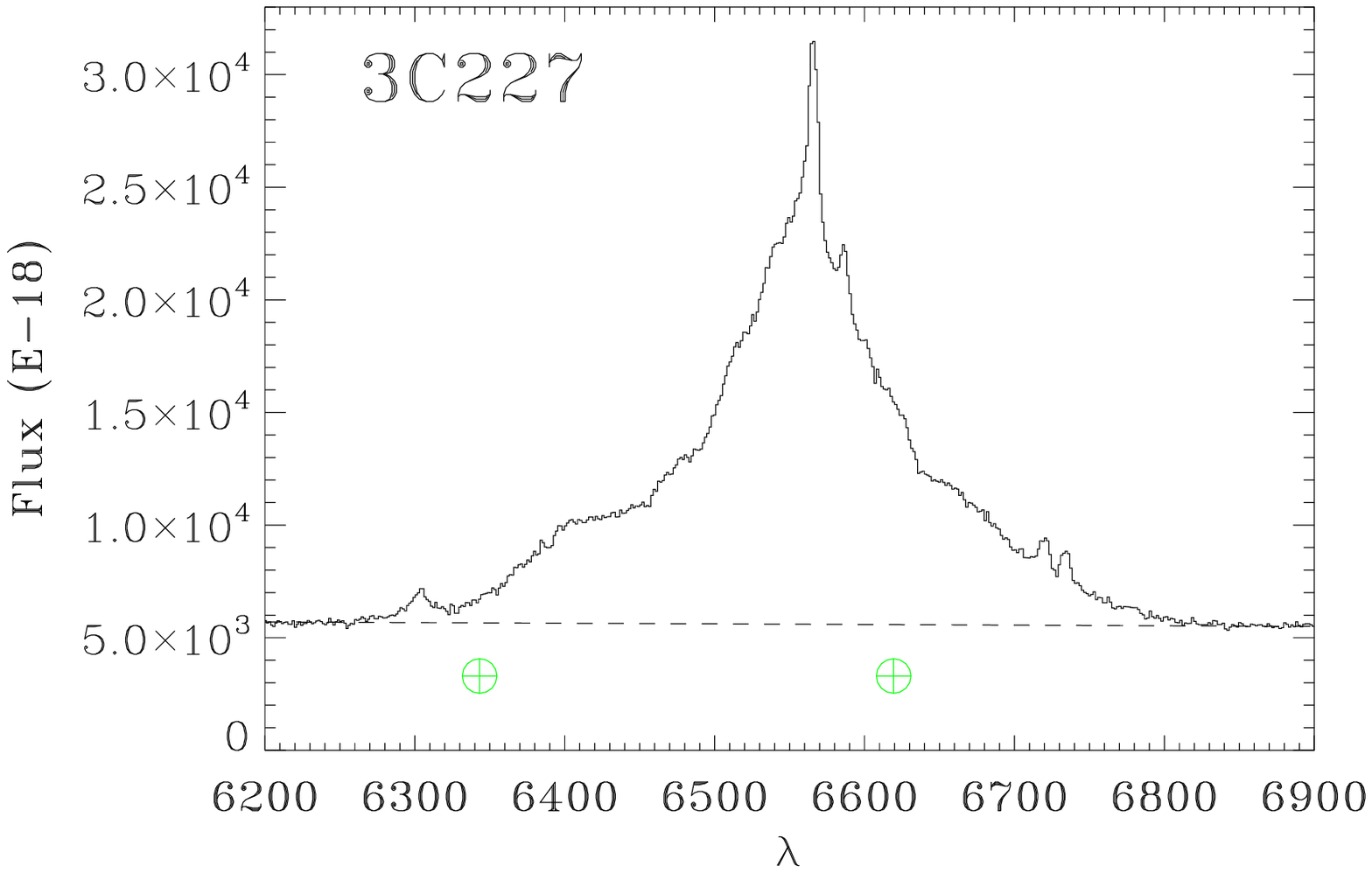,width=0.30\linewidth}
    \psfig{figure=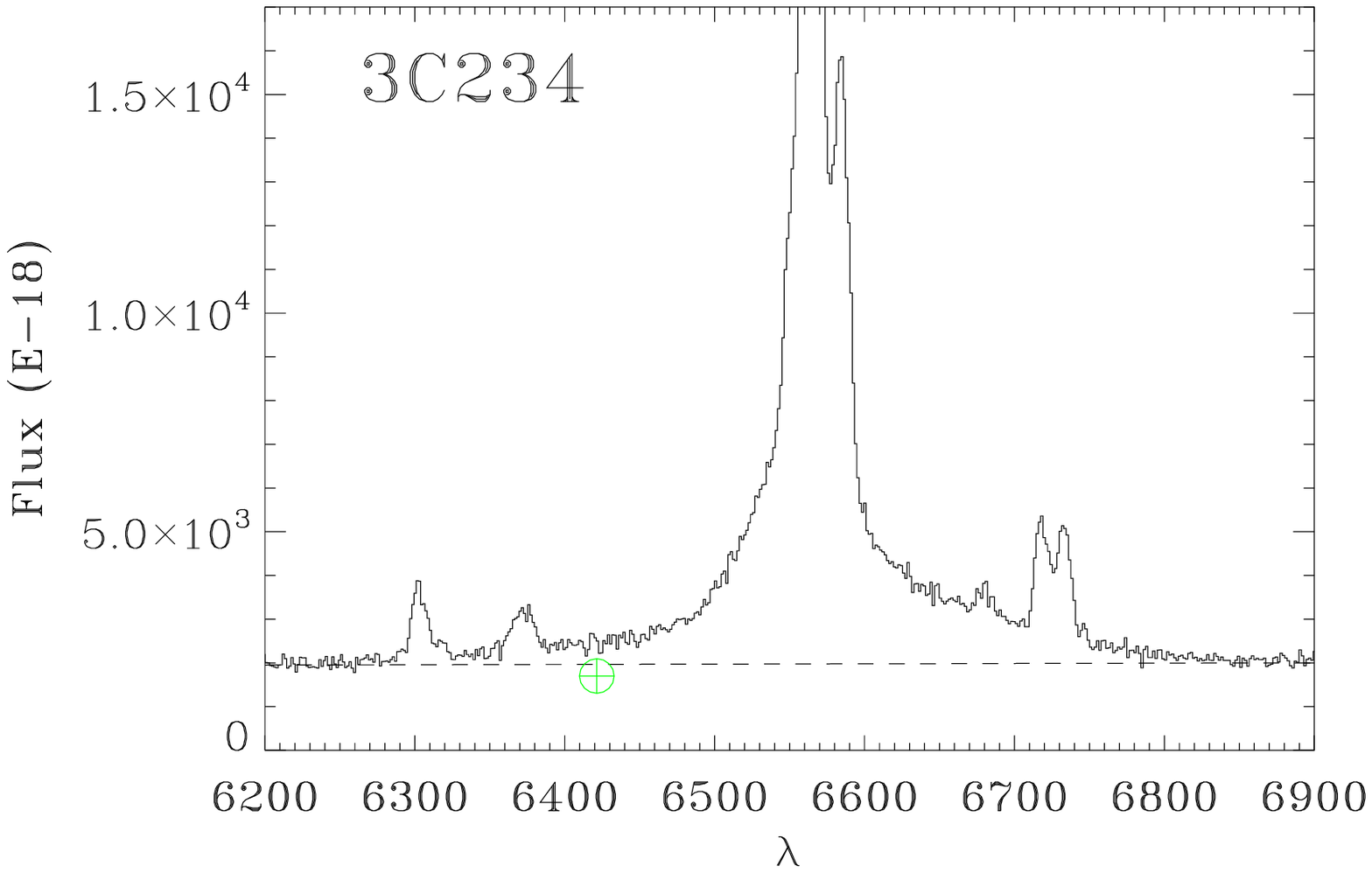,width=0.30\linewidth}}

  \centerline{ 
    \psfig{figure=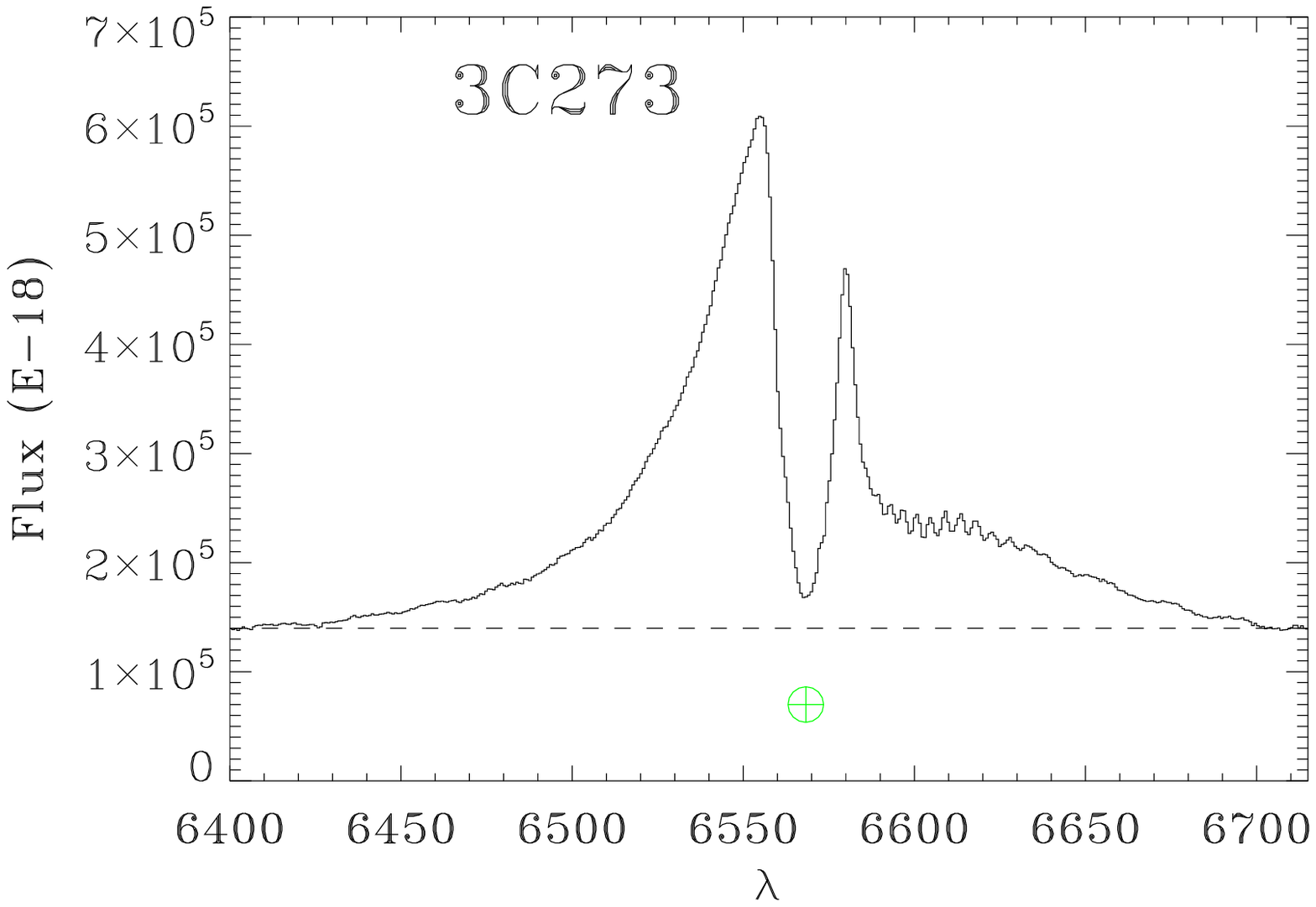,width=0.30\linewidth} 
    \psfig{figure=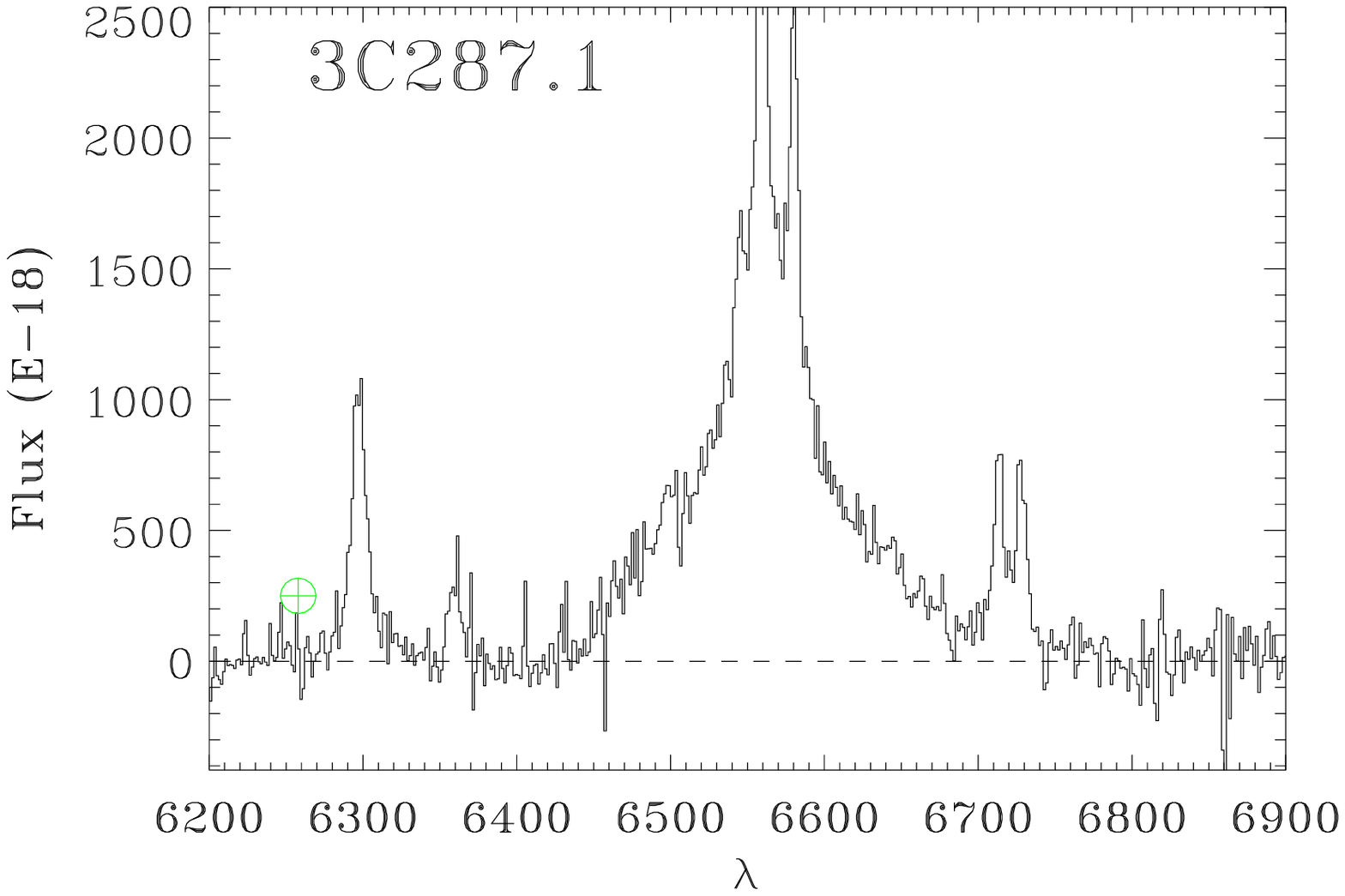,width=0.30\linewidth} 
    \psfig{figure=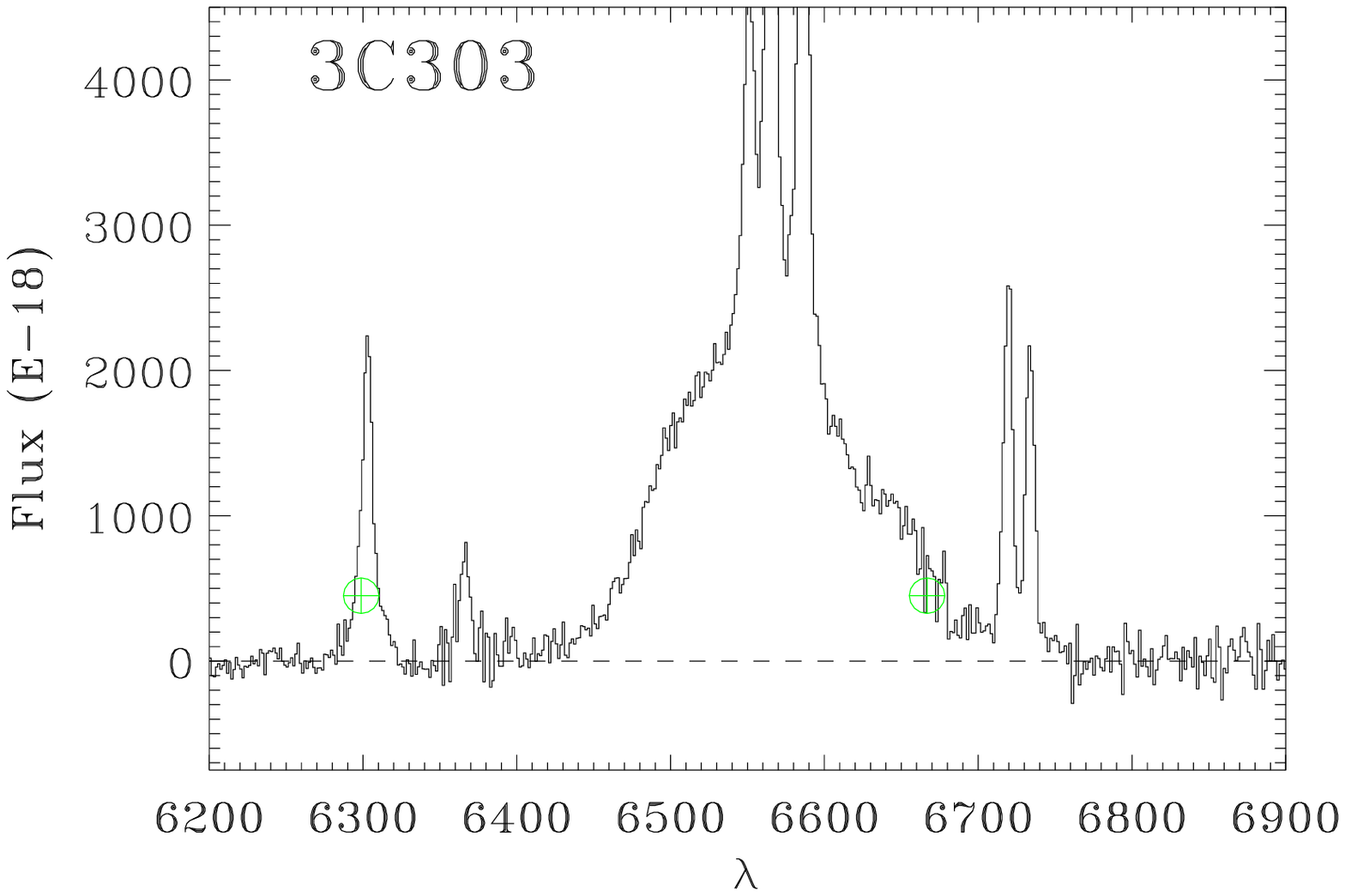,width=0.30\linewidth}}

  \centerline{ 
    \psfig{figure=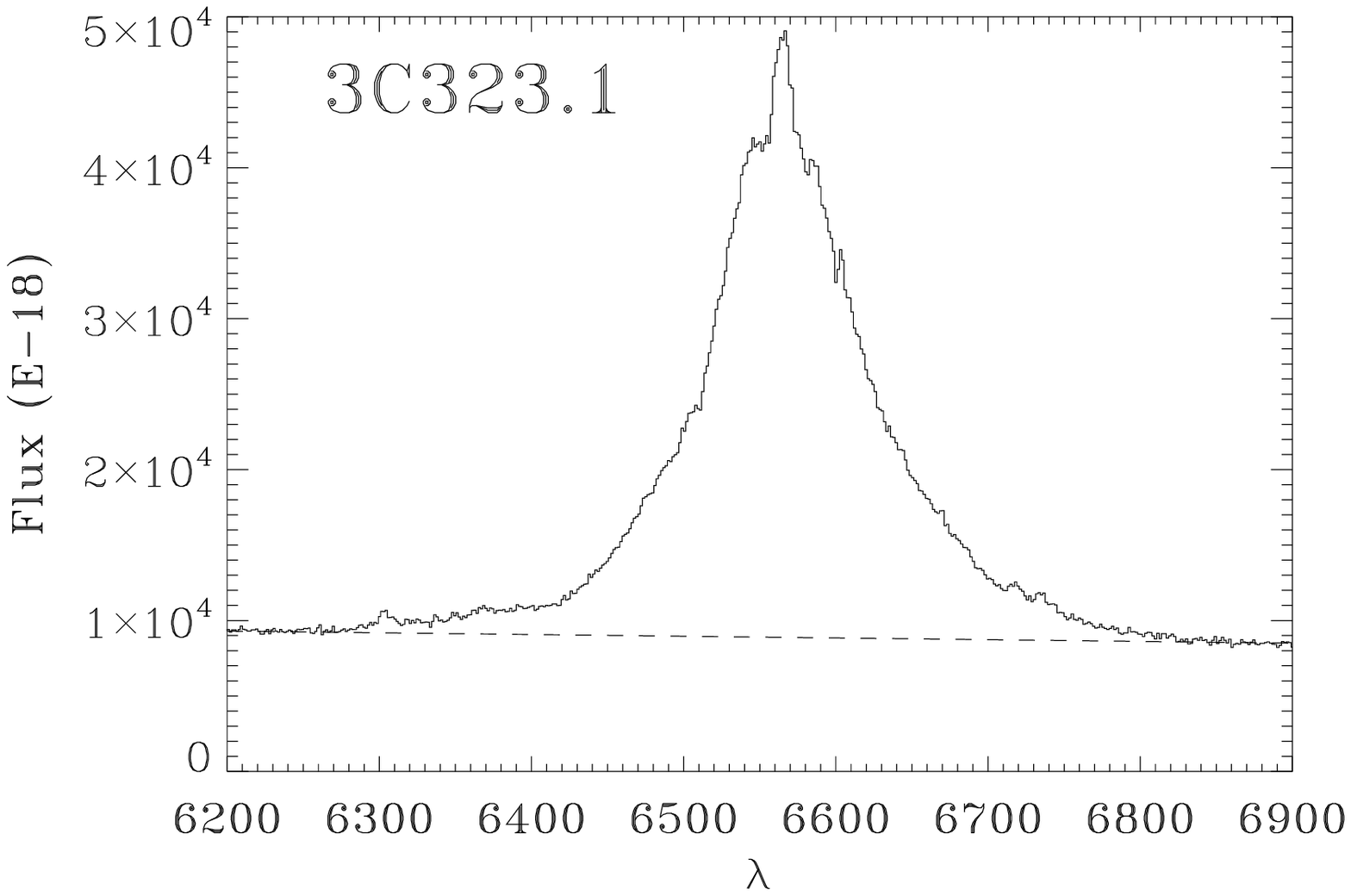,width=0.30\linewidth} 
    \psfig{figure=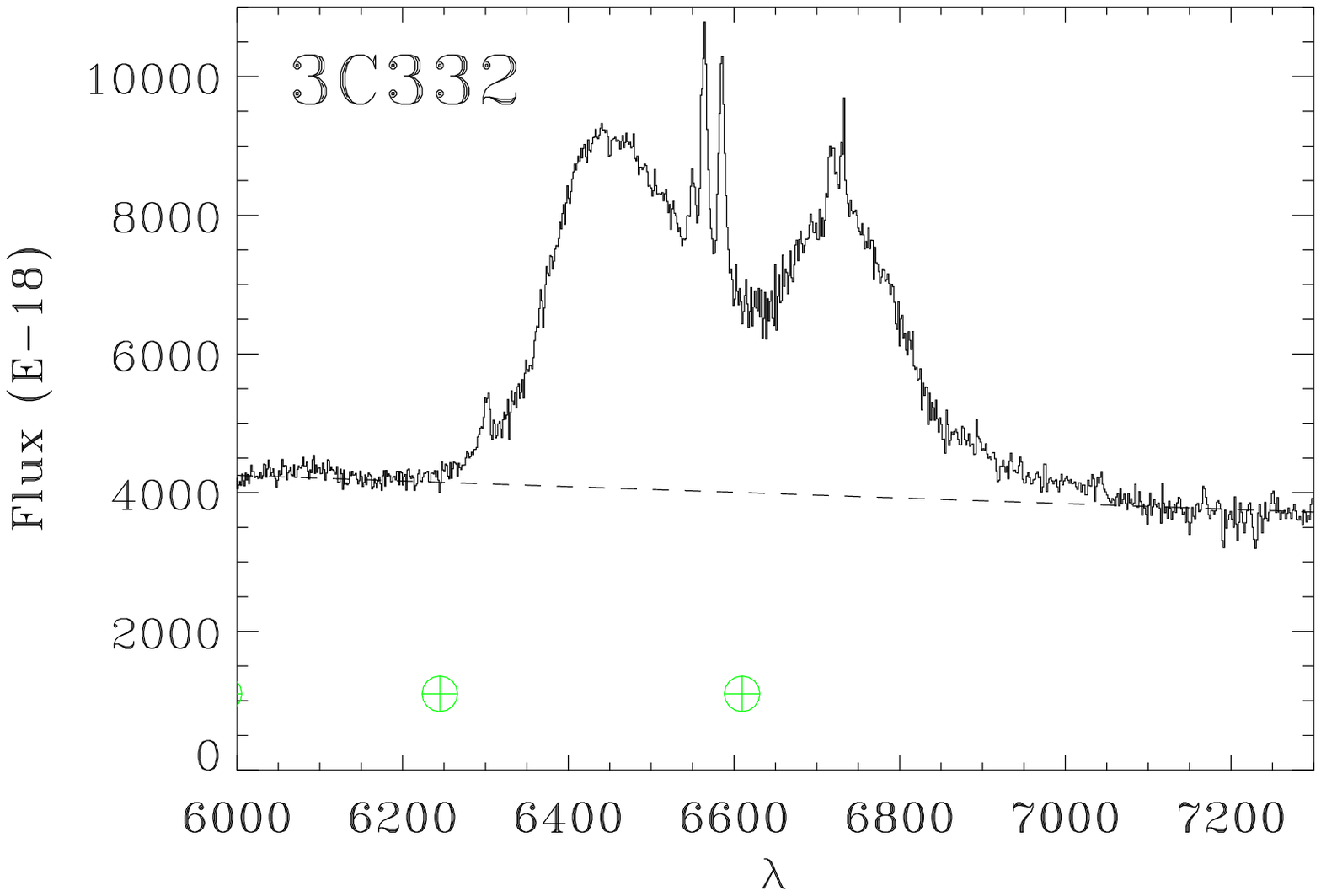,width=0.30\linewidth} 
    \psfig{figure=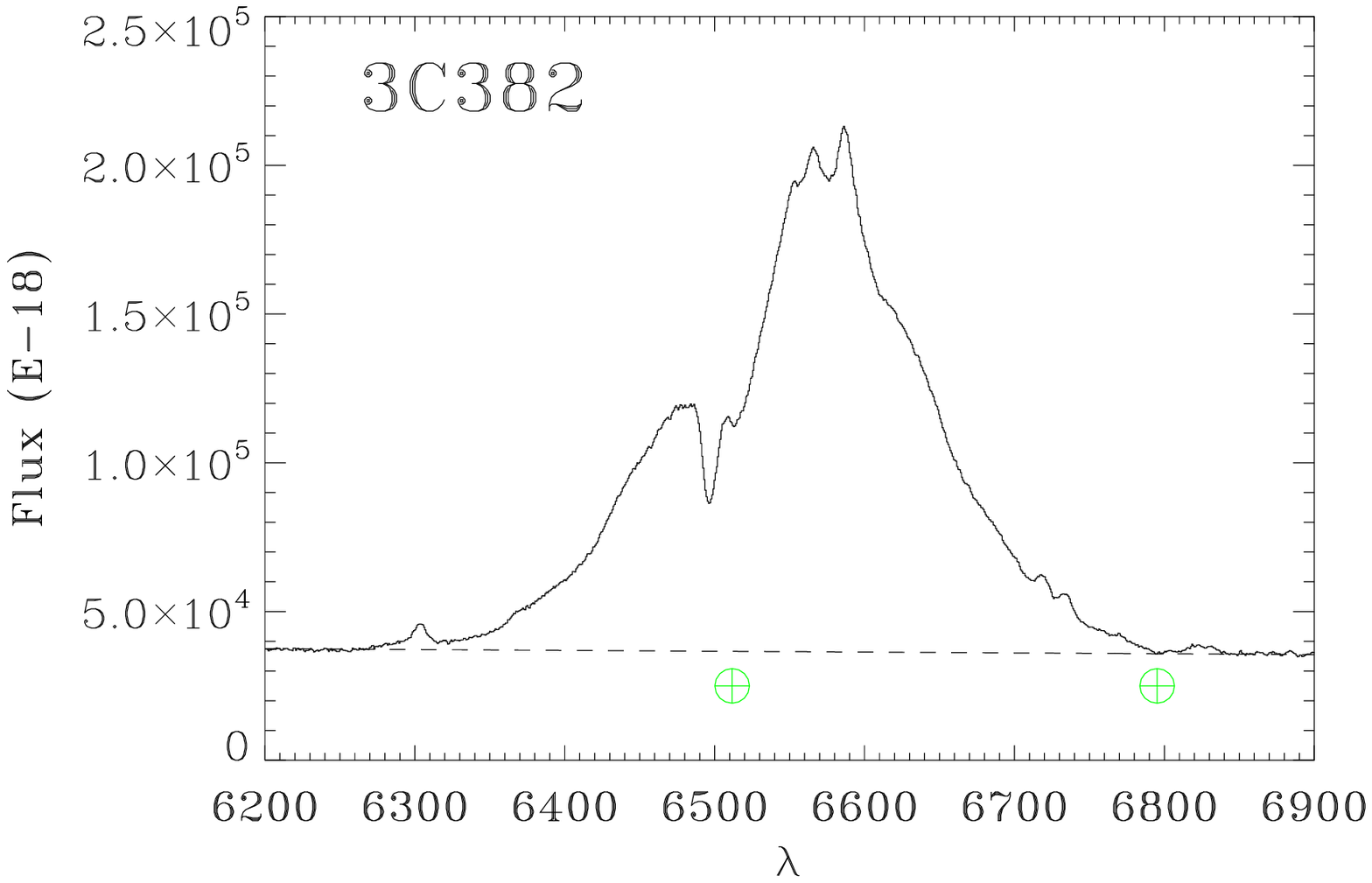,width=0.30\linewidth}}

  \centerline{ 
    \psfig{figure=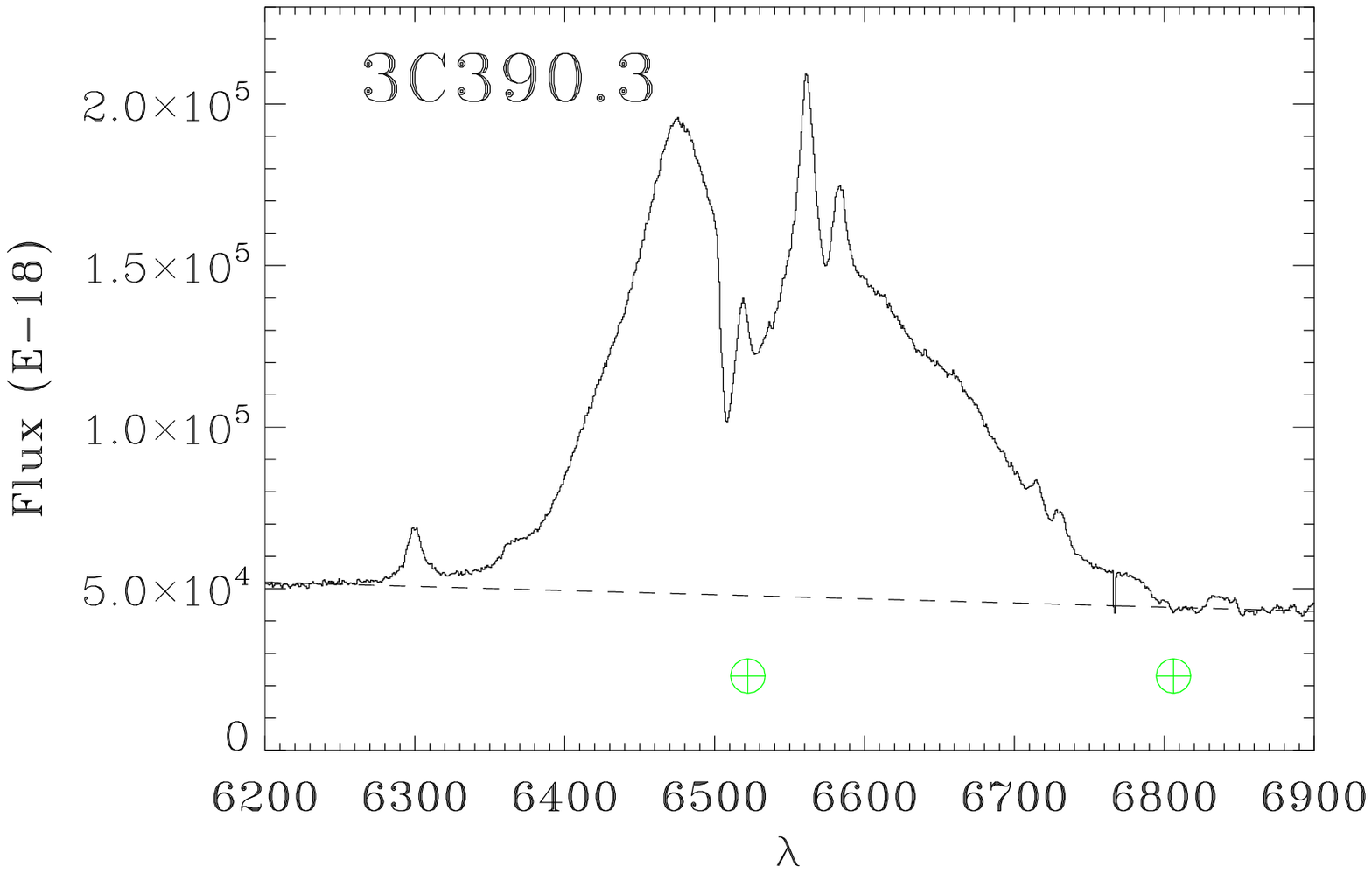,width=0.30\linewidth} 
    \psfig{figure=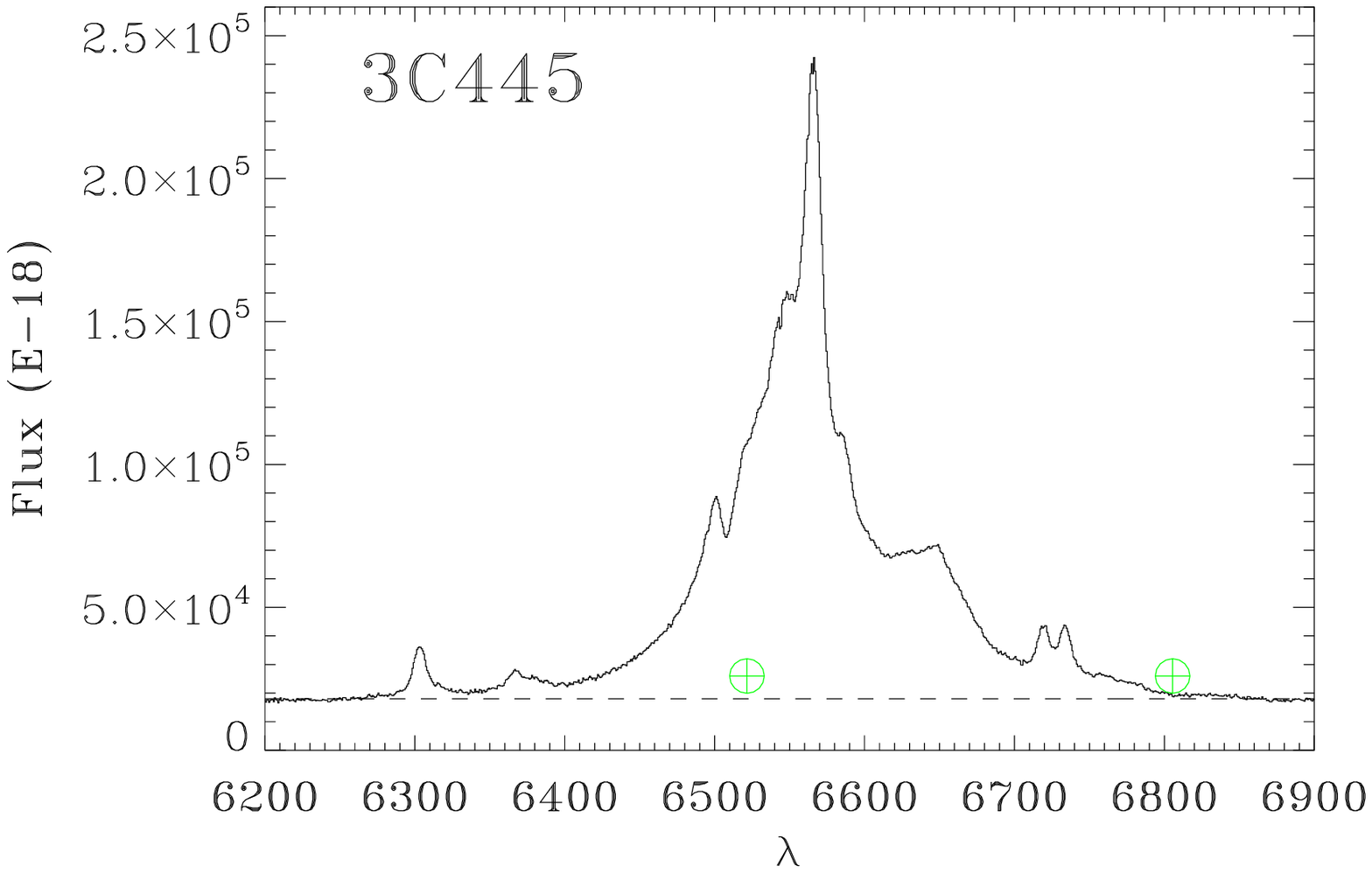,width=0.30\linewidth}
    \psfig{figure=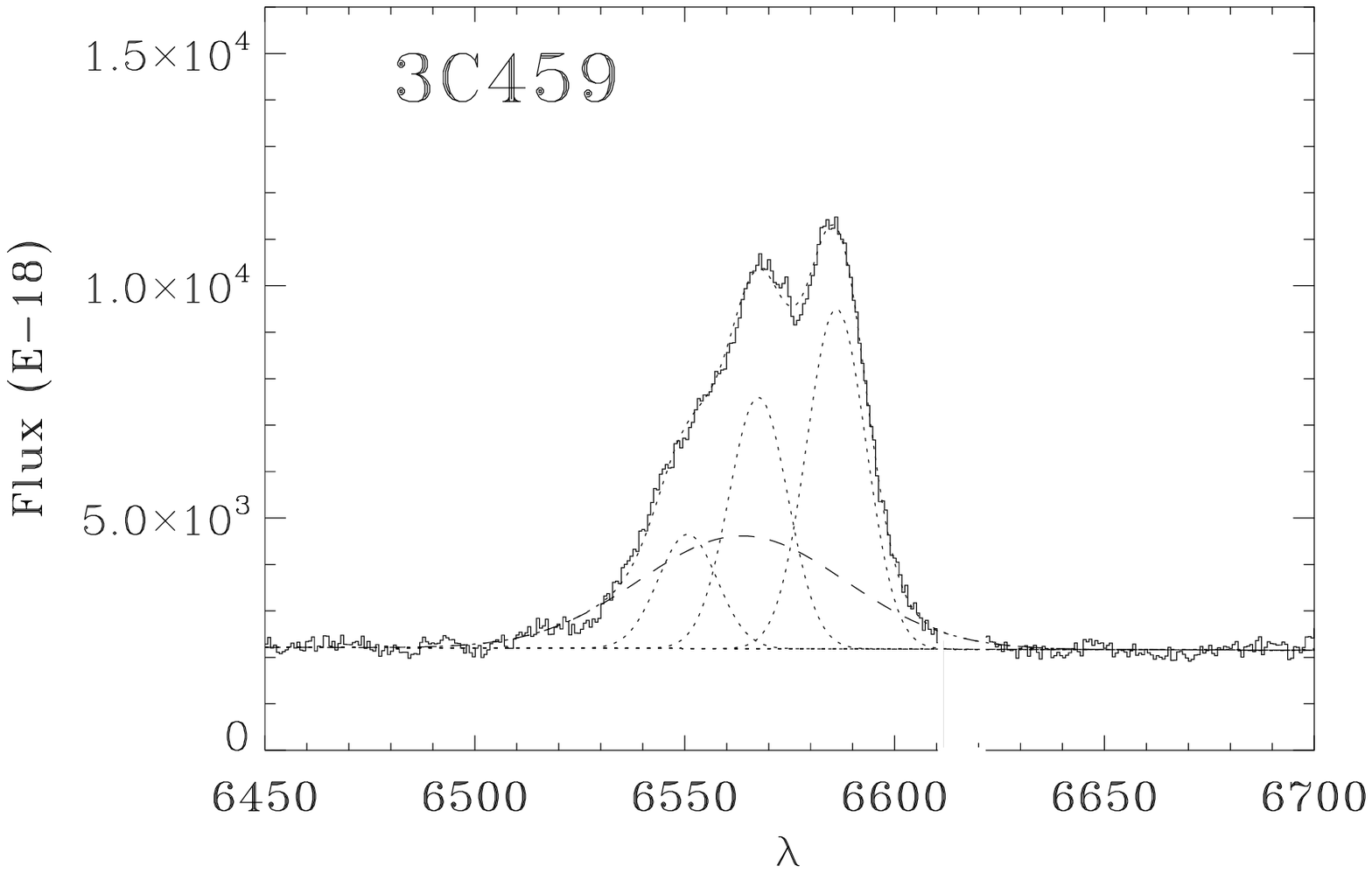,width=0.30\linewidth}}
  \caption{\label{blr} Radio galaxies with a broad H$\alpha$ emission line. 
The dashed line represents the continuum emission level. A null continuum
level implies that the continuum has been removed during the process of
starlight subtraction. The flux is measured in
    $10^{-18}$ erg cm$^{-2}$ s$^{-1}$ \AA$^{-1}$ while the wavelength is in
    \AA\ units.}
\end{figure*}

In 18 sources, all of them FR~II galaxies, we detected a broad \Ha\
component. Broad lines can be very strong, as in 3C~273, or, as in 3C~184.1,
they become clearly visible only after the subtraction of the stellar light.
Fig. \ref{blr} shows all spectra showing the H$\alpha$ broad components: the
dashed line, representing the continuum emission, corresponds to the zero flux
level when the SSP has been subtracted, otherwise it represents the AGN power
law emission.

In some cases they are symmetric and centered on the narrow component
(3C~287.1), but more often they are highly red or blue shifted (3C~184.1,
3C~227). In most objects the \Ha\ line presents a highly irregular profile,
with `bumps' in the line wings. For example, in 3C~197.1 and 3C~445 the broad
component has a step-shaped emission in the red wing, while in 3C~227 the
step-shaped emission is on the blue side of the line. We also confirm the
`double-humped' shape of the BLR of 3C~332 \citep{eracleous94}.

The technique used to measure the narrow line fluxes (described in
Sect. \ref{line}) 
provides us also with flux estimates for the broad components.
In the cases where no satisfactory fit was obtained with multiple gaussians, we
measured the flux by directly integrating the area defined by the broad
component. The \Ha\ broad line fluxes are reported in the last column of 
Table \ref{bigtable}. 

We will explore in more detail the issue of the presence and properties of
broad lines in a forthcoming paper.

\section{Summary}
\label{summary}

We presented a homogeneous and 92 \% complete dataset of optical nuclear
spectra of the 113 3CR radio sources with redshift $<$ 0.3. The data were
obtained mostly from observations performed at the Telescopio Nazionale
Galileo and are complemented with SDSS spectra, available for 18 galaxies.
For each source we obtained a low (20 \AA) resolution spectrum covering the
$\sim$3500-8000 \AA\ spectral range and a higher (5 \AA) resolution spectrum
centered on the \Ha\ line and $\sim$1700 \AA\ wide.  From the two dimensional
calibrated data, we extracted spectra over a 2$\arcsec$ x 2$\arcsec$ nuclear
region. A complete atlas of the optical nuclear spectra is presented.

We separated the contribution of the emission produced
by the active nucleus from the galaxy starlight. This has been achieved by
subtracting the best fit model of stellar emission from a grid of single
stellar population templates. We then measured and tabulated the fluxes of the
main emission lines by fitting gaussians to the residual spectra.  We
discussed the effects of the possible mis-match between the galaxies stellar
population and the adopted templates; the conclusion of this analysis is that
they do not affect significantly the line measurements, with only very few
exceptions where the estimate of the \Hb\ flux is compromised.

Only in three radio galaxies did we fail to detect any emission line. 
In most cases ($\sim$90 \% of the sample) 
we were able to measure all the brightest lines, 
e.g. \Hb, [O~III]$\lambda\lambda$4959,5007,
[O~I]$\lambda\lambda$6300,64, \Ha, [N~II]$\lambda\lambda$6548,84,
and [S~II]$\lambda\lambda$6716,31. 

In addition, a broad \Ha\ line is seen in 18 galaxies. They show a wide
variety in terms of line profiles, from well behaved symmetric lines centered
onto the narrow \Ha, to `double humped' lines, but also include highly
irregular shapes, with secondary peaks and steps.
 
The data presented here will be used in follow up papers to address the
connection between the optical spectral characteristics and the
multiwavelength properties of the sample.

\bigskip
{\sl Acknowledgments.} We thank the referee, Katherine Inskip, for her useful
comments and suggestions.
SB and ACe acknowledge the Italian MIUR for financial
support. ACa acknowledges COFIN-INAF-2006 grant financial
support. This research has made use of the NASA/IPAC
Extragalactic Database (NED) which is operated by the Jet Propulsion
Laboratory. California institute of Technology, under contract with the
National Aeronautics and Space Administration.  This research has made use of
NASA's Astrophysics Data System (ADS).  This research has made use of the SDSS
Archive, funding for the creation and distribution of which was provided by
the Alfred P. Sloan Foundation, the Participating Institutions, the National
Aeronautics and Space Administration, the National Science Foundation, the
U.S. Department of Energy, the Japanese Monbukagakusho, the Max Planck
Society, and The Higher Education Funding Council for England. The official
SDSS Web site is www.sdss.org.  The SDSS is managed by the Astrophysical
Research Consortium for the Participating Institutions. The Participating
Institutions are the American Museum of Natural History, Astrophysical
Institute Potsdam, University of Basel, University of Cambridge, Case Western
Reserve University, University of Chicago, Drexel University, Fermilab, the
Institute for Advanced Study, the Japan Participation Group, Johns Hopkins
University, the Joint Institute for Nuclear Astrophysics, the Kavli Institute
for Particle Astrophysics and Cosmology, the Korean Scientist Group, the
Chinese Academy of Sciences (LAMOST), Los Alamos National Laboratory, the
Max-Planck-Institute for Astronomy (MPIA), the Max-Planck-Institute for
Astrophysics (MPA), New Mexico State University, Ohio State University,
University of Pittsburgh, University of Portsmouth, Princeton University, the
United States Naval Observatory, and the University of Washington.

\end{document}